\documentclass[usenatbib]{mnras}
\usepackage{amsmath}
\usepackage{natbib}
\usepackage{epsfig}
\usepackage{graphicx}
\usepackage{amssymb}
\usepackage{url}

\def\gtsima{$\; \buildrel > \over \sim \;$}
\def\ltsima{$\; \buildrel < \over \sim \;$}
\def\gtrsim{\lower.5ex\hbox{\gtsima}}
\def\lesssim{\lower.5ex\hbox{\ltsima}}

\begin{document}

\title[Rotation in young massive star clusters]{Rotation in young massive star clusters}
\author[Michela Mapelli]{Michela Mapelli$^{1}$\thanks{E-mail: michela.mapelli@oapd.inaf.it} 
\\
$^{1}$INAF-Osservatorio Astronomico di Padova, Vicolo dell'Osservatorio 5, I--35122, Padova, Italy {\tt michela.mapelli@oapd.inaf.it}}

\maketitle \vspace {7cm }
\bibliographystyle{mnras}
 
 
\begin{abstract}
Hydrodynamical simulations of turbulent molecular clouds show that star clusters form from the hierarchical merger of several sub-clumps. 
 We run smoothed-particle hydrodynamics simulations of turbulence-supported molecular clouds with mass ranging from $1700$ to $43000$ M$\odot$. We study the kinematic evolution of the main cluster that forms in each cloud. We find that the parent gas acquires significant rotation, because of large-scale torques during the process of hierarchical assembly. The stellar component of the embedded star cluster inherits the rotation signature from the parent gas. Only star clusters with final mass $<$ few $\times{}100$ M$_\odot$  do not show any clear indication of rotation. Our simulated star clusters have high ellipticity ($\sim{}0.4-0.5$ at $t=4$ Myr) and are  subvirial ($Q_{\rm vir}\lesssim{}0.4$). The signature of rotation is stronger than radial motions due to subvirial collapse.  Our results suggest that rotation is  common in embedded massive ($\gtrsim{}1000$ M$_\odot{}$) star clusters. This  might provide a key observational test for the hierarchical assembly scenario.
\end{abstract}



%

\section{Introduction}
Hydrodynamical simulations of turbulent molecular clouds show that star clusters  form from the hierarchical merger of several sub-clumps (e.g. \citealt{bonnell2003,bate2009,federrath2010,bonnell2011,dale2011,girichidis2011}). These sub-clumps, composed of stars and gas, develop preferentially along dense filaments, and then aggregate into dense regions at the junction between filaments. 

Whether real-life star clusters develop hierarchically or monolithically  is still matter of debate. The simulations by \cite{banerjee2015} support a monolithic formation or a prompt ($<1$ Myr) assembly for the young dense star cluster NGC~3603. In contrast, the {\it Hubble Space Telescope} data of 30 Doradus indicate the existence of two distinct stellar populations, which might be the signature of an ongoing merger \citep{sabbi2012}. Overall, the properties of simulated star clusters deserve to be further investigated and compared against observations, to probe the robustness of the hierarchical formation scenario.

In this paper, we discuss a set of high-resolution simulations of star cluster formation from turbulent molecular clouds.  In our simulations, star clusters form from the hierarchical assembly of several clumps. We show that most star clusters, especially the most massive ones, develop rotation as a consequence of the torques arising from hierarchical merging.

Rotation was found to be common in globular clusters \citep{pryor1986,vanleeuwen2000,anderson2003,pancino2007,anderson2010,bellazzini2012,bianchini2013,fabricius2014,kimming2015,lardo2015,lee2015}. Several theoretical studies investigate the origin and discuss the importance of rotation for models of globular clusters \citep{bekki2010,mastrobuono2013,vesperini2014,bianchini2015,gavagnin2016,mastrobuono2016}. In particular, \cite{vesperini2014} show that globular cluster rotation might arise as a consequence of violent relaxation in the tidal field of the host galaxy, while \cite{gavagnin2016} suggest that it may be the result of the merger between proto-globular clusters.

Signatures of rotation have been claimed also in few young  (e.g. R136 in the Large Magellanic Cloud, \citealt{henault2012}) and intermediate-age star clusters (e.g. NGC~1846, \citealt{mackey2013}, GLIMPSE-CO1, \citealt{davies2011}). Other observations highlight the existence of velocity gradients in young star forming regions, which might  be connected with rotation  (e.g. \citealt{andre2007,rosolowsky2008,tobin2009,cottaar2015,tsitali2015,rigliaco2016}).

In agreement with other recent theoretical studies (e.g. \citealt{leehennebelle2016a,leehennebelle2016b}), our results show that rotation should be very common in the first stages of star cluster formation, and suggest future observational campaigns should look for this signature. This paper is organized as follows. In Section~2, we describe the methodology; our results are presented in Section~3; in Section~4 we discuss our results, and in Section~5 we draw our conclusions.

\begin{figure}
  \center{
    \epsfig{figure=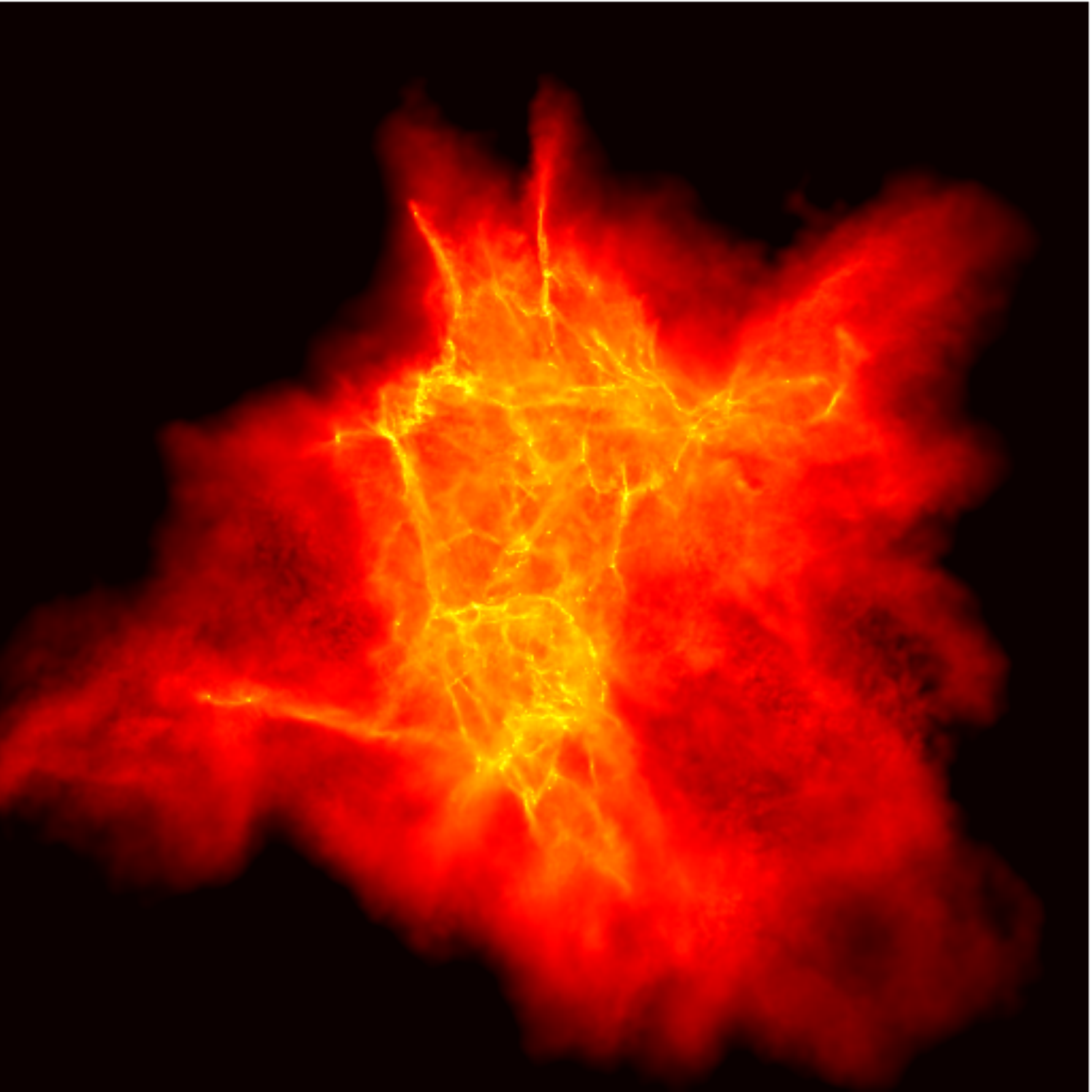,width=8.0cm} 
    \epsfig{figure=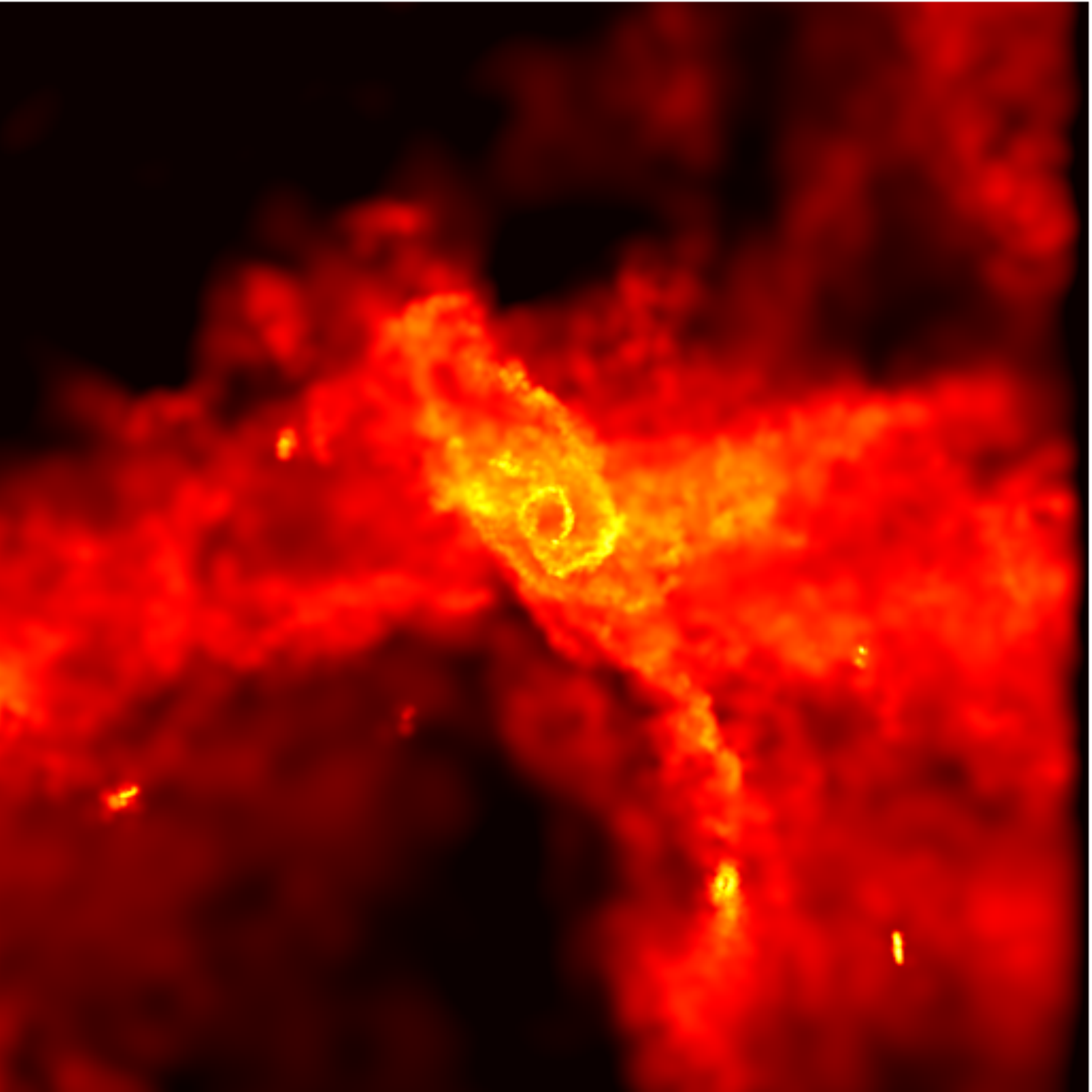,width=8.0cm} 
  \caption{  \label{fig:fig1}
Projected density of gas in run~A at $t=2.5$ Myr. Top: entire cloud (the box measures $40\times{}40$ pc). The colour-coded map (in logarithmic scale) ranges from $2.2\times{}10^{-4}$ to 22.2 M$_\odot$ pc$^{-3}$. Bottom: zoom of the region where the main star cluster is going to form (the box measures $0.8\times{}0.8$ pc). The colour-coded map ranges from $2.2\times{}10^{-2}$ to 70.2 M$_\odot$ pc$^{-3}$. 
}}
\end{figure}

\begin{figure*}
  \center{
    \epsfig{figure=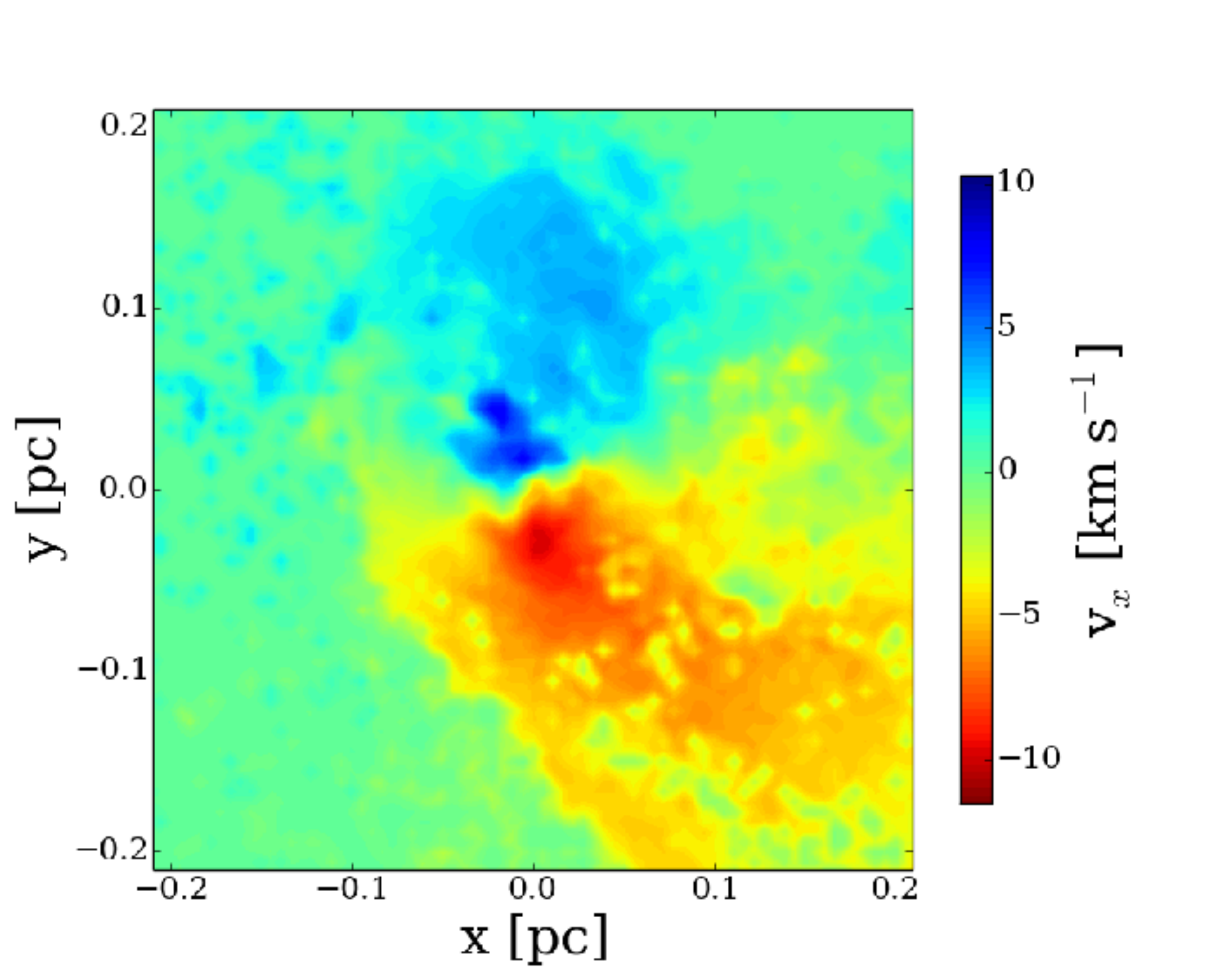,width=5.5cm} 
    \epsfig{figure=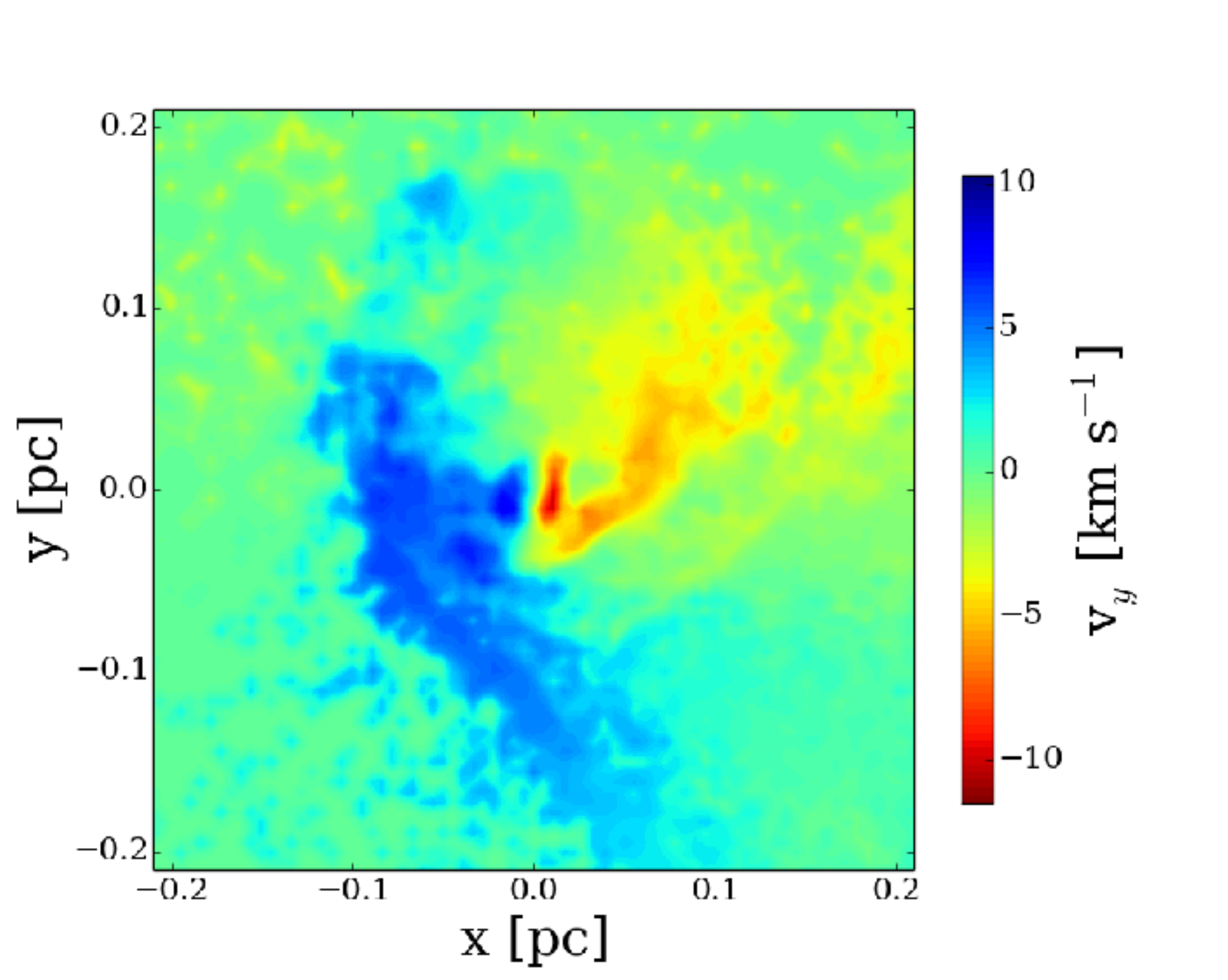,width=5.5cm}
    \epsfig{figure=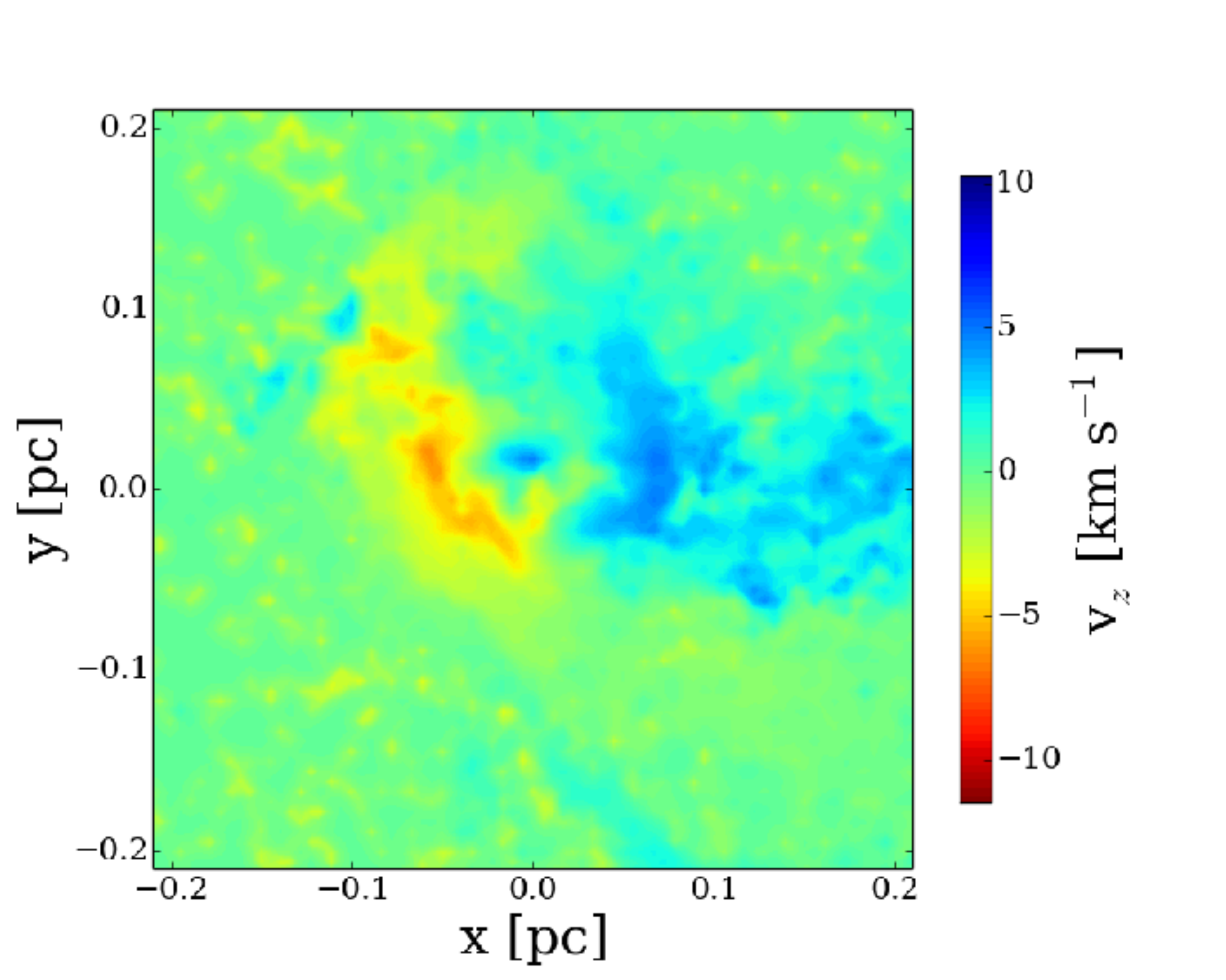,width=5.5cm}
    \epsfig{figure=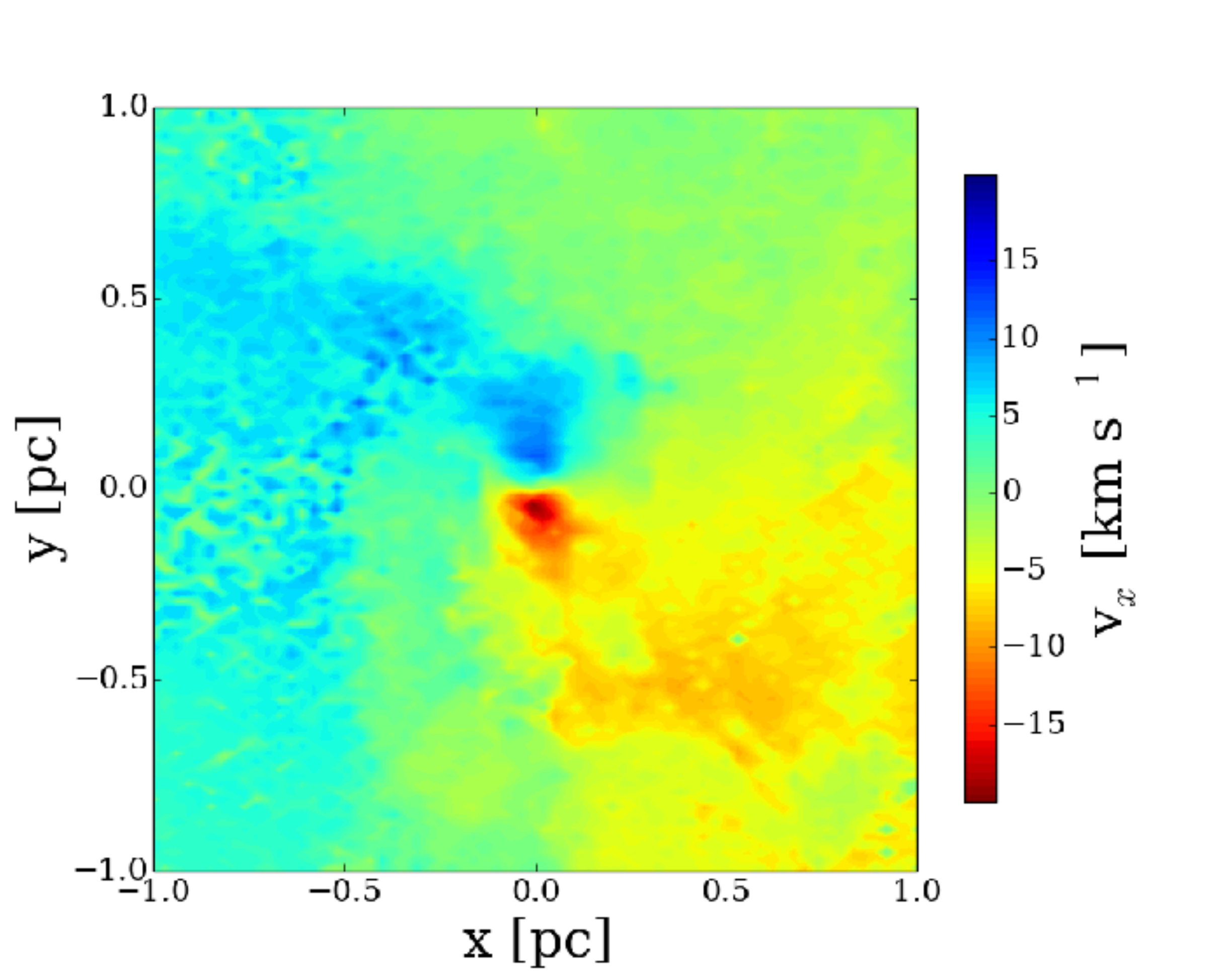,width=5.5cm} 
    \epsfig{figure=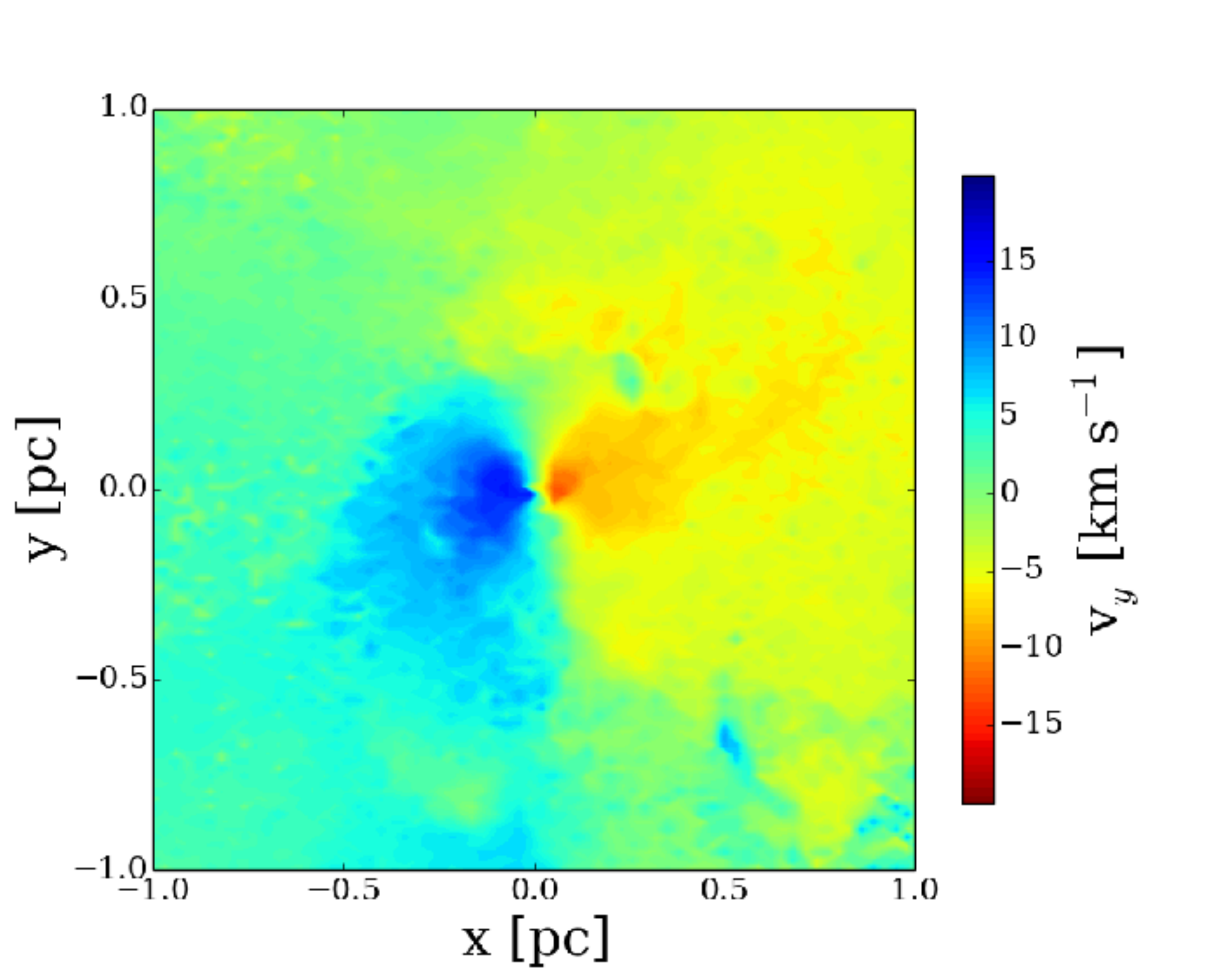,width=5.5cm}
    \epsfig{figure=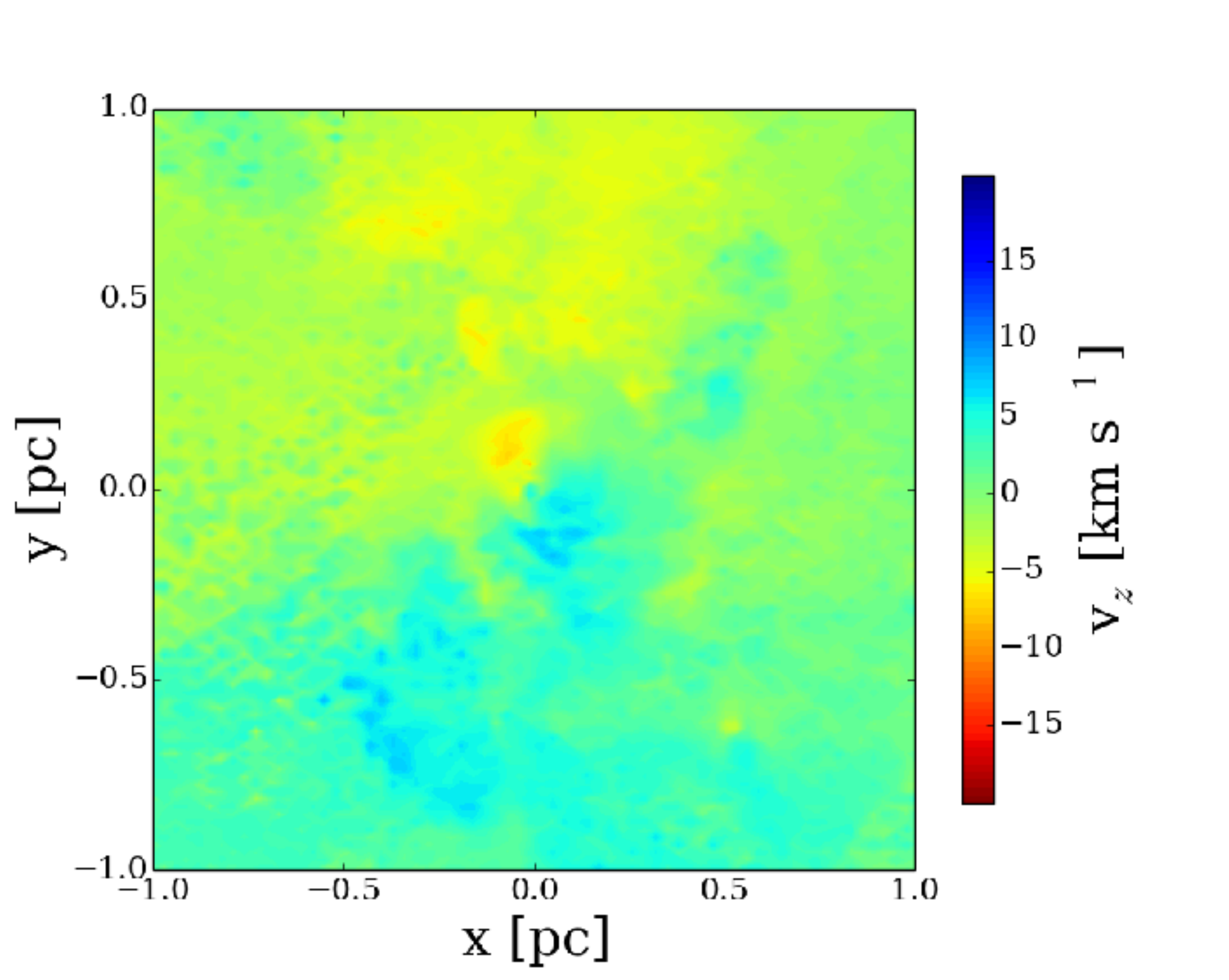,width=5.5cm}
  \caption{  \label{fig:fig2}
Contour plot of the velocity of gas particles in run~A at $t=2.5$ Myr (top panels) and $t=4.0$ Myr (bottom panels). The $xy$ plane is defined as the plane perpendicular to the total angular momentum vector of stars. From left to right: the colour-coded map shows the component of the gas velocity along the $x$, $y$, and $z$ axis  (where the $z$ axis is the direction of the angular momentum vector), respectively. 
}}
\end{figure*}

\section{Methods}
We generate models of turbulence-supported molecular clouds. The clouds are initially spherical, with homogeneous density. They are seeded with supersonic turbulent velocities and marginally self-bound (see \citealt{hayfield2011,mapelli2012,mapelli2013,mapelli2016}). To simulate interstellar turbulence, the velocity field of the cloud is generated on a grid as a divergence-free Gaussian random field with an imposed power spectrum $P(k)$, varying as $k^{-4}$. This yields a velocity dispersion $\sigma{}(l)$, varying as $l^{1/2}$, chosen to agree with the \cite{larson1981} scaling relations. 

\begin{table}
\begin{center}
\caption{\label{tab:tab1} Initial conditions.} \leavevmode
\begin{tabular}[!h]{llll}
\hline
Run
&
$M$ [M$_\odot{}$]
&
$R$ [pc]
&
$N_{\rm gas}$ $[\times{}10^6]$
\\
\hline
A & $4.3\times{}10^4$ & 8.8 & 10\\
B & $4.3\times{}10^4$ & 8.8 & 10\\
C & $1.0\times{}10^4$ & 5.4 & 10\\
D & $1.7\times{}10^3$ & 3.0 & 2\\

\hline
\end{tabular}
\end{center}
\begin{flushleft}
\footnotesize{Column~1: run name; column~2: total mass of gas ($M$); column~3: cloud radius ($R$); column~4: number of equal-mass gas particles ($N_{\rm gas}$).}
\end{flushleft}
\end{table}

\begin{figure*}
  \center{
    \epsfig{figure=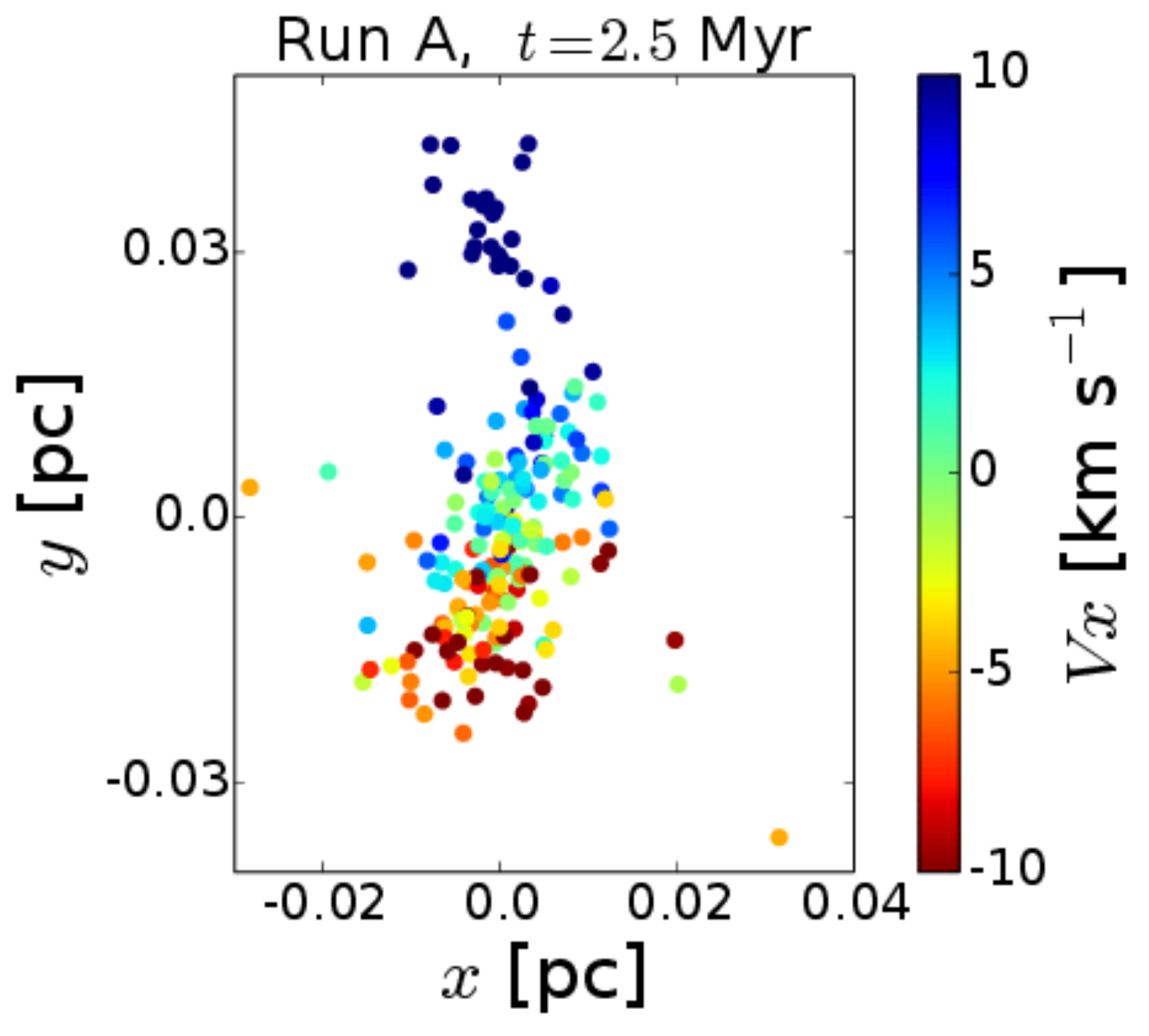,width=5.0cm} 
    \epsfig{figure=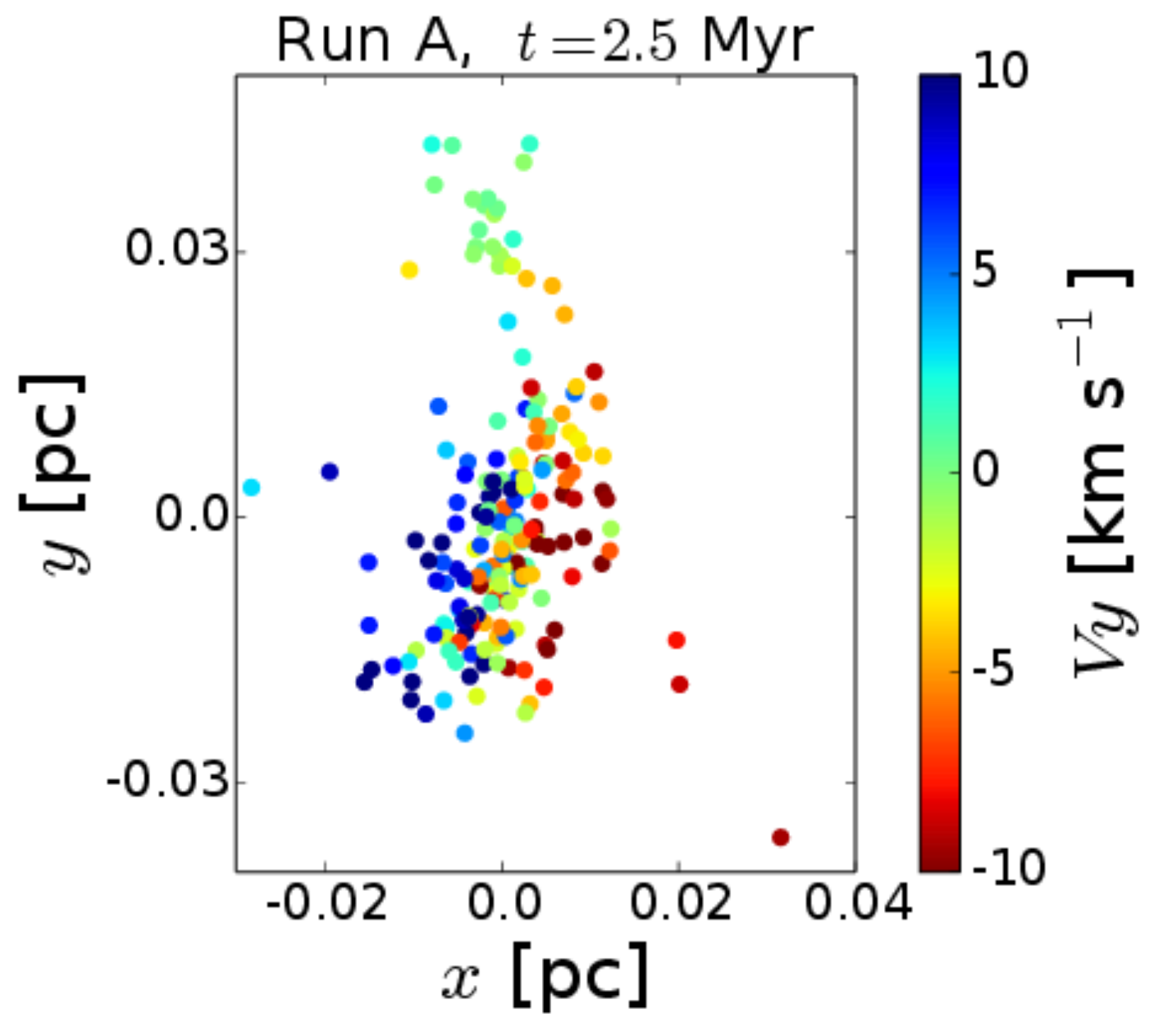,width=5.0cm}
    \epsfig{figure=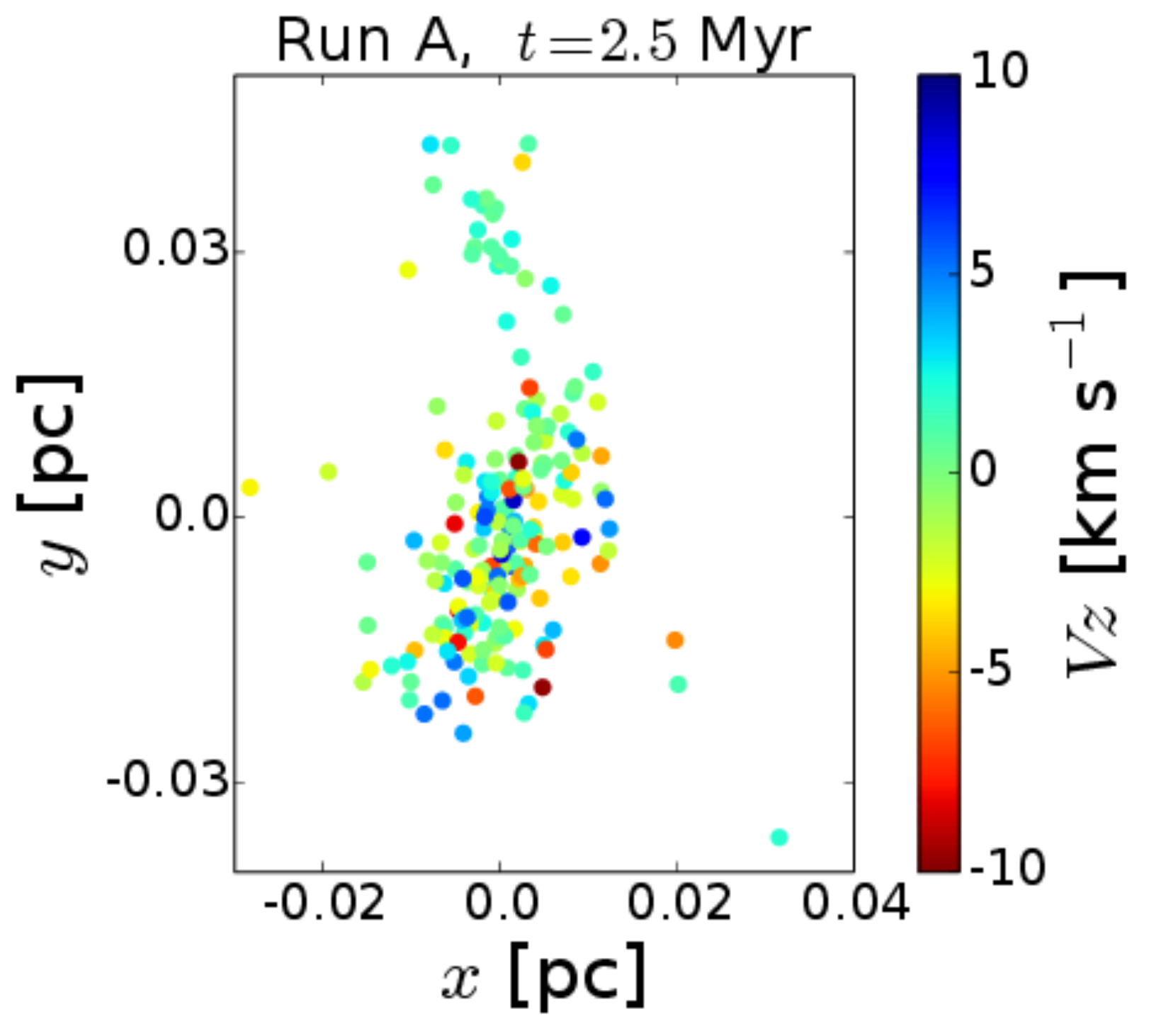,width=5.0cm}  
    \epsfig{figure=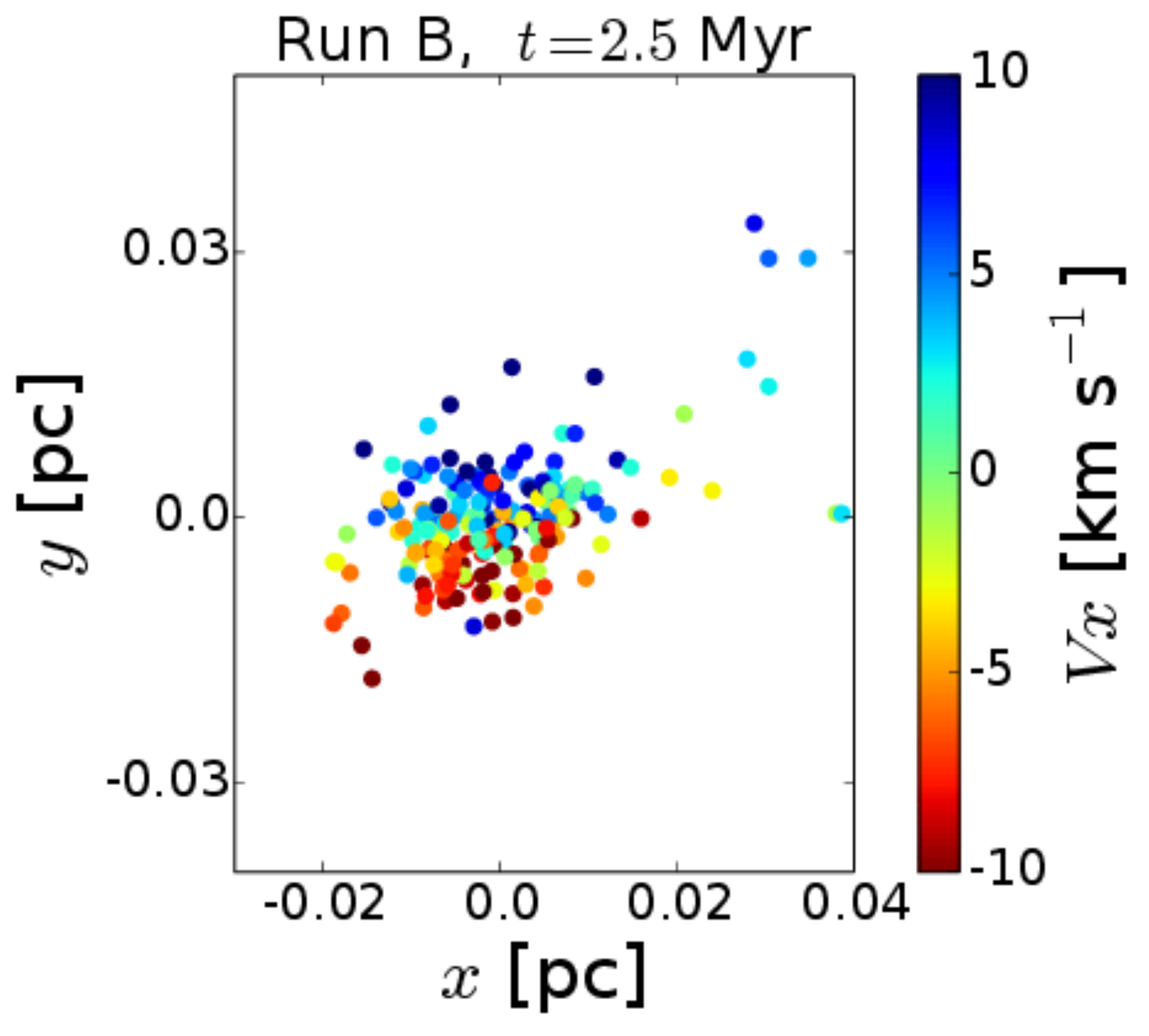,width=5.0cm} 
    \epsfig{figure=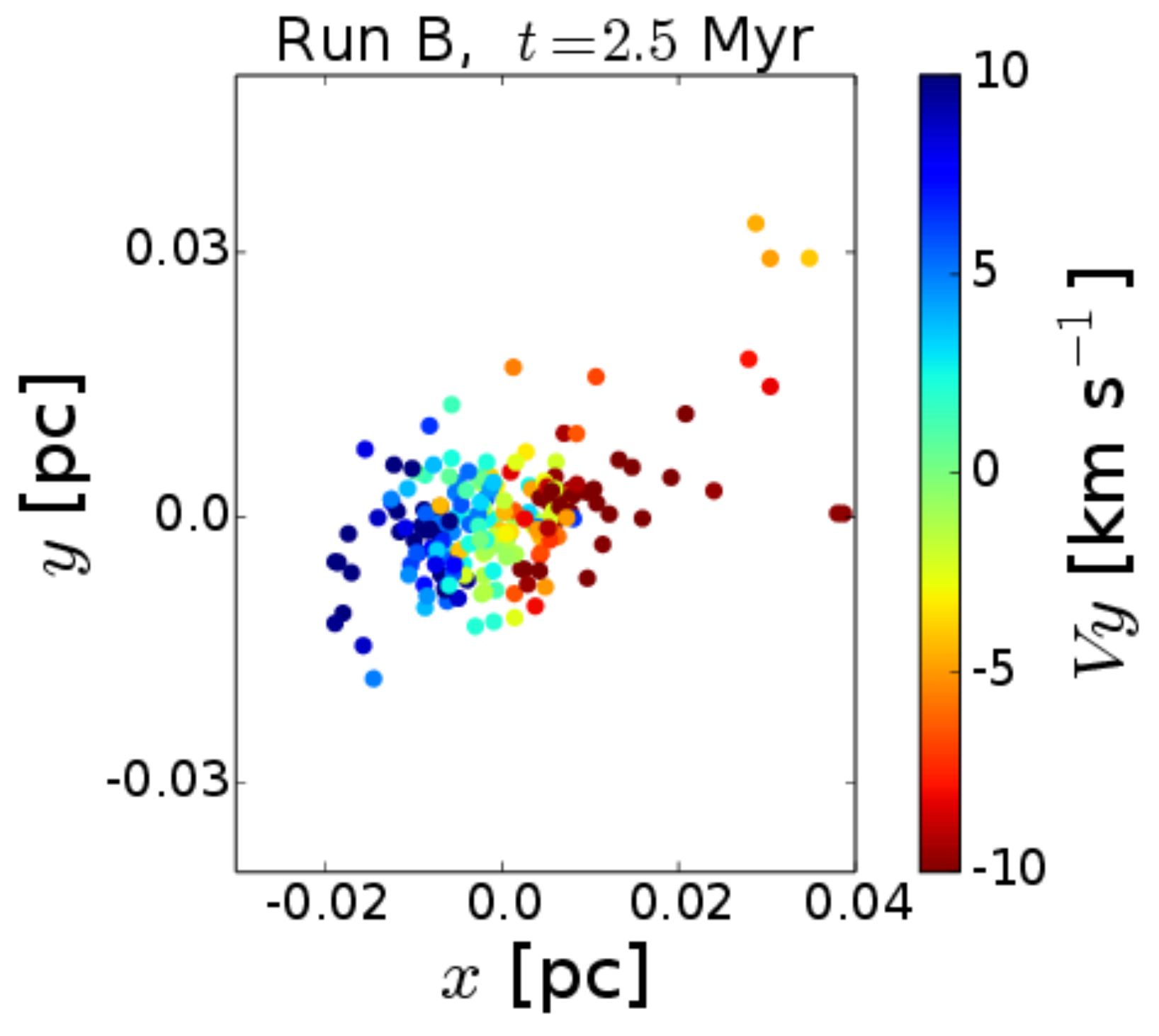,width=5.0cm}
    \epsfig{figure=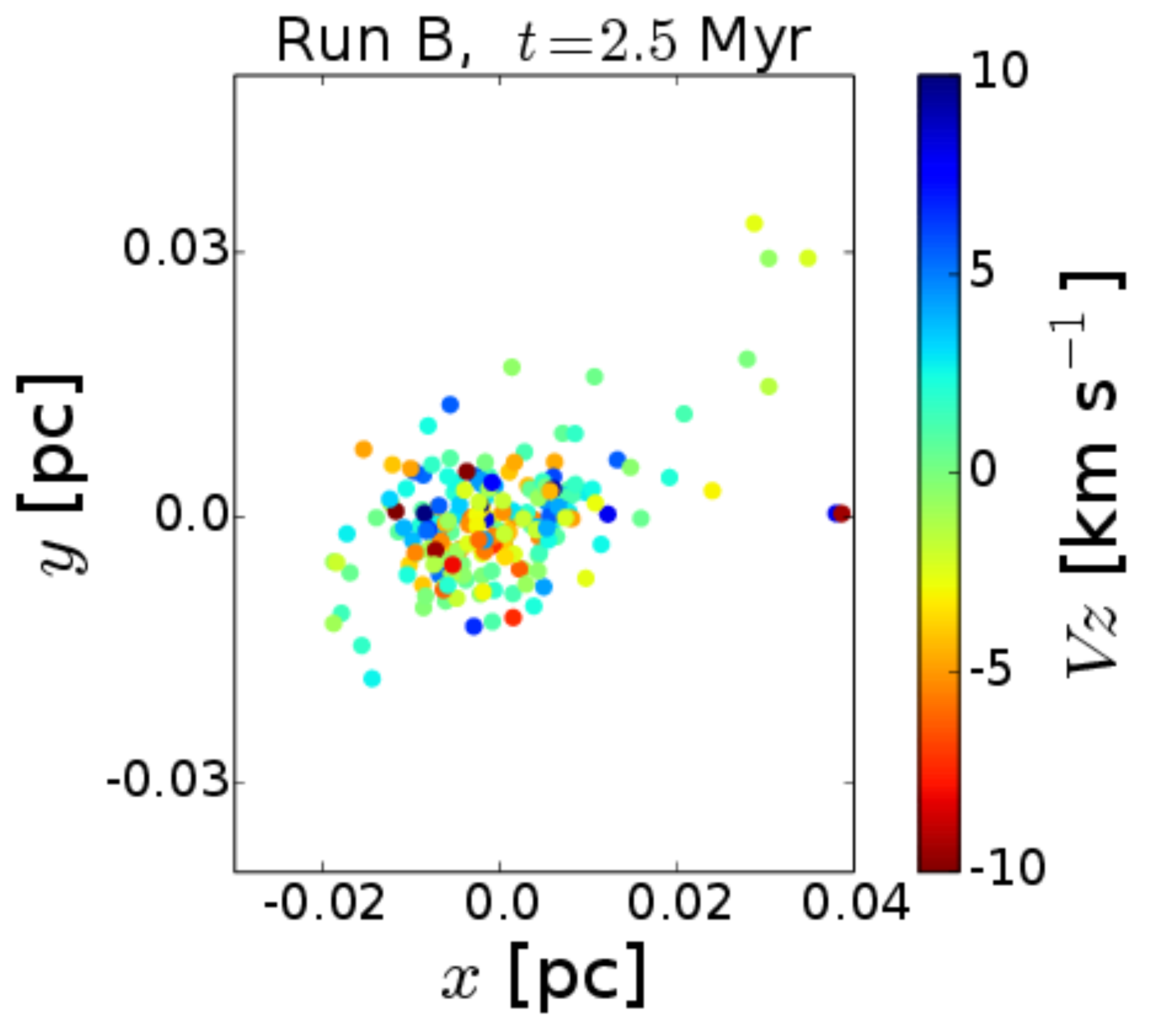,width=5.0cm}  
    \epsfig{figure=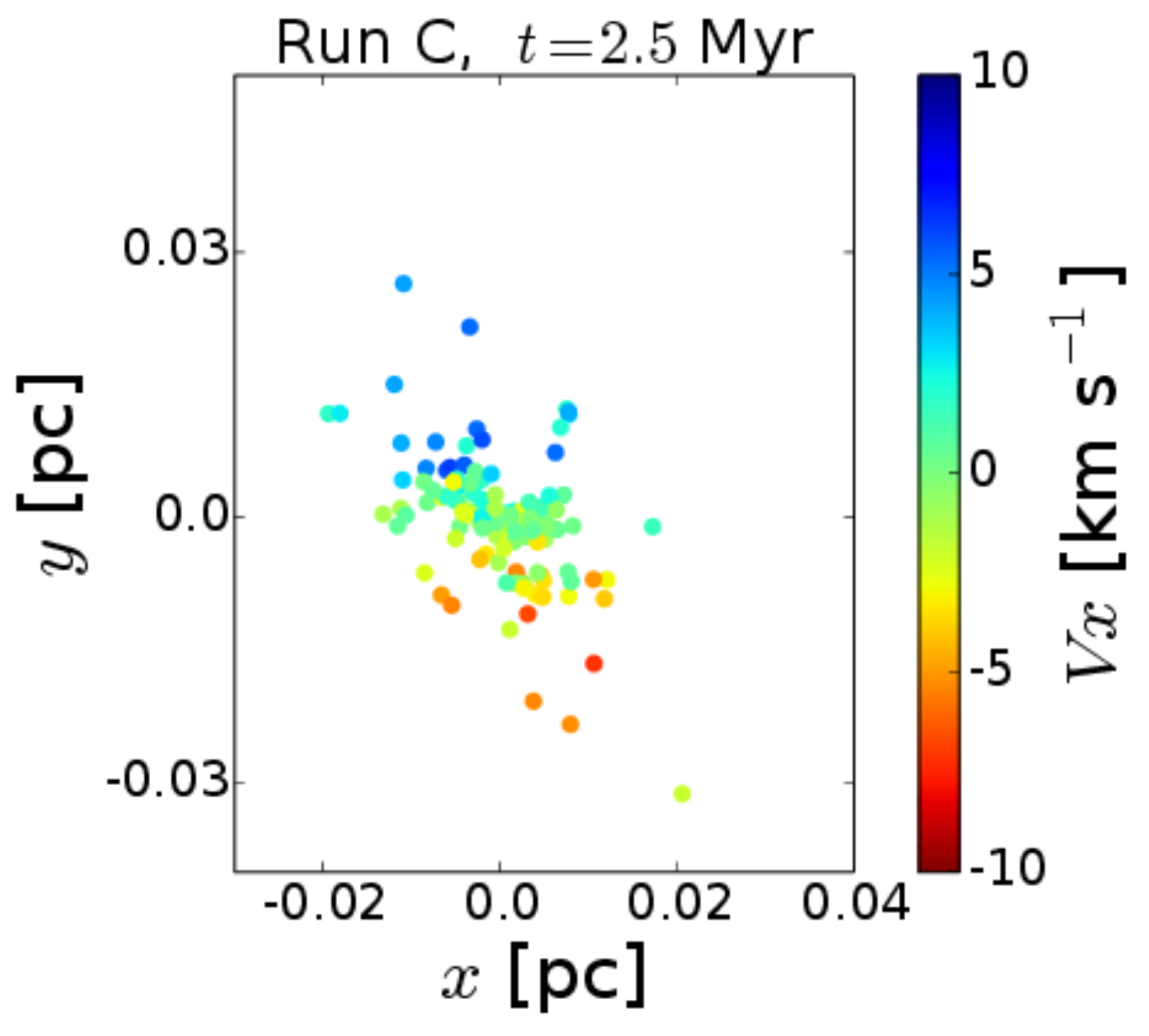,width=5.0cm} 
    \epsfig{figure=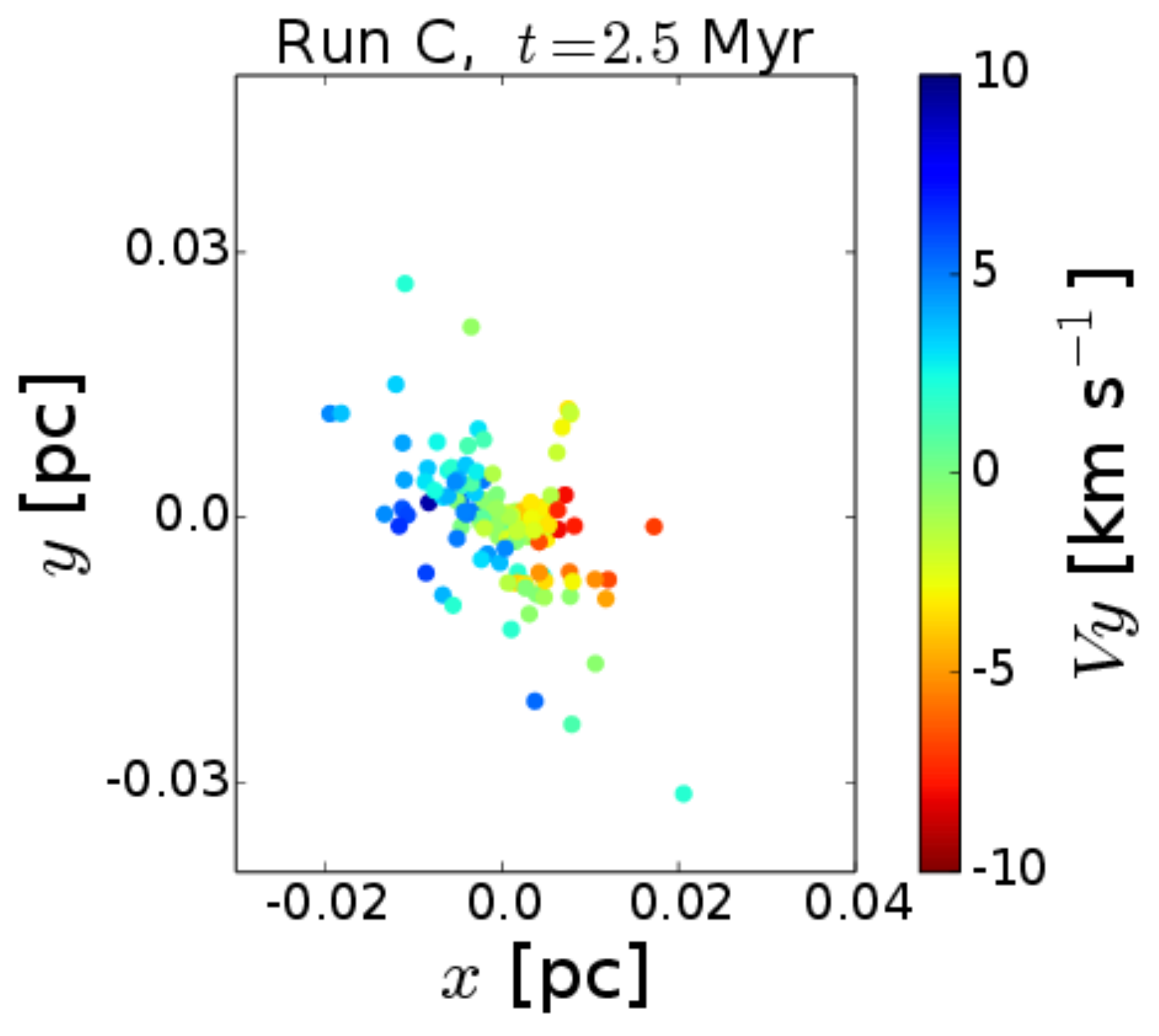,width=5.0cm}
    \epsfig{figure=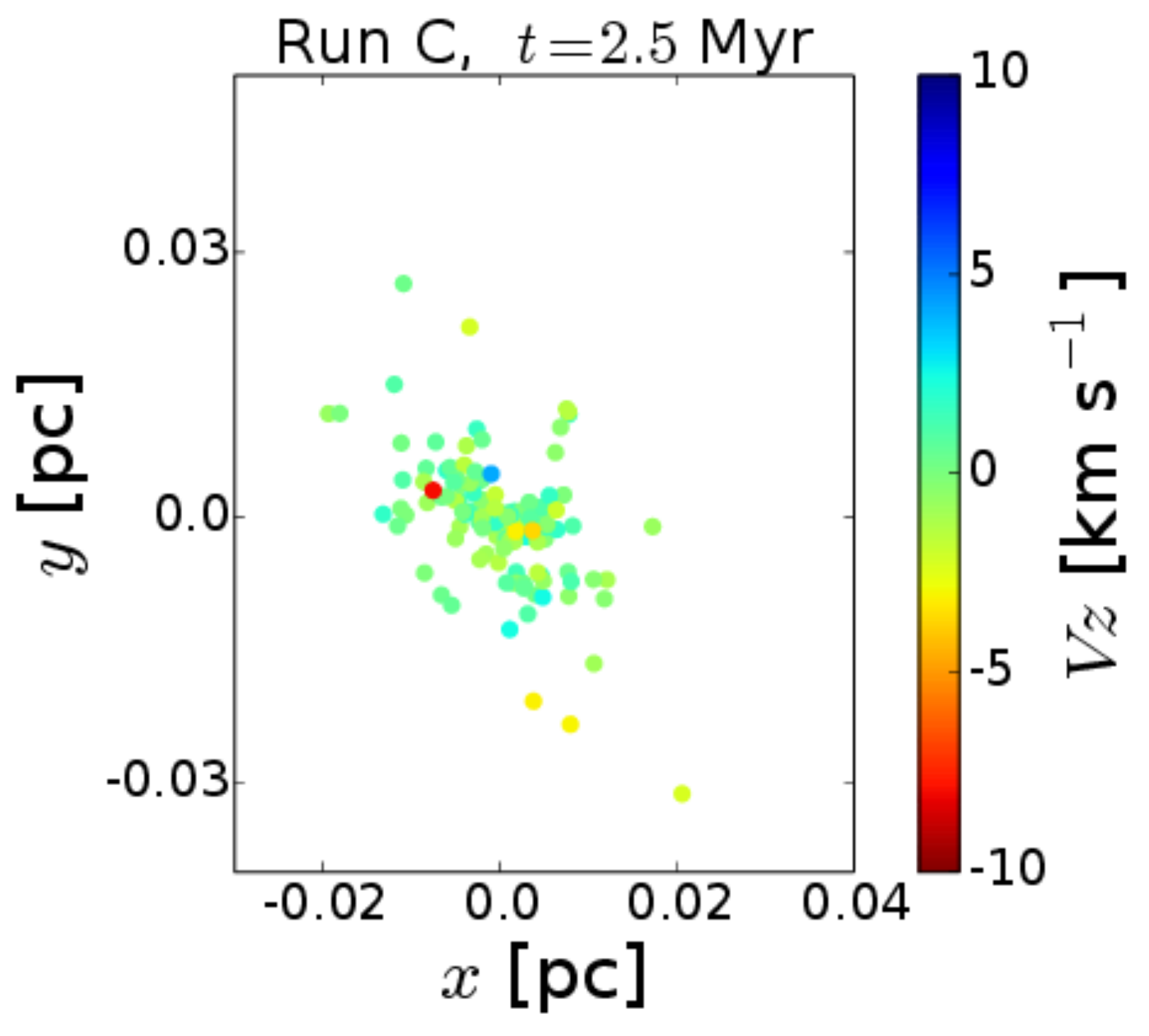,width=5.0cm}

  \caption{  \label{fig:fig3}
Simulated star particles (i.e. sink particles) in the $xy$ plane at $t=2.5$ Myr. The $xy$ plane is defined as the plane perpendicular to the total angular momentum vector. From left to right: the colour-coded map shows the component of the stellar velocity along the $x$, $y$, and $z$ axis (where the $z$ axis is the direction of the angular momentum vector), respectively. From top to bottom: run~A, B, and  C. Run~D is not shown because the number of stars at $t=2.5$ Myr is too small to define a star cluster.
}}
\end{figure*}

\begin{figure}
  \center{
    \epsfig{figure=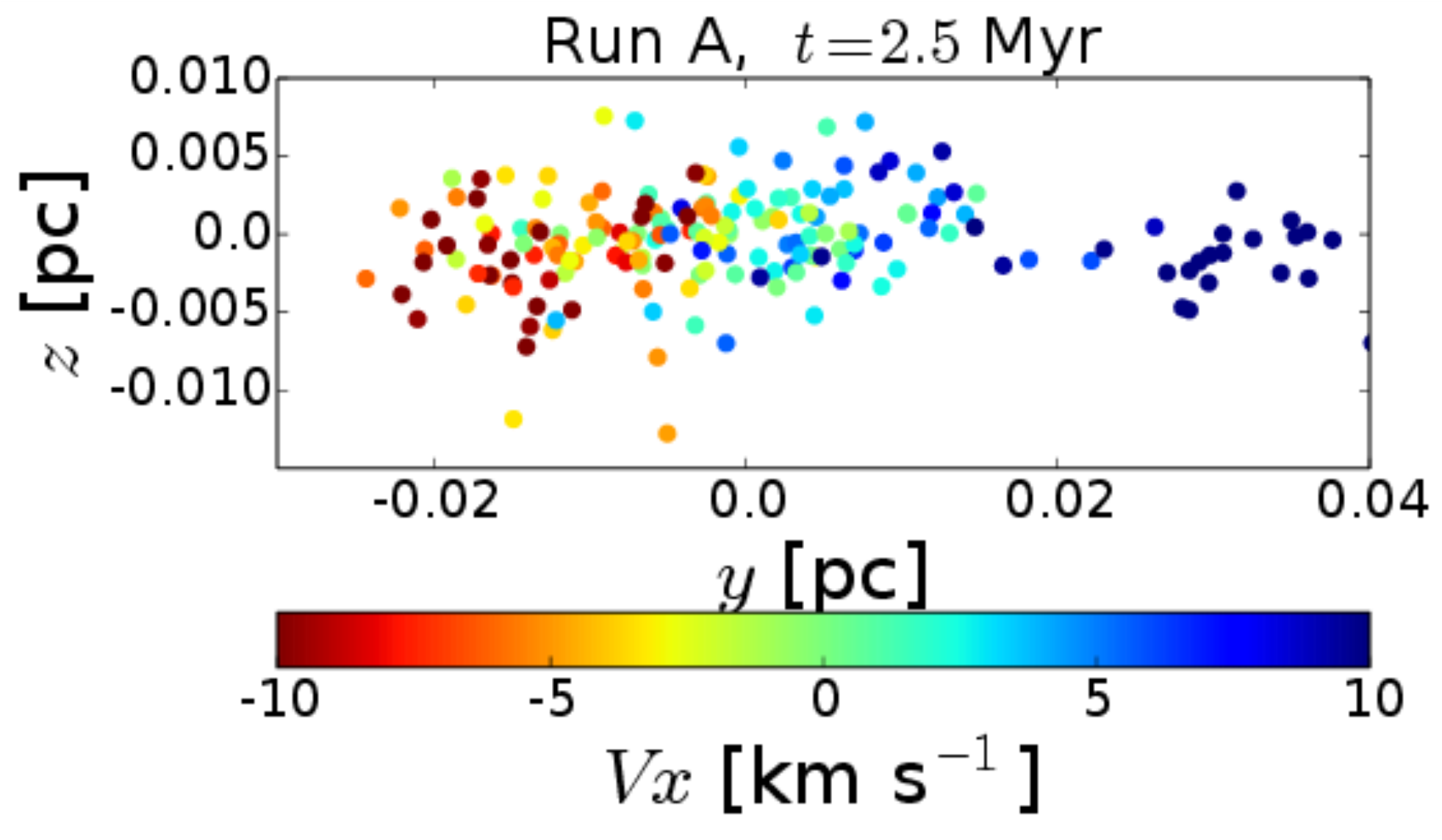,width=6.0cm} 
    \epsfig{figure=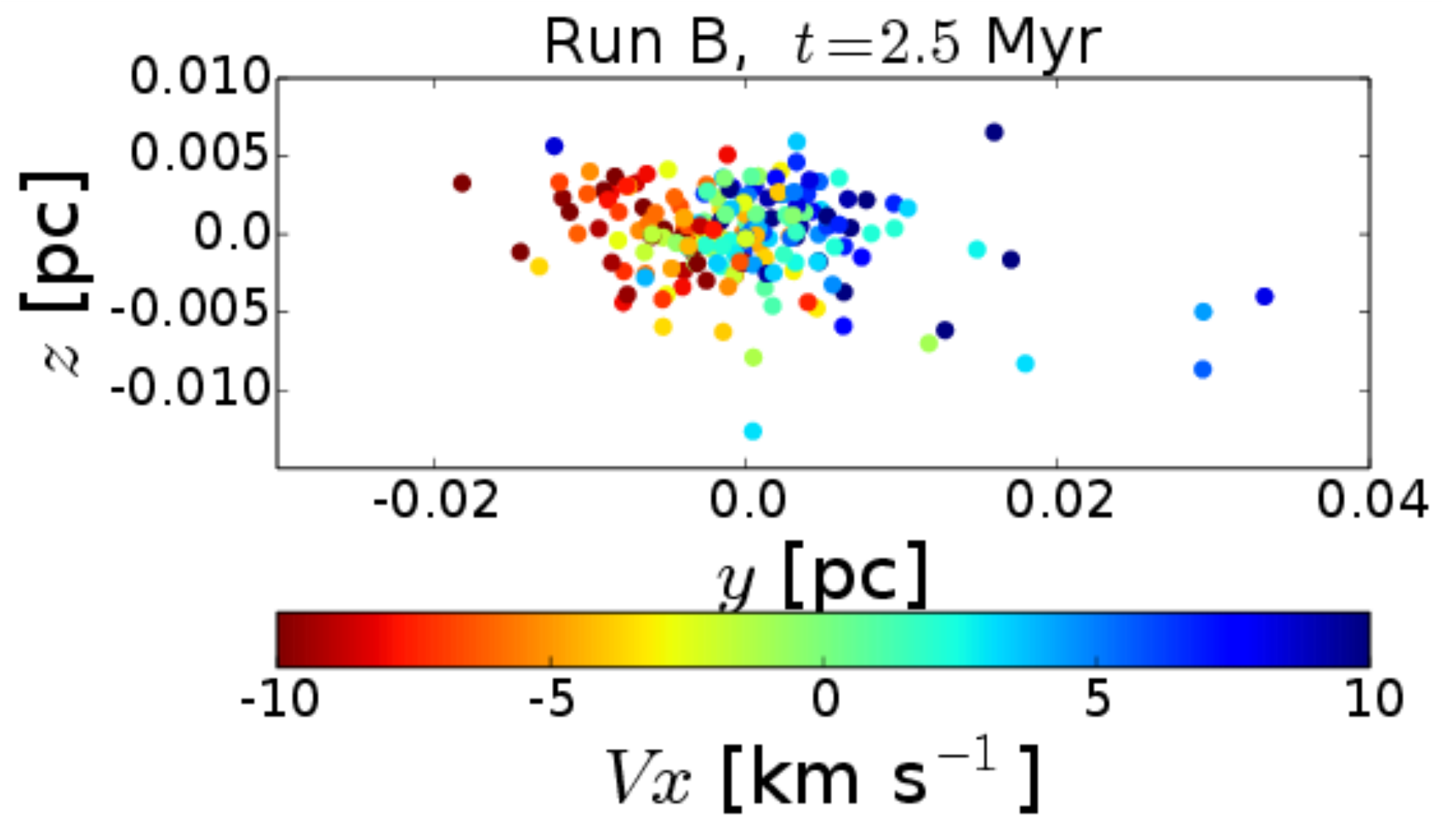,width=6.0cm}
    \epsfig{figure=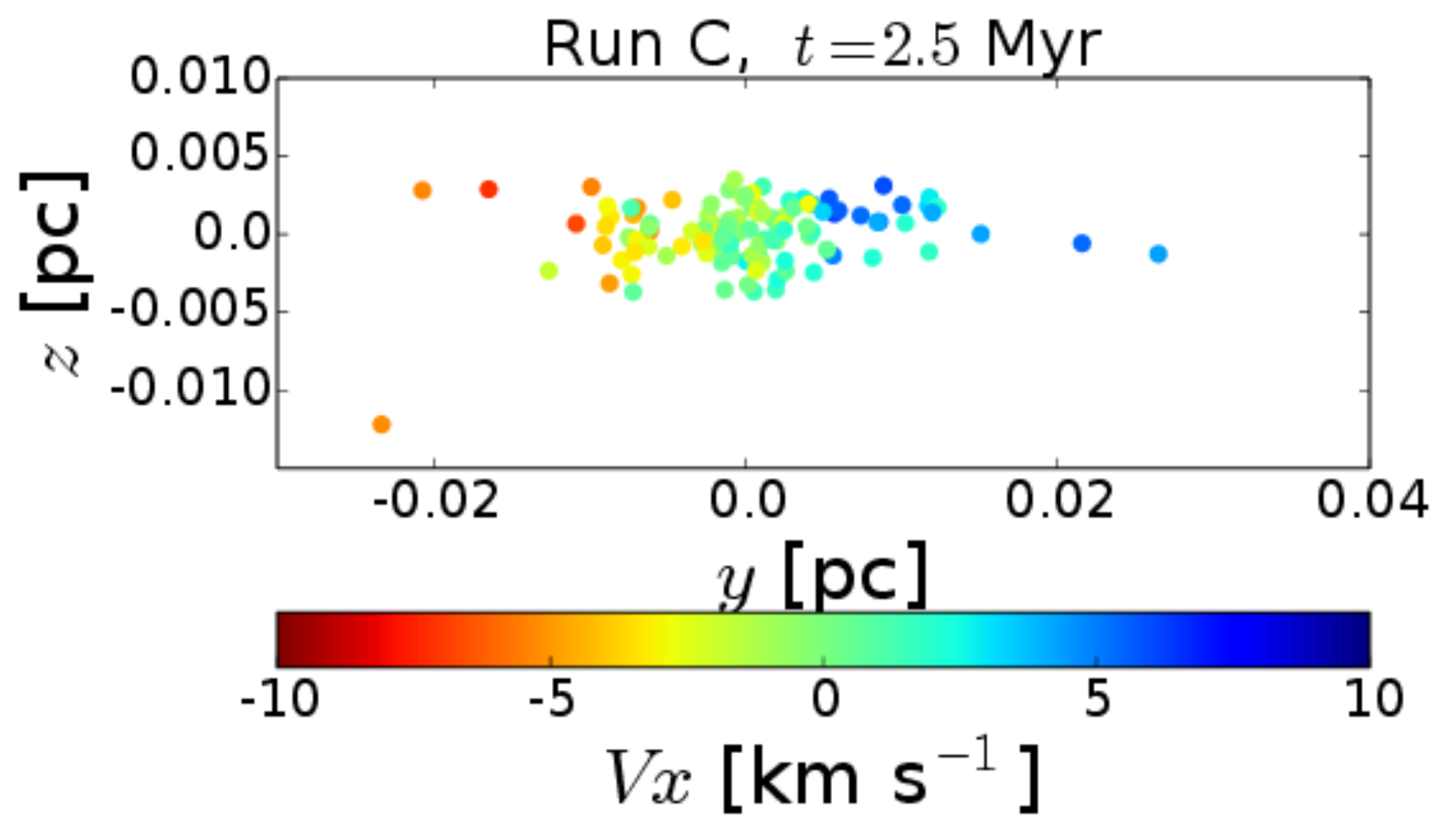,width=6.0cm}
  \caption{  \label{fig:fig4}
Simulated star particles (sink particles) in the $yz$ plane at $t=2.5$ Myr. The colour-coded map shows the component of the stellar velocity along the $x$ axis. From top to bottom: run~A, B, and C. Run~D is not shown because the number of stars is too small to define a star cluster.
}}
\end{figure}

The clouds are simulated with the smoothed-particle hydrodynamics (SPH) code {\sc gasoline} \citep{wadsley2004}, upgraded with the \cite{read2010} optimized SPH (OSPH) modifications, to address the SPH limitations outlined by \cite{agertz2007}. We include radiative cooling in all our simulations. The radiative cooling algorithm is the same as that described in \citet{boley2009} and in \citet{boley2010}.  According to this algorithm, the divergence of the flux is $\nabla{}\cdot{}F=-(36\,{}\pi{})^{1/3}\,{}s^{-1}\sigma{}\left({\rm T}^4-{\rm T}^4_{\rm irr}\right)\,{}(\Delta{}\tau{}+1/\Delta{}\tau{})^{-1}$, where $\sigma{}=5.67\times{}10^{-5}$ erg cm$^{-2}$ s$^{-1}$ K$^{-4}$ is the Stefan's constant, ${\rm T}_{\rm irr}$ is the incident irradiation, $s=\left(m/\rho{}\right)^{1/3}$ and $\Delta{}\tau{}=s\,{}\kappa{}\,{}\rho{}$, for the local opacity $\kappa{}$, particle mass $m$ and density $\rho{}$.
 
\citet{dalessio2001} Planck and Rosseland opacities are used, with a 1~$\mu{}$m maximum grain size.  Such opacities are appropriate for temperatures in the range of a few Kelvins up to thousands of Kelvins. In our simulations, the irradiation temperature is $T_{\rm irr}=10$~K everywhere. The only feedback mechanism we account for is compressional heating of gas. We include no treatment for photoionisation, stellar winds or supernovae, to keep our analysis as simple as possible. The impact of stellar winds on star cluster formation was found to be relatively minor \citep{dale2013,dale2015}. 
Moreover, several studies suggest that star formation proceeds on a timescale which is much faster than  the timescale for photoionising radiation or supernovae to remove the gas \citep{kruijssen2012a,kruijssen2012b,dale2011,dale2012,iffrig2015}. The impact of feedback on our results will be assessed in a follow-up study.

In our simulations, star formation is modelled through the sink-particle technique. Sink particles have been implemented according to the criteria described in \citealt{bate1995} (hereafter B95). We adopt a sink accretion radius $r_{\rm acc}=2\,{}\tilde{\epsilon{}}$, where $\tilde{\epsilon}$ is the softening length ($\tilde{\epsilon}=10^{-4}-10^{-3}$ pc, depending on the resolution of the simulation, see e.g. \citealt{mapelli2012}). A gas particle is considered a sink candidate if its density is above the threshold $\rho{}_{\rm th}=10^{-17}$ g cm$^{-3}$, corresponding to $n_{\rm th}\sim{}10^7$ atoms cm$^{-3}$ (we checked that the sink mass function does not depend on this choice significantly). 
If gas particles inside the accretion radius of the sink candidate satisfy B95 criteria\footnote{B95 criteria for converting gas particles into sink particles require i) that the thermal energy of particles inside $r_{\rm acc}$ is  $E_{\rm th}\le{}0.5\,{}E_{\rm g}$, where $E_{\rm g}$ is the magnitude of the gravitational energy of the particles; ii) that $E_{\rm th}/E_{\rm g} + E_{\rm r}/E_{\rm g}\le{}1$, where $E_{\rm r}$ is the rotational energy of particles; iii) that the total energy of particles  inside $r_{\rm acc}$ is negative.}, then the candidate becomes a sink particle. A similar procedure is followed to decide whether a particle which has already become a sink will accrete gas particles\footnote{According to B95 criteria a gas particle within $r_{\rm acc}$ will be accreted by the sink particle if i) the gas particle is bound to the sink; ii) the specific angular momentum of the particle about the sink is less than required to form a circular orbit at $r_{\rm acc}$; iii) the gas particle is more tightly bound to the considered sink particle than to other sink particles.}. The whole procedure is clearly affected by the resolution of the simulation. We checked that this dependence has no effect on the results of this paper (i.e. the feature of rotation), by running the same simulation with lower resolution. 

We simulate four molecular clouds, with initial mass $M$ spanning from 1700 to $4.3\times{}10^4$ M$_\odot$, and initial radius $R$ in the $3.0-8.8$ pc range (see Table~\ref{tab:tab1}). $R$ and $M$ were chosen so that the initial density of the cloud is always the same $\rho{}\sim{}10^{-21}$ g cm$^{-3}$ (i.e. $n\sim{}250$ cm$^{-3}$ when assuming molecular weight $\mu{}=2.46$). Thus, the free fall time-scale of the simulated clouds is $t_{\rm ff}=[3\,{}\pi{}/(32\,{}G\,{}\rho{})]^{0.5}\sim{}2$ Myr, where $G$ is the gravity constant. Each run was integrated for $\sim{}2\,{}t_{\rm ff}=4$ Myr.

Cloud~A and B are two different realizations (with two different random seeds for the turbulence) of the same molecular cloud, with $M=4.3\times{}10^4$ M$_\odot$ and $R=8.8$ pc. Clouds C and D have $M=10^4$ M$_\odot$ and $1.7\times{}10^3$ M$_\odot$ ($R=5.4$ and 3.0 pc), respectively. Clouds A, B and C were simulated with $10^7$ equal-mass gas particles, while the low-mass cloud D was simulated with $2\times{}10^6$ particles. Thus, the particle mass is always $\lesssim{}4\times{}10^{-3}$ M$_\odot$.

\begin{figure*}
  \center{
    \epsfig{figure=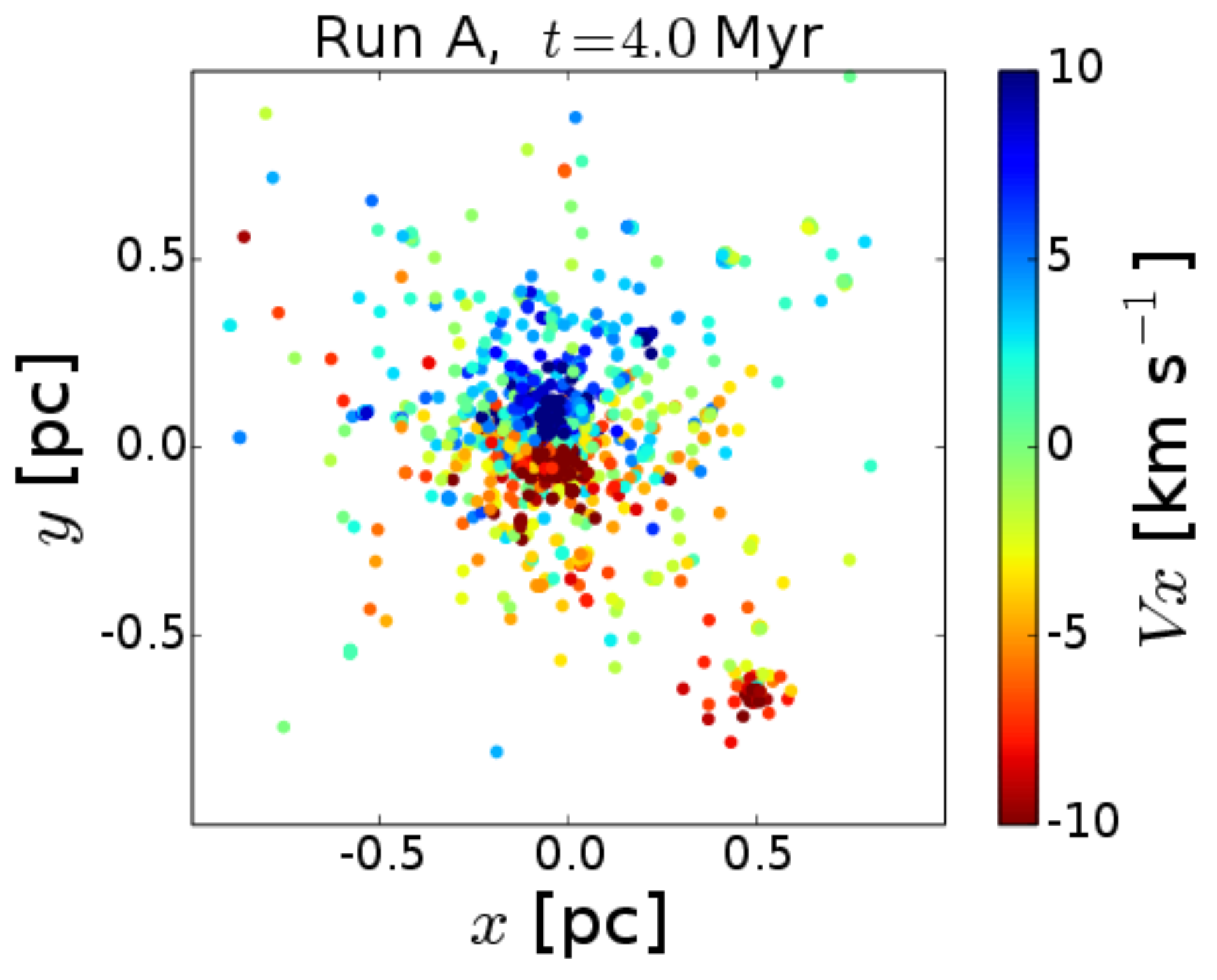,width=5.0cm} 
    \epsfig{figure=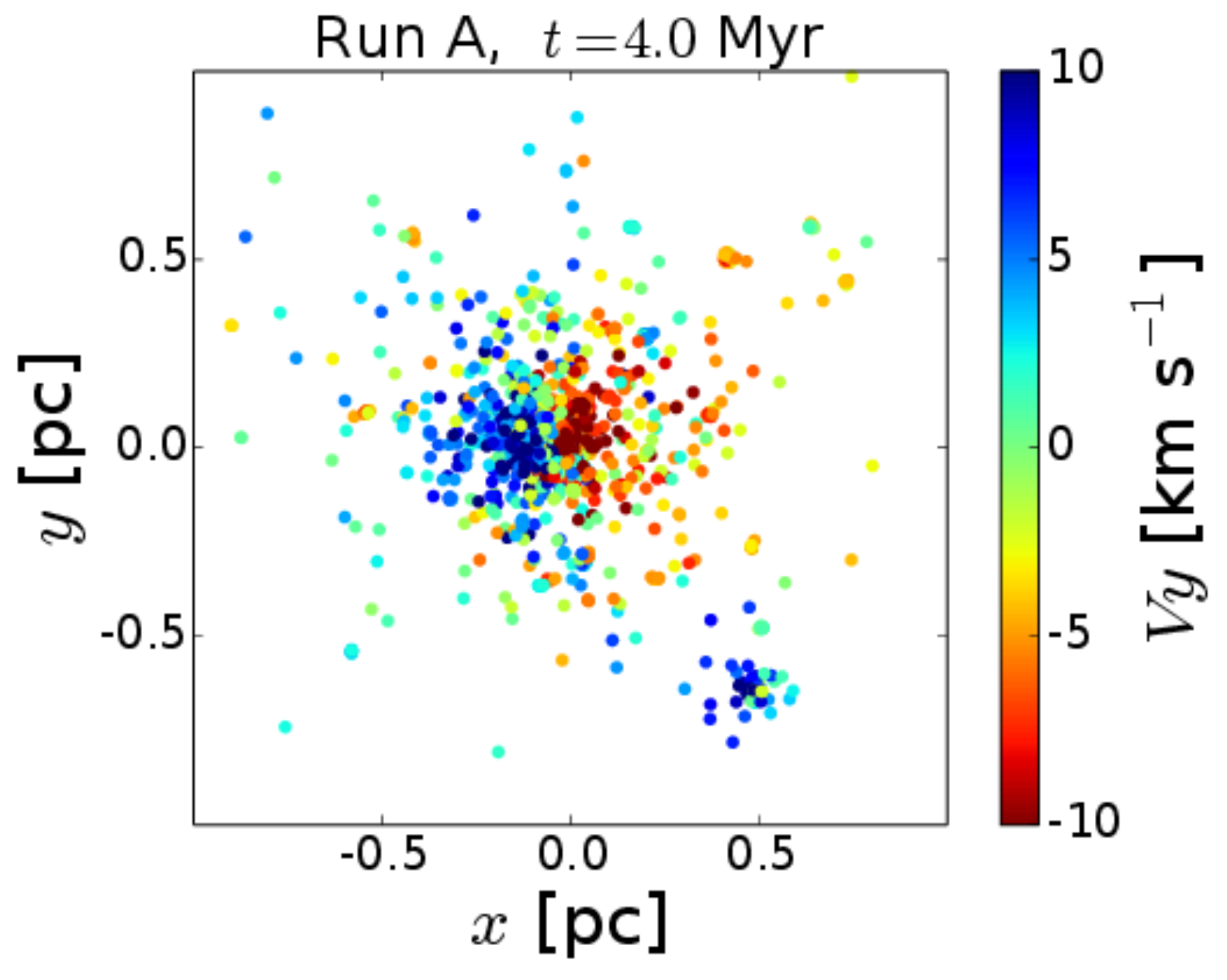,width=5.0cm}
    \epsfig{figure=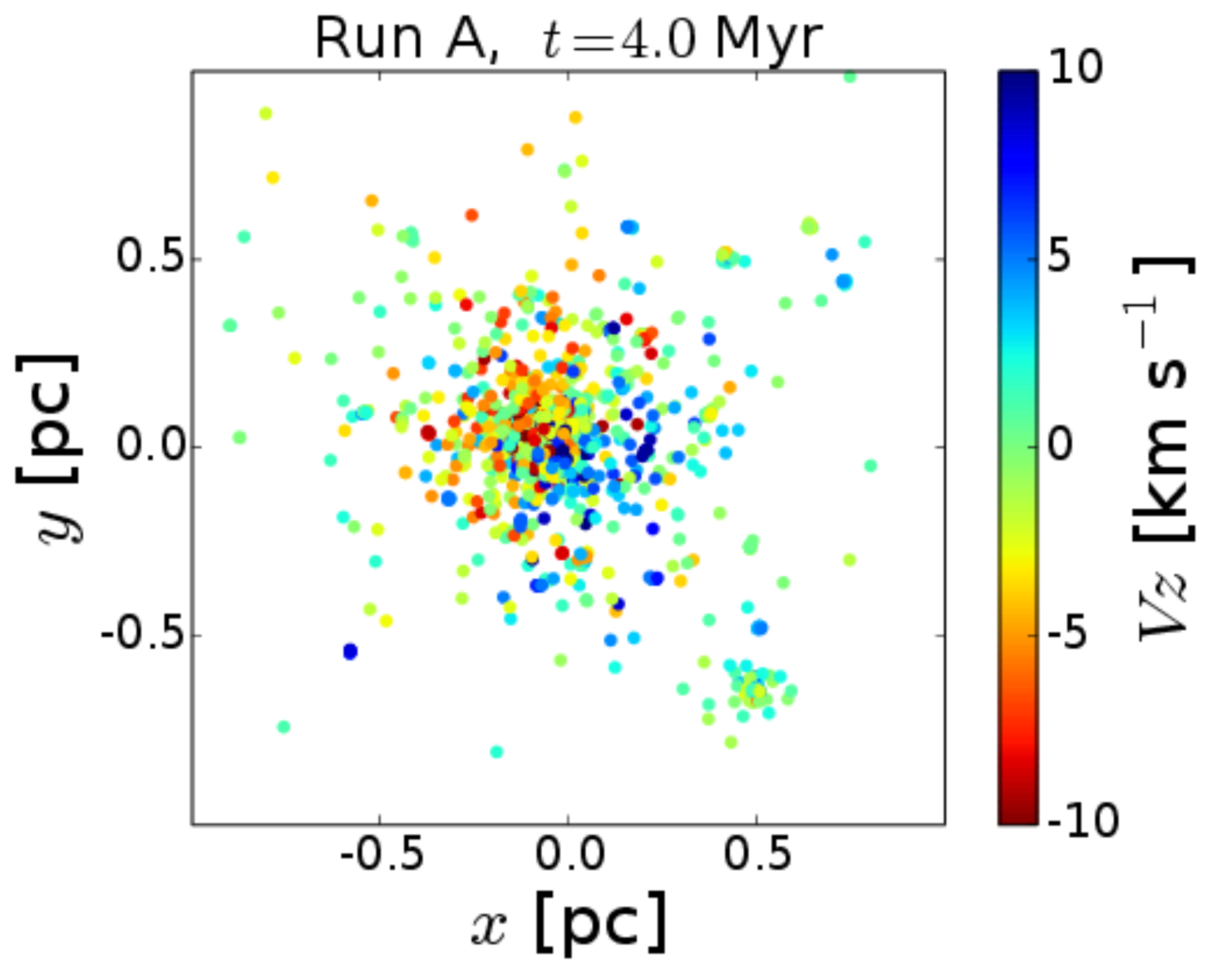,width=5.0cm}  
    \epsfig{figure=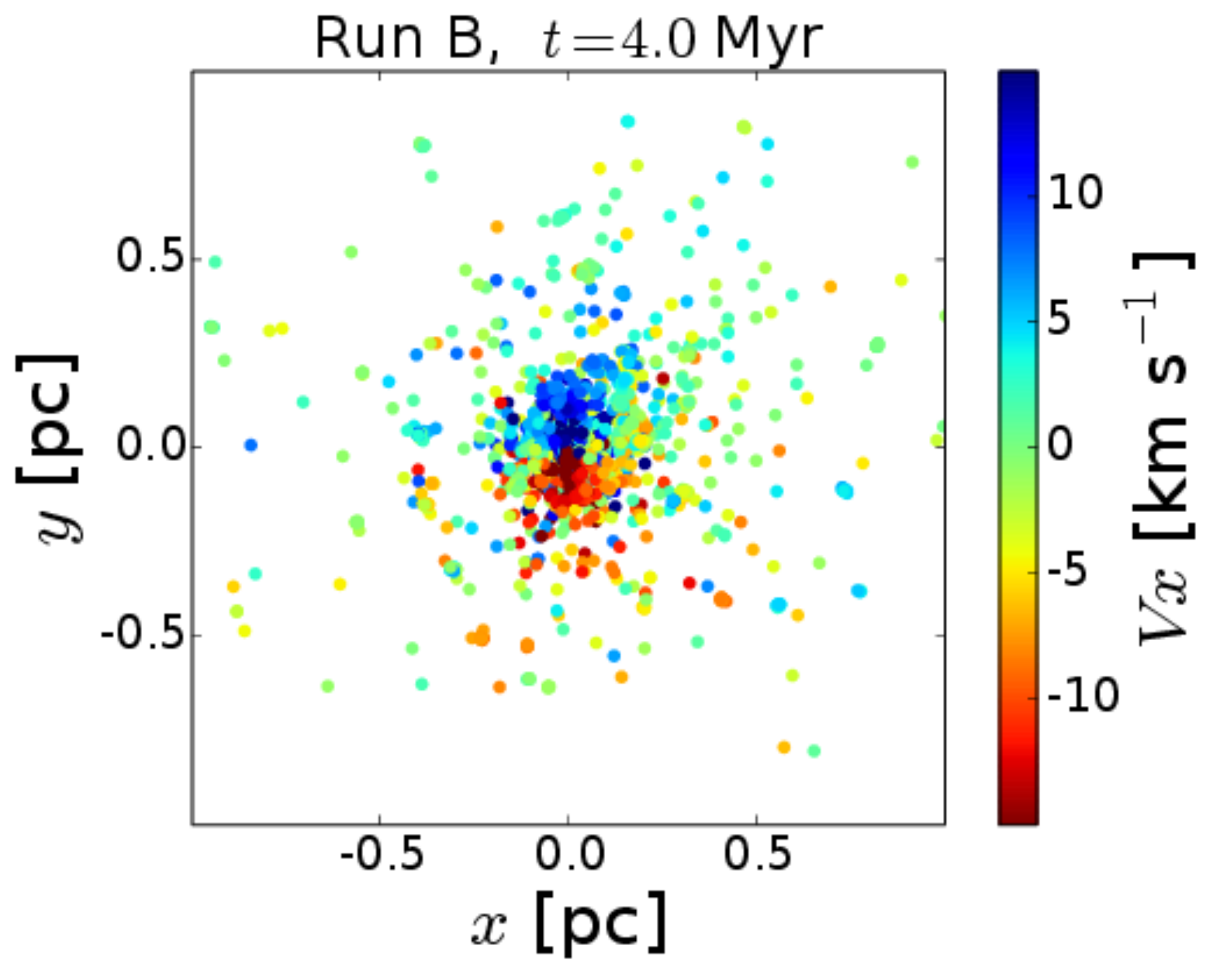,width=5.0cm} 
    \epsfig{figure=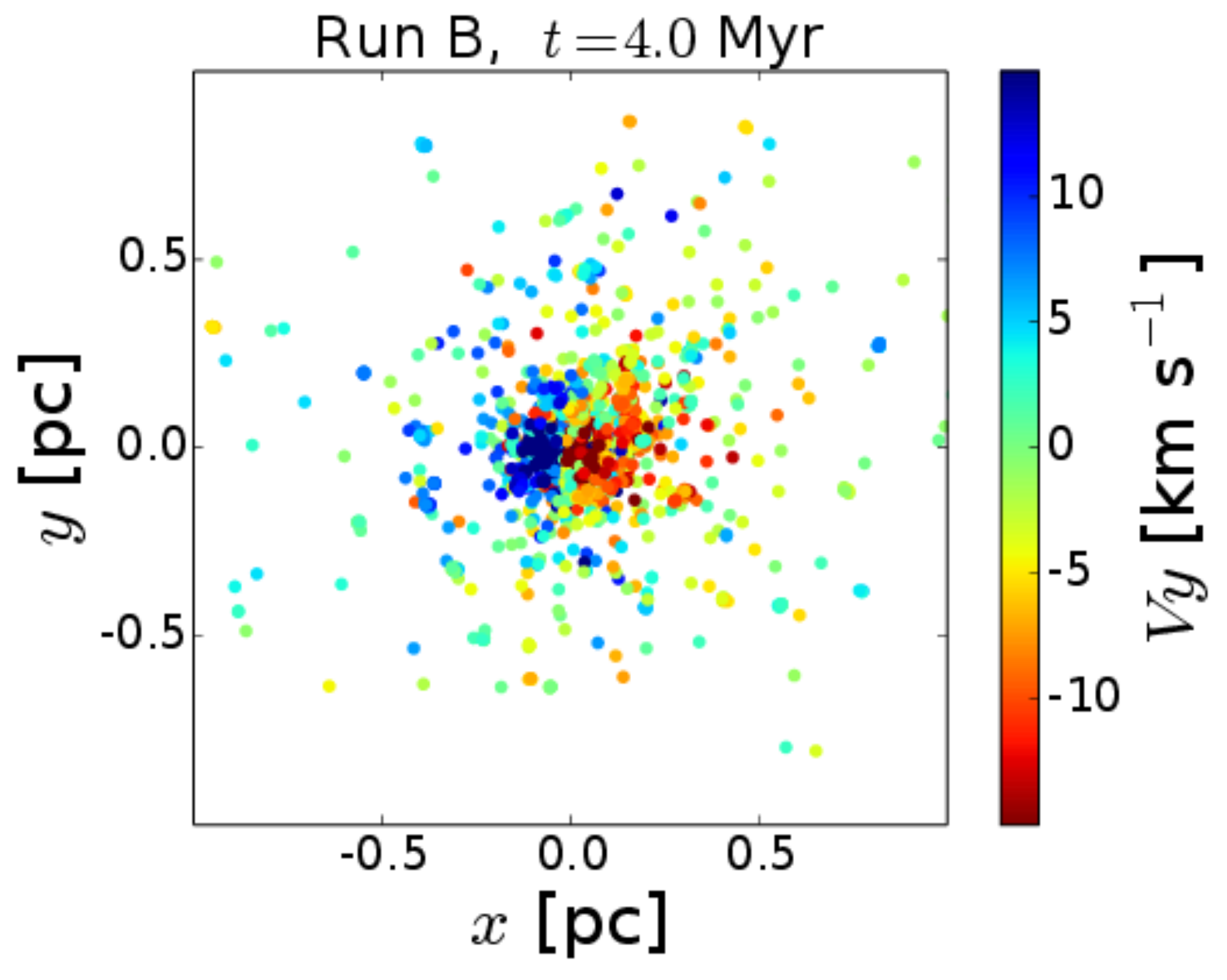,width=5.0cm}
    \epsfig{figure=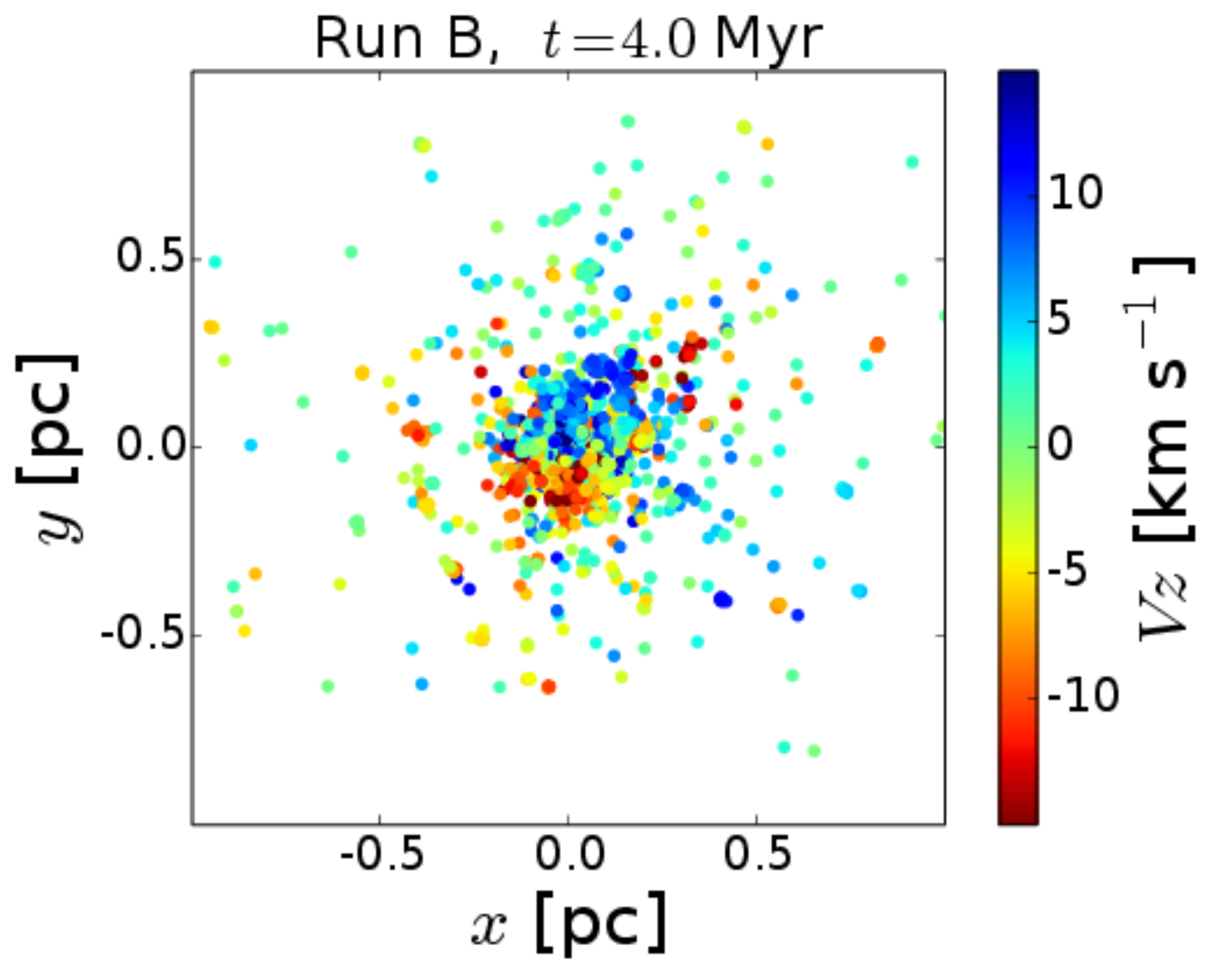,width=5.0cm}  
    \epsfig{figure=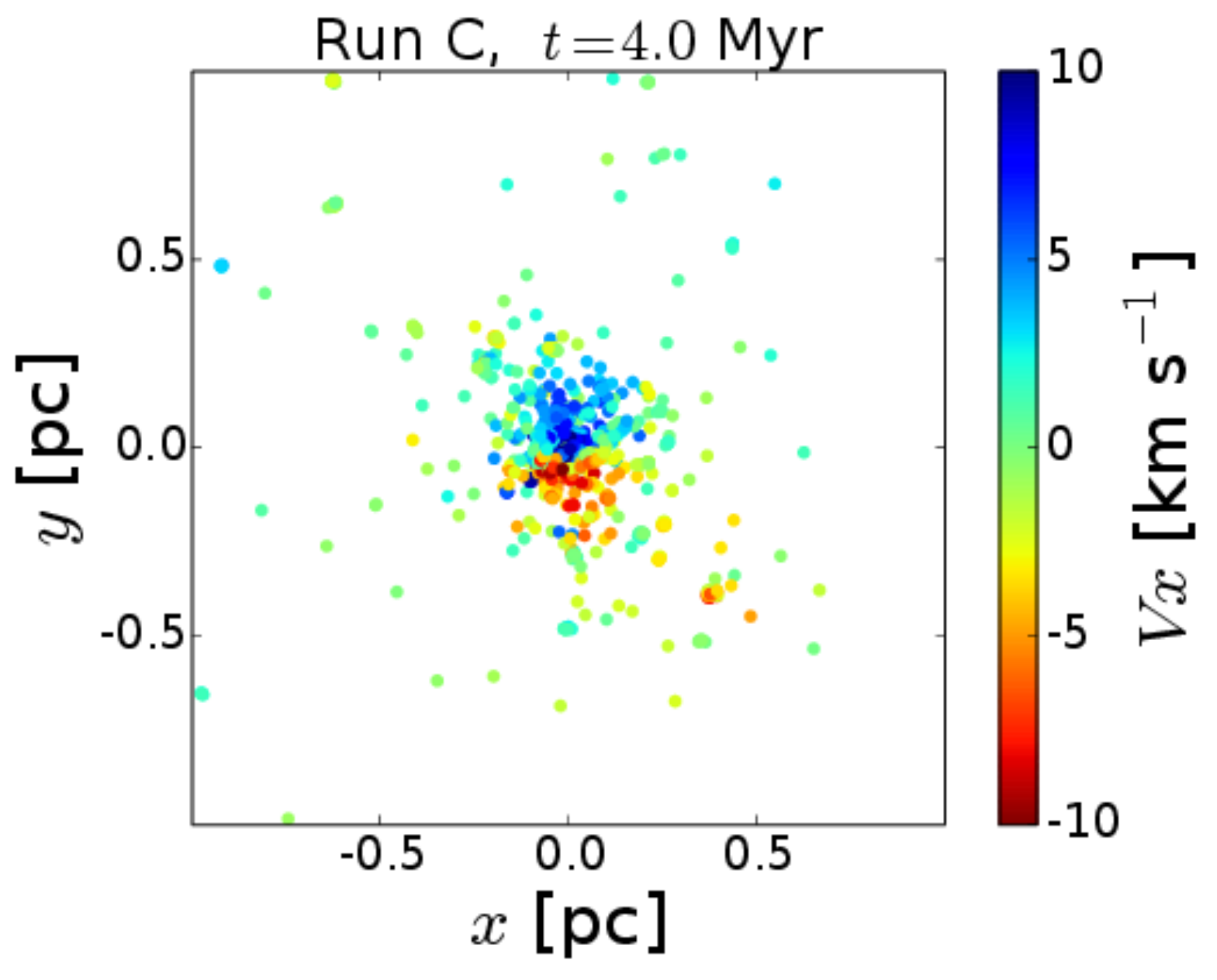,width=5.0cm} 
    \epsfig{figure=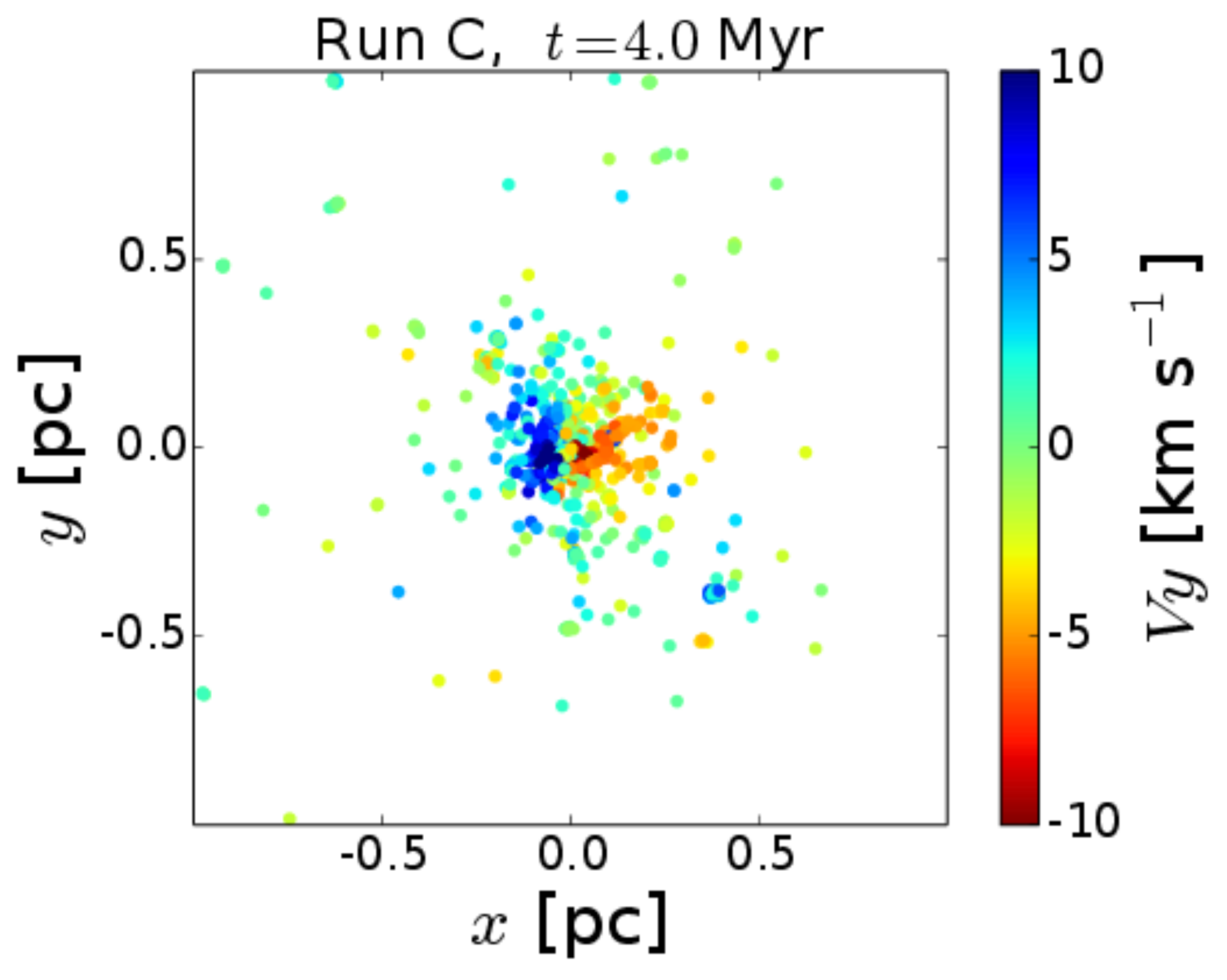,width=5.0cm}
    \epsfig{figure=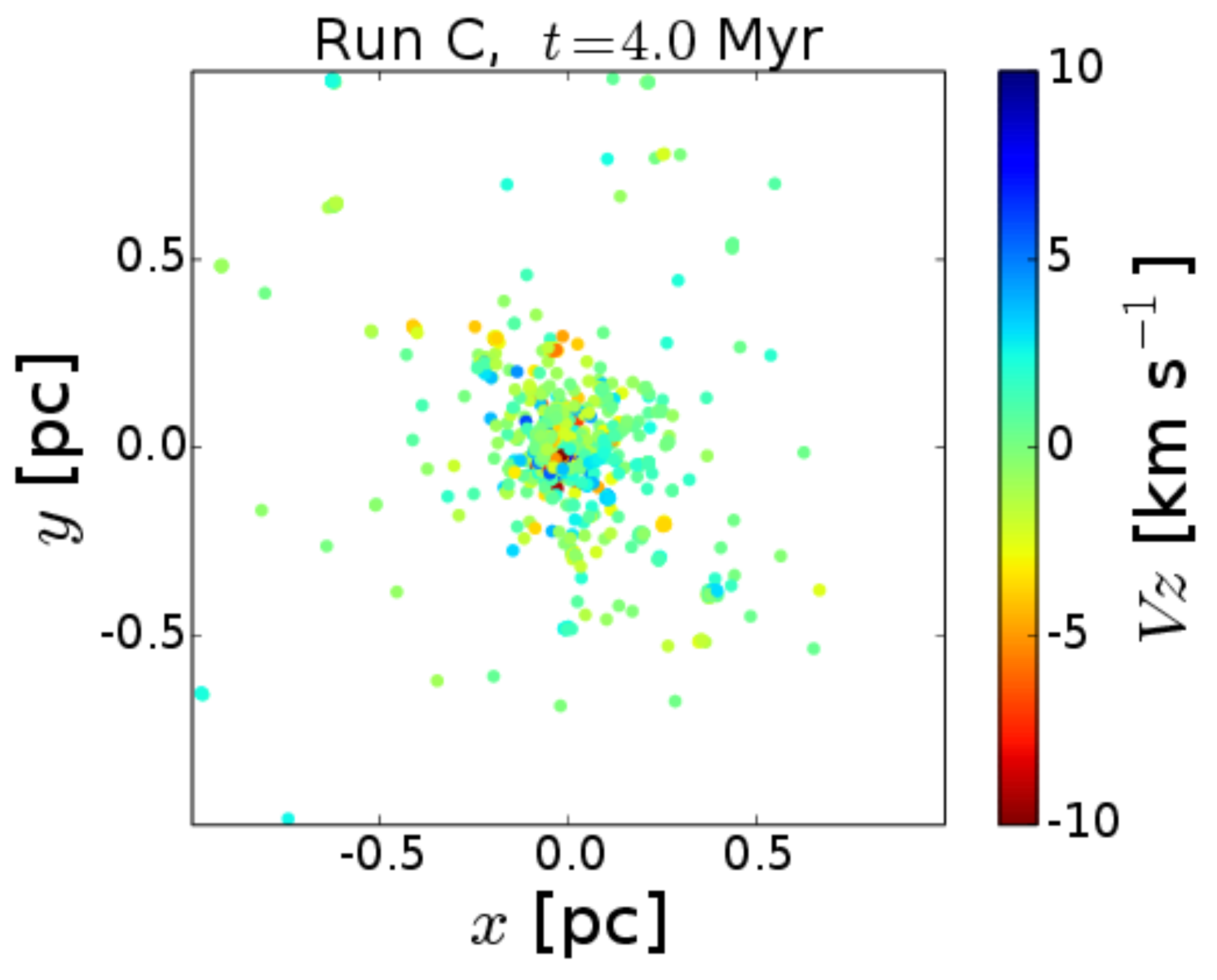,width=5.0cm}  
    \epsfig{figure=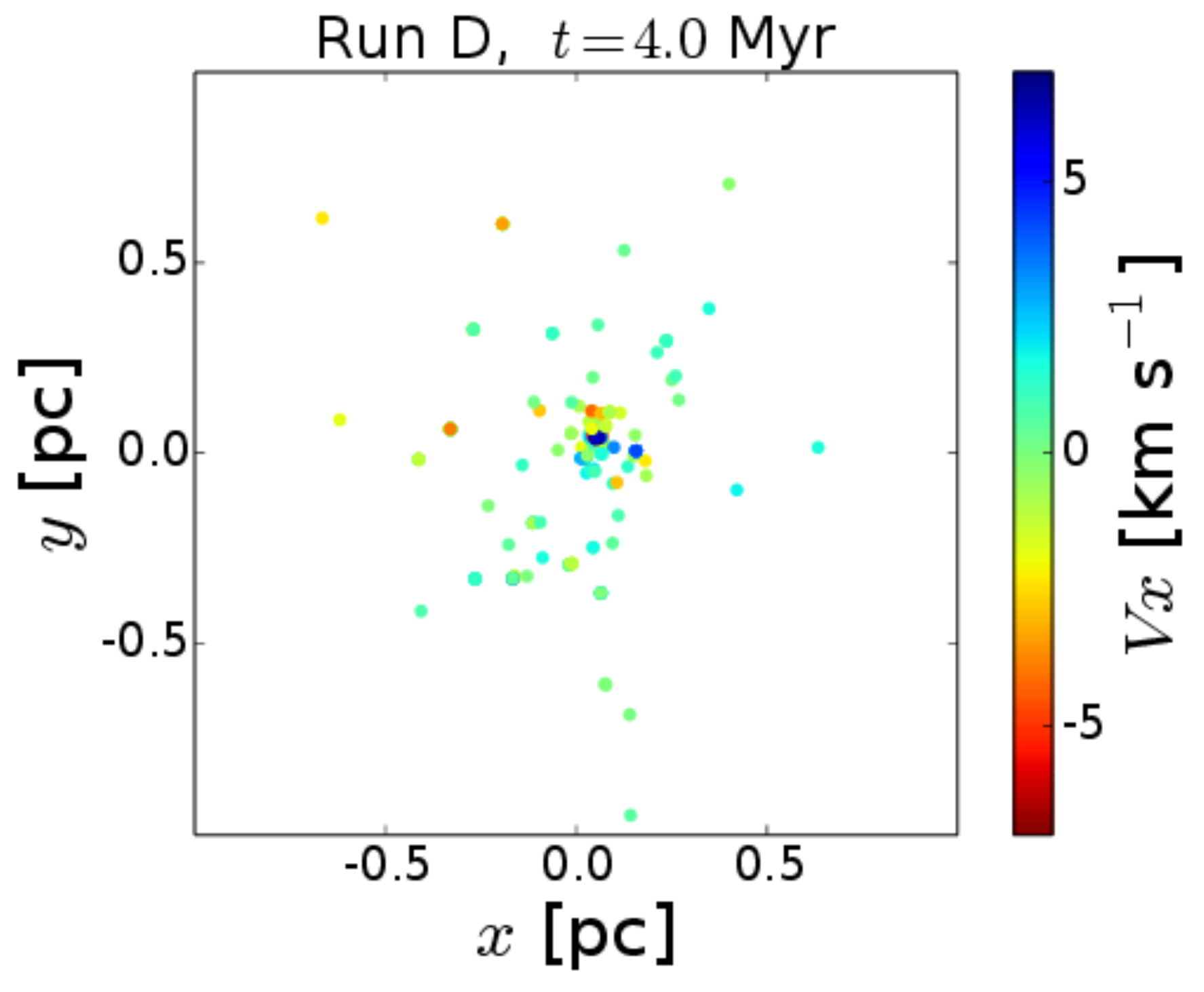,width=5.0cm} 
    \epsfig{figure=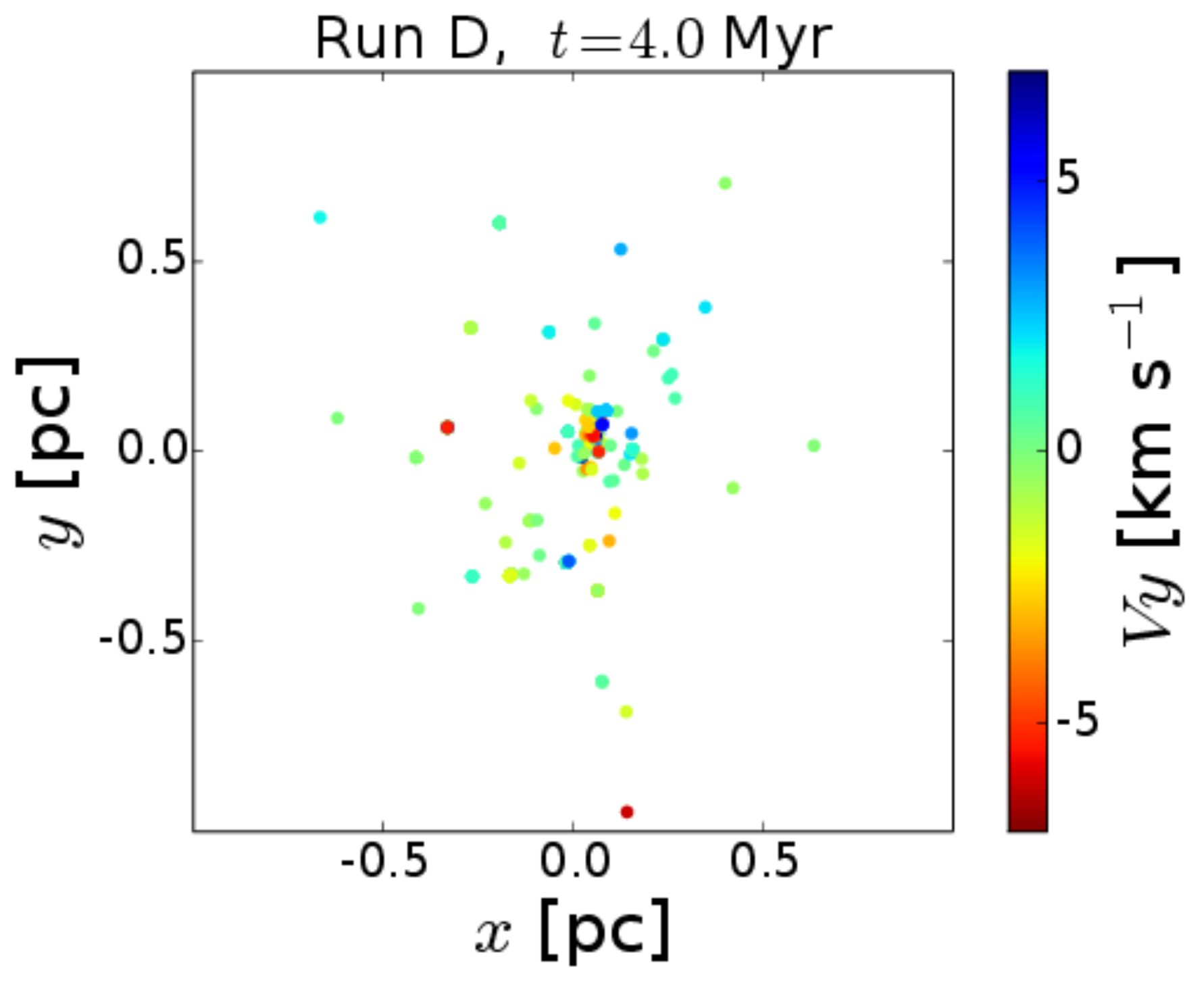,width=5.0cm}
    \epsfig{figure=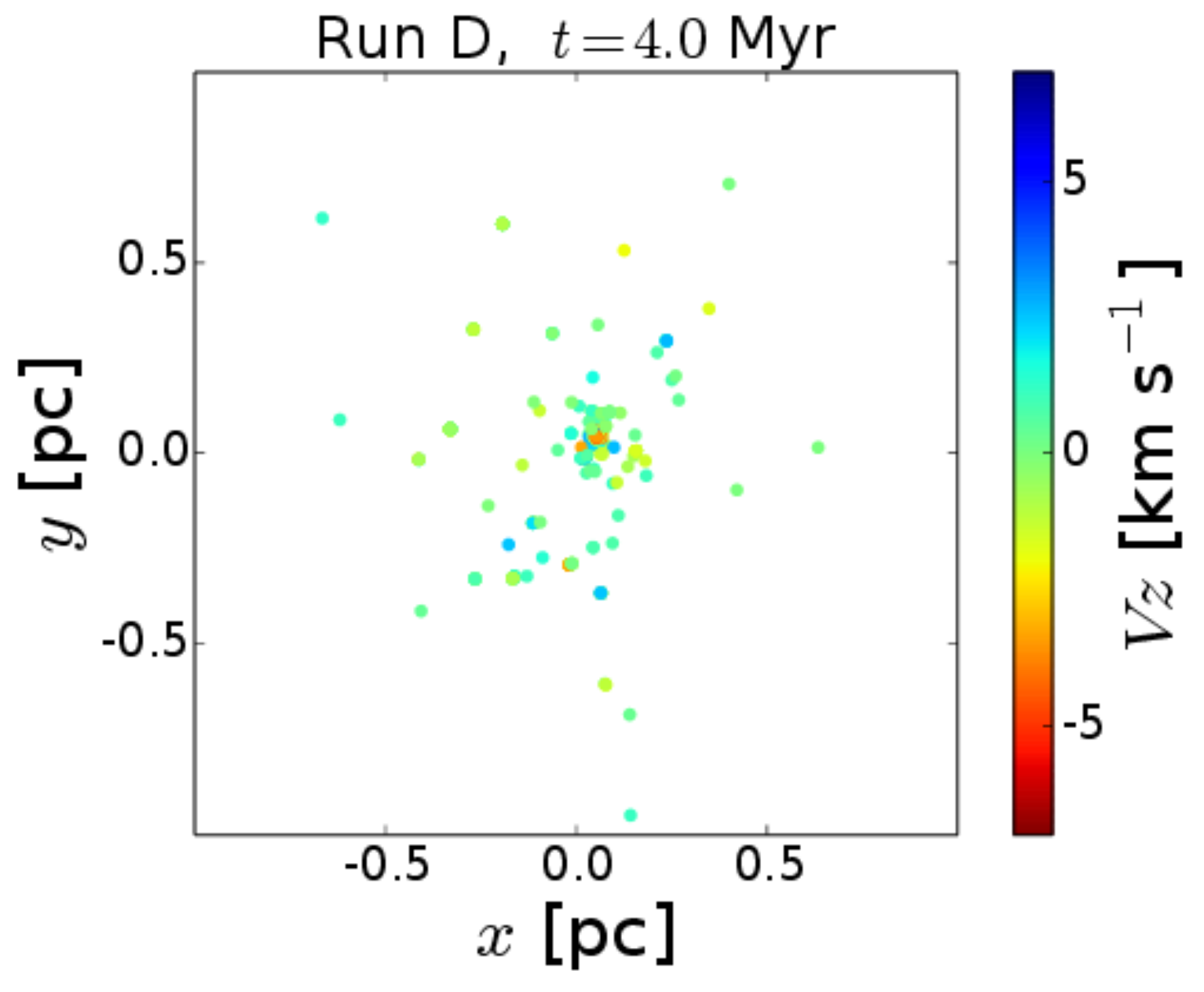,width=5.0cm}  
  \caption{  \label{fig:fig5}
Same as Fig.~\ref{fig:fig3}, but for time $t=4.0$ Myr. From top to bottom: run~A, B, C, and D. 
}}
\end{figure*}

\begin{figure}
  \center{
    \epsfig{figure=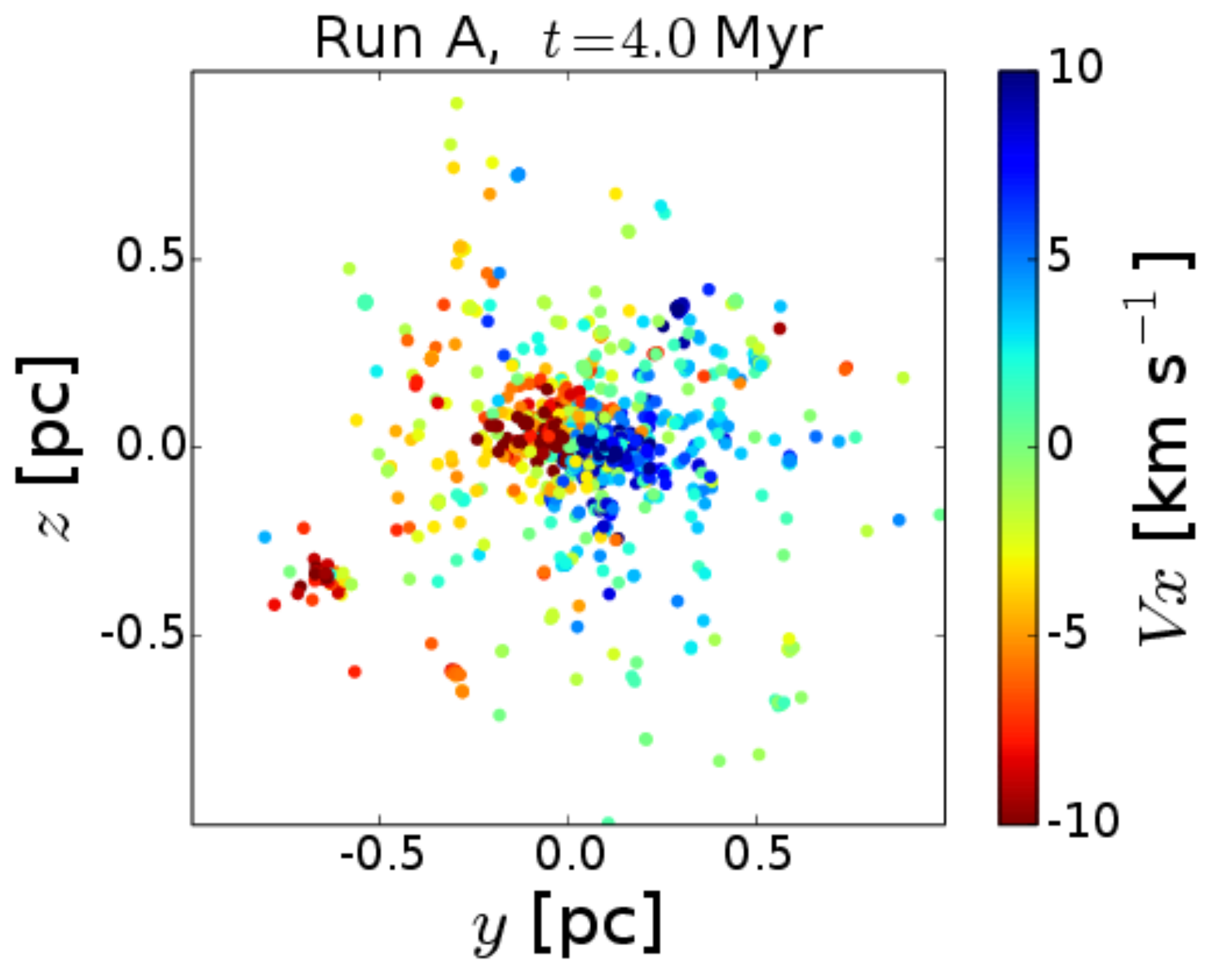,width=6.0cm} 
    \epsfig{figure=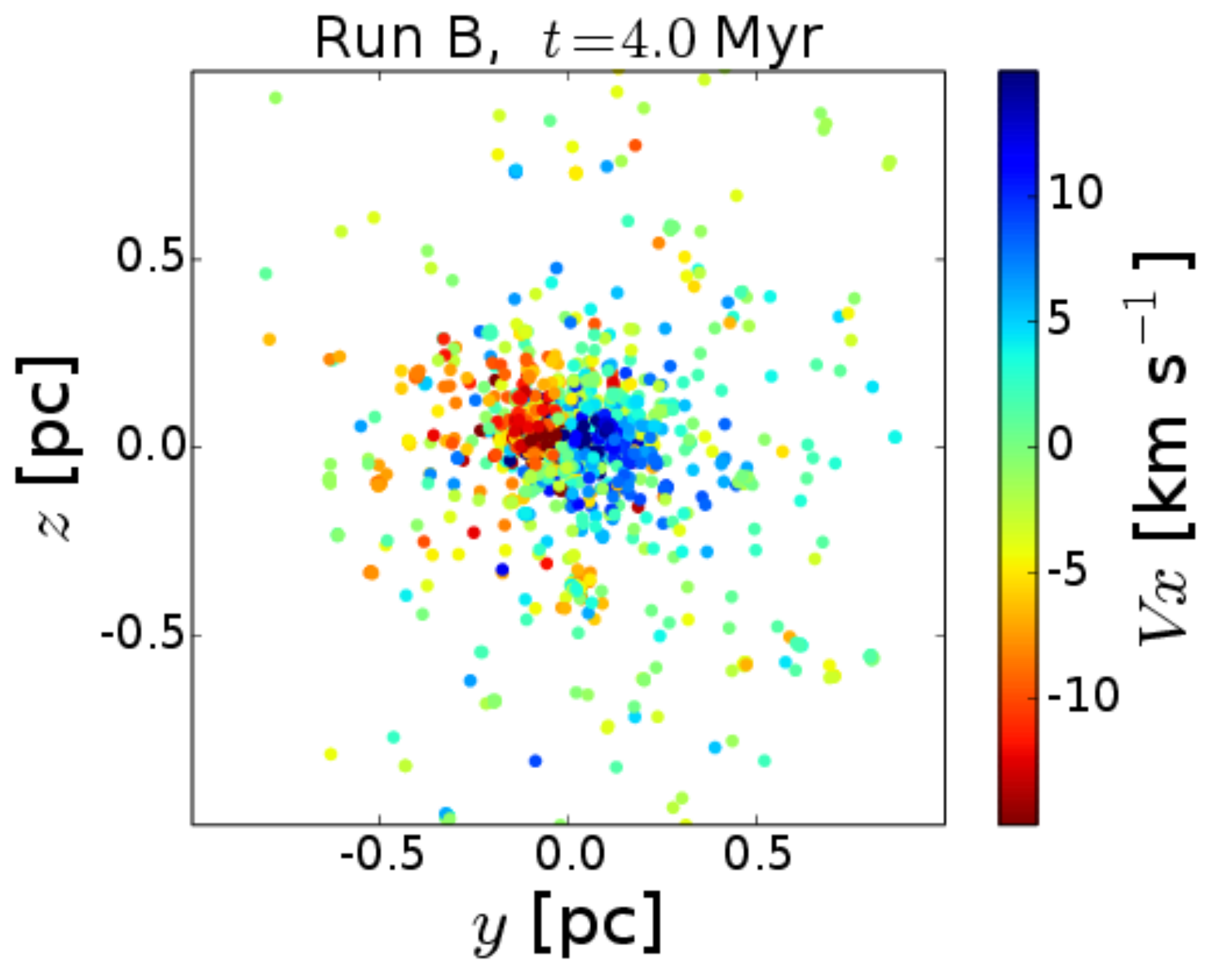,width=6.0cm}
    \epsfig{figure=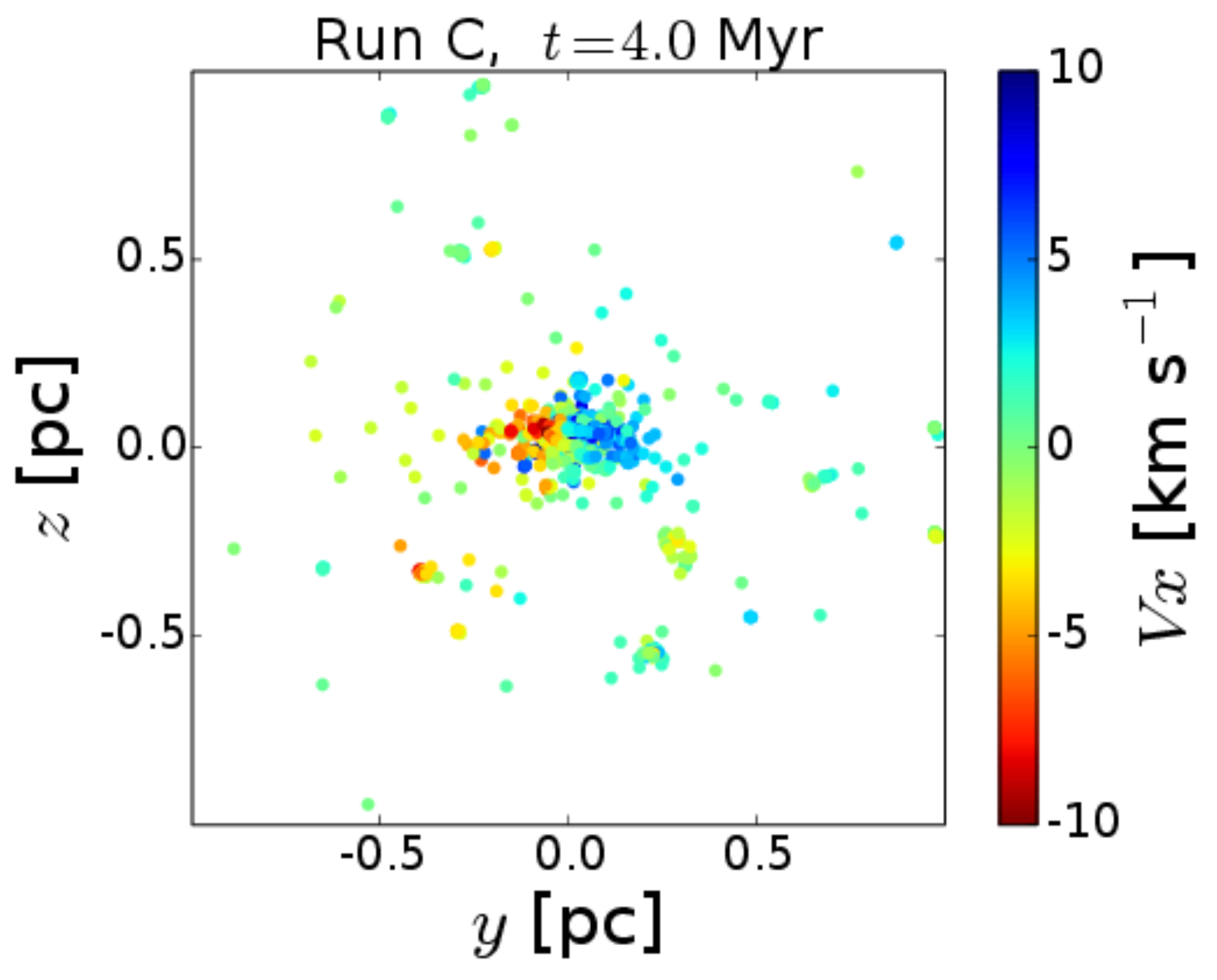,width=6.0cm}
    \epsfig{figure=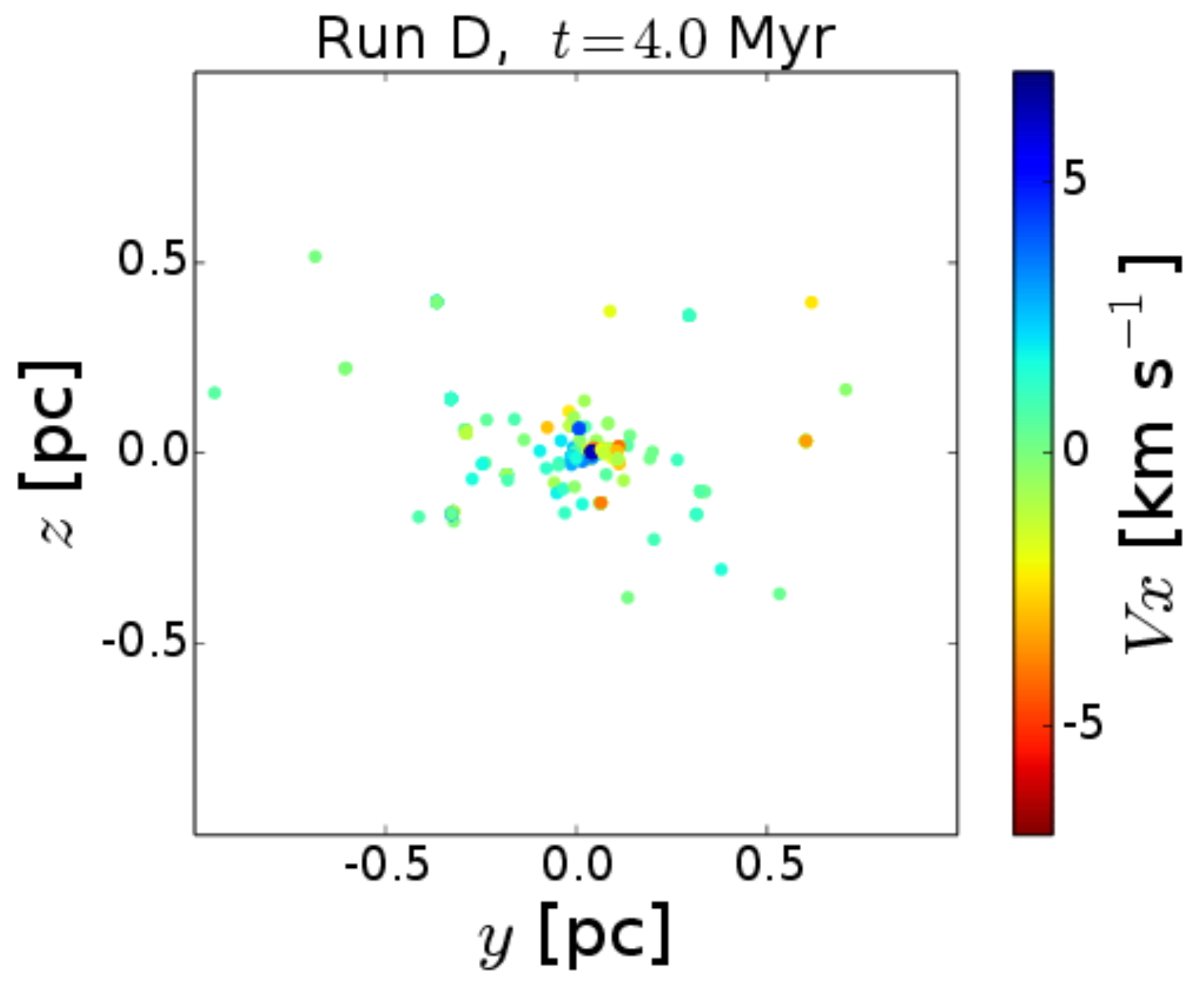,width=6.0cm}
  \caption{  \label{fig:fig6}
Same as Fig.~\ref{fig:fig4}, but for time $t=4.0$ Myr. From top to bottom: run~A, B, C, and D. 
}}
\end{figure}

\begin{figure}
  \center{
    \epsfig{figure=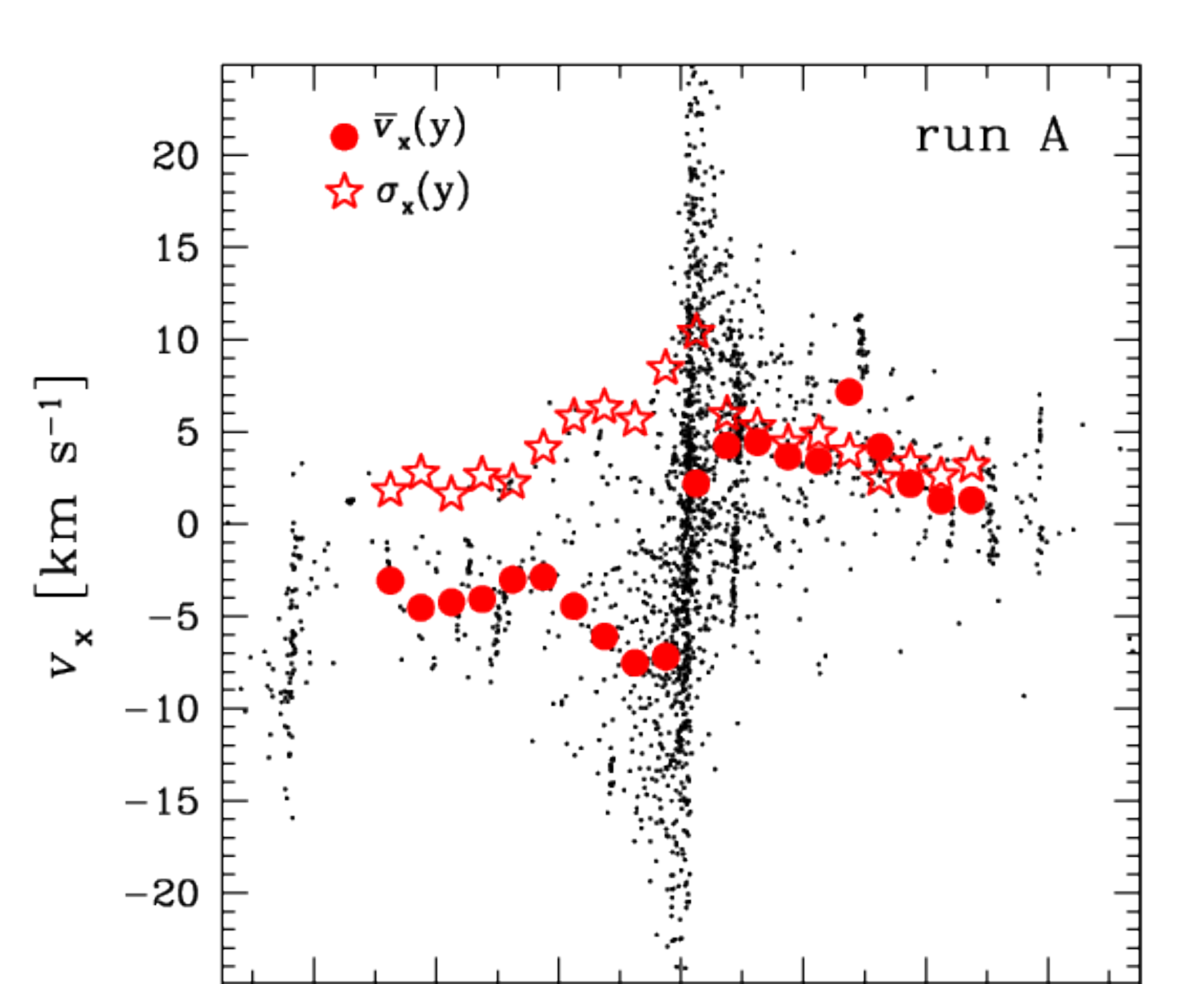,width=6.0cm}
    \epsfig{figure=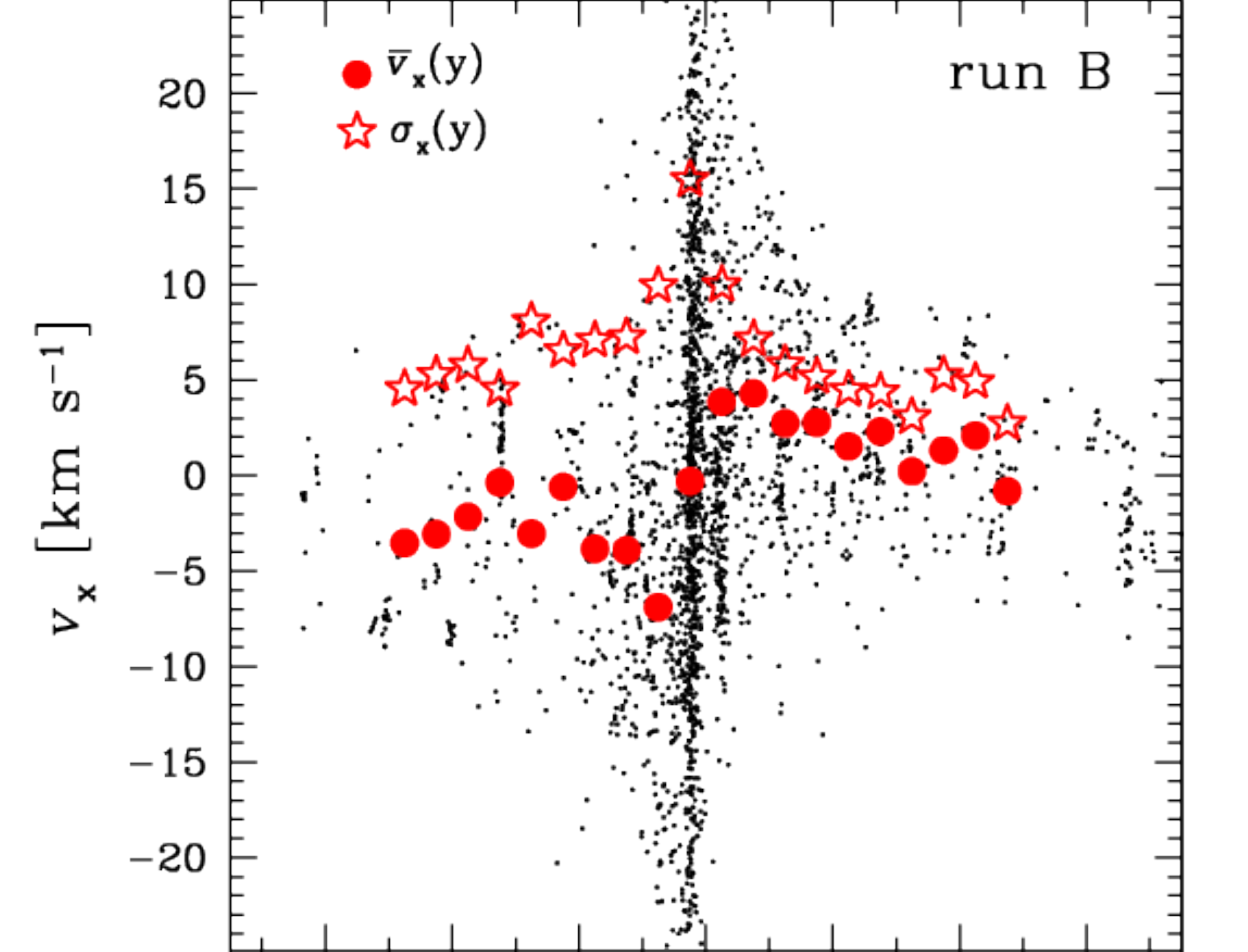,width=6.0cm}
    \epsfig{figure=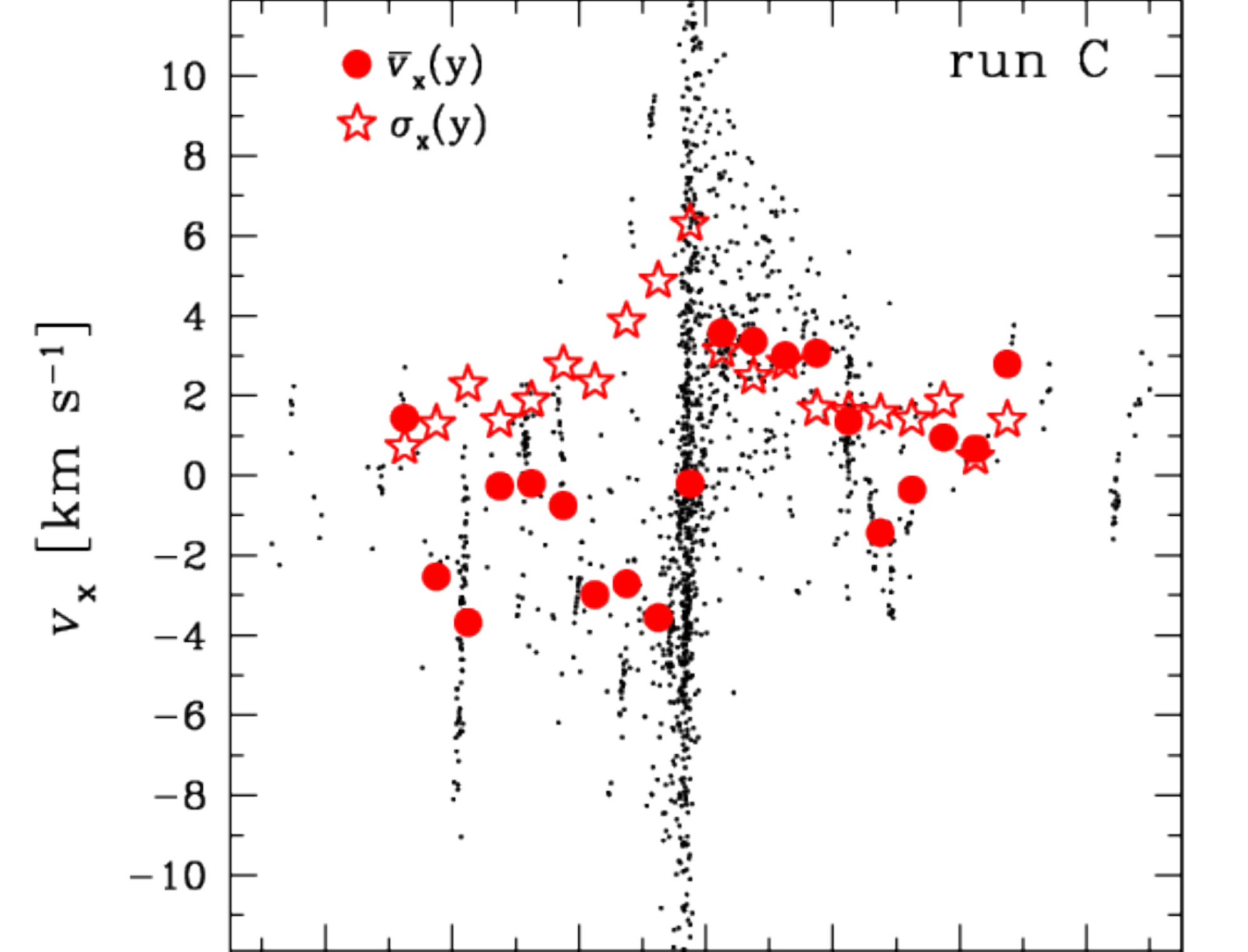,width=6.0cm}
    \epsfig{figure=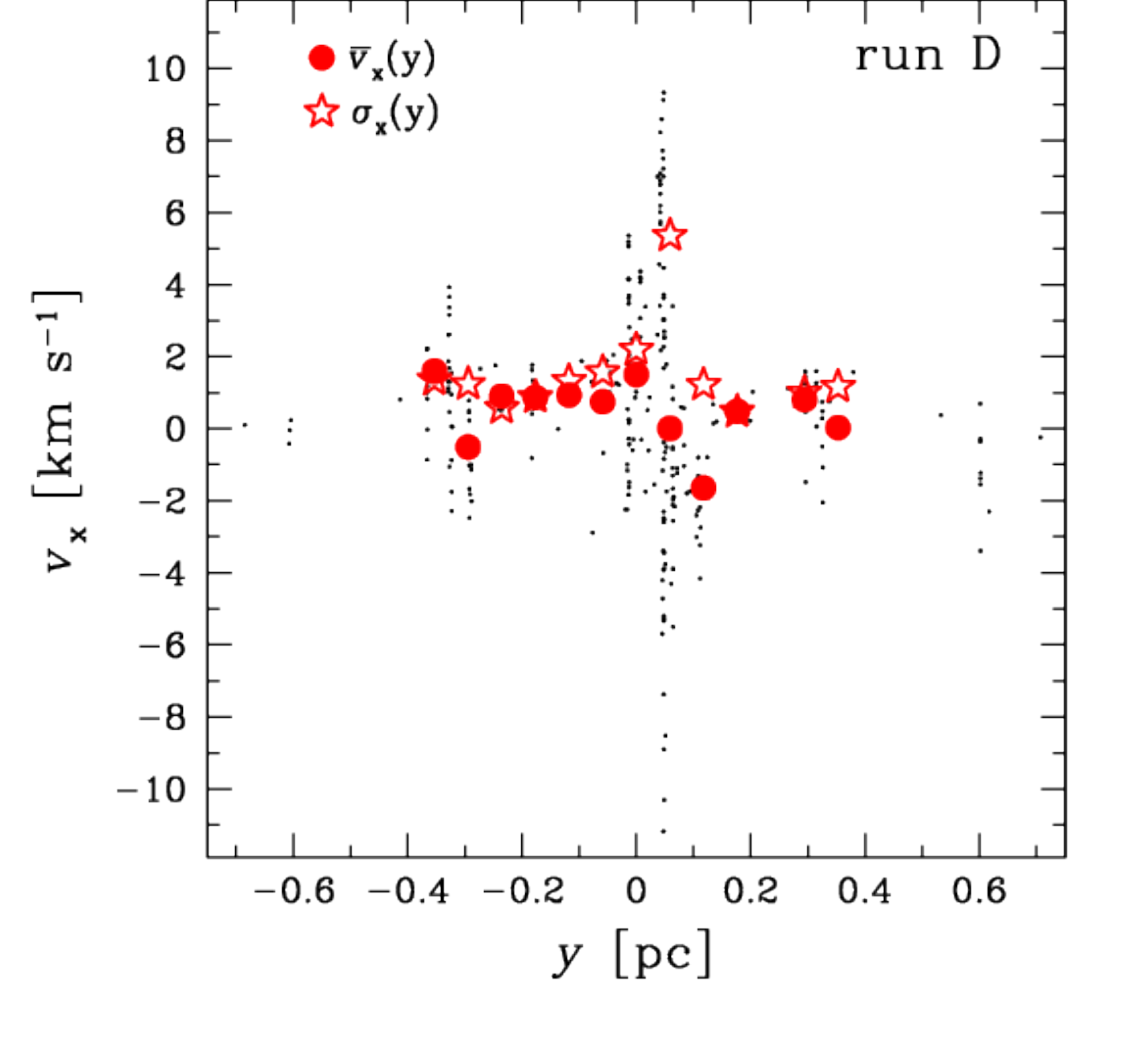,width=6.0cm}
  \caption{  \label{fig:fig7}
Rotation curve of the simulated star clusters at $t=4$ Myr.  Small black dots ($v_{\rm x}$): velocity along the $x$ axis of the individual stars (i.e. sink particles) as a function of the position on the $y$ axis. The angular momentum vector is along the $z$ axis. Filled red circles ($\bar{v}_{\rm x}(y)$): average one-dimensional stellar velocity along the $x$ axis as a function of the position on the $y$ axis. The values of $\bar{v}_{\rm x}(y)$ are averaged over  stars with position $y-\Delta{}y/2\le{}y_i<y+\Delta{}y/2$ (where $y$ and $\Delta{}y$ are the middle point and size of the bin, respectively). Open red stars ($\sigma{}_{\rm x}(y)$): one-dimensional stellar velocity dispersion along the $x$ axis as a function of the position on the $y$ axis. The values of $\sigma{}_{\rm x}(y)$ are calculated as the standard deviation relative to $\bar{v}_{\rm x}(y)$. From top to bottom: run~A, B, C, and D.
}}
\end{figure}

\begin{table*}
\begin{center}
\caption{\label{tab:tab2} Main properties of simulated star clusters.} \leavevmode
\begin{tabular}[!h]{lllllllllll}
\hline
Run
&
$t$ 
& 
$t_\ast$ 
&
$M_\ast$ 
&
$r_{\rm hm}$ 
&
$Q_{\rm vir}$
&
$r_{1},$ $r_{2},$ $r_{3}$ 
&
$\epsilon{}$
&
$\sigma_{\rm x}$ 
&
${\langle{}|v_{\rm x}|\rangle{}}/{\sigma{}_{\rm x}}$ 
& $L_{\rm z}$ \\
&
[Myr]
&
[Myr]
&
[M$_\odot{}$]
&
[pc]
&
&
[pc], [pc], [pc]
&
&
[km s$^{-1}$]
&
& [M$_\odot$ km$^2$ s$^{-1}$] \\
\hline
A & 2.5 & $0.7\pm{}0.2$ & 865   & 0.012 & 0.31  & 0.037, 0.022, 0.009 & 0.75 & 6.9   & 0.82 & $2.8\times{}10^{15}$ \\
A & 4.0 & $1.3\pm{}0.6$ & 7968  & 0.085 & 0.36  & 0.55, 0.39, 0.32    & 0.42 & 8.9   & 0.77 & $1.6\times{}10^{17}$  \\
B & 2.5 & $0.7\pm{}0.2$ & 1060  & 0.008 & 0.22  & 0.024, 0.013, 0.007 & 0.71 & 6.5   & 0.85 & $2.7\times{}10^{15}$ \\
B & 4.0 & $1.6\pm{}0.6$ & 11874 & 0.049 & 0.36  & 0.70, 0.35, 0.33    & 0.53 & 11.8  & 0.73 & $1.8\times{}10^{17}$ \\
C & 2.5 & $0.8\pm{}0.1$ & 202   & 0.006 & 0.20  & 0.018, 0.011, 0.004 & 0.75 & 2.9   & 0.77 &  $2.0\times{}10^{14}$ \\
C & 4.0 & $1.3\pm{}0.7$ & 2057  & 0.070 & 0.32  & 0.65, 0.47, 0.40    & 0.39 & 5.0   & 0.76 &  $1.2\times{}10^{16}$ \\
D & 2.5 & $0.3\pm{}0.1$ & 15    & $3\times{}10^{-4}$ & 0.11 & 0.0010, 0.0005, 0.0003  & 0.71 & -- & -- & $5.4\times{}10^{11}$ \\ 
D & 4.0 & $0.7\pm{}0.5$ & 277   & 0.074 & 0.16  & 0.51, 0.41, 0.34    & 0.51 & 3.5   & 0.70 & $6.5\times{}10^{14}$ \\
\hline
\end{tabular}
\end{center}
\begin{flushleft}
\footnotesize{Column~1: run name; column~2: time elapsed since the beginning of the simulation ($t$); column~3: average stellar age ($t_{\ast}$) and standard deviation  (the age of an individual star in our simulations is defined as the time elapsed since that individual sink particle was created); column~4: total star cluster mass ($M_{\ast}$); column~5: half-mass radius ($r_{\rm hm}$); column~6: virial ratio of the star cluster $Q_{\rm vir}=K/|W|$, where $K$ and $W$ are the kinetic and potential energy, respectively; column~7: star cluster size in three orthogonal directions ($r_1,\,{}r_2,\,{}r_3$), obtained through the rotational inertia matrix (see text); column~8: star cluster ellipticity defined as $\epsilon{}=1-\sqrt{\lambda{}_3/\lambda{}_1}$, where $\lambda{}_1$ and $\lambda{}_3$ are the maximum and minimum eigenvalue of the rotational inertia matrix, respectively (see text); column~9: velocity dispersion along the $x$ axis ($\sigma{}_{\rm x}=\sqrt{(N-1)^{-1}\,{}\sum_i(v_{\rm x}(i)-\langle{}v_{\rm x}\rangle{})^2}$, where $v_{\rm x}(i)$ is the velocity along the $x$ axis of the $i-$th star,  $N$ is the number of stars in the star cluster, and $\langle{}v_{\rm x}\rangle{}=\sum_i v_{\rm x}(i)/N$ is the average velocity  along the $x$ axis); column~10: average of the absolute value of the  velocity along the $x$ axis (${\langle{}|v_{\rm x}|\rangle{}}=\sum_i |v_{\rm x}(i)|/N$) over velocity dispersion  along the $x$ axis ($\sigma{}_{\rm x}$).  Column~11 ($L_{\rm z}$): total angular momentum of the system (this is also the component of the angular momentum along the $z$ axis, because the angular momentum vector is aligned with the $z$ axis).  }
\end{flushleft}
\end{table*}

\section{Results}
Star formation begins at time $t=1.3\,{}{\rm Myr}\sim{}0.6\,{}t_{\rm ff}$ in clouds A and B, at $t=1.5\,{}{\rm Myr}\sim{}0.7\,{}t_{\rm ff}$ in cloud C, and at $t=1.9\,{}{\rm Myr}\sim{}0.9\,{}t_{\rm ff}$ in cloud D. 
 The top panel of Figure~\ref{fig:fig1} shows the projected density of gas in simulation~A at $t=2.5$ Myr (i.e. $\sim{}1.2\,{}t_{\rm ff}$). 
The cloud has developed the well-known filamentary structures \citep{bonnell2003, bonnell2011}, as an effect of the interplay between self-gravity and turbulence \citep{moeckel2015,federrath2016}. The overall behaviour is similar in the other simulated clouds. 

The bottom panel of Figure~\ref{fig:fig1} is a zoom of the region where the main star cluster is assembling. 
The gas velocity maps reported in Fig.~\ref{fig:fig2} indicate that this region rotates. Rotation is still apparent in the gas even at the end of the simulation ($t=4$ Myr). In our simulations, rotation arises from the torques exerted by gas filaments and clumps while they merge with the main proto-cluster structure.

The rotation of the parent molecular gas leaves a strong imprint on the star cluster. Fig.~\ref{fig:fig3} shows the velocity field of the main star cluster in simulations A, B, and C at time $t=2.5$ Myr. Each point in the Figure is a sink particle. The star clusters are projected in the $xy$ plane, that is defined as the plane perpendicular to the angular momentum vector (which is thus aligned along the $z$ axis).  Simulation D is not shown because the main star cluster is not well defined at $t=2.5$ Myr (it has a total mass of only $\sim{}15$ M$_\odot$). All star clusters in Fig.~\ref{fig:fig3} show a clear signature of rotation in $v_{\rm x}$ and $v_{\rm y}$.

Similarly, Fig.~\ref{fig:fig4} shows  the velocity field of the main star cluster in simulations A, B, and C, at time $t=2.5$ Myr, when projected in the $yz$ plane. With respect to this projection, the velocity along the $x$ axis ($v_{\rm x}$), shown by the colour-coded map, can be interpreted as the line-of-sight velocity. Again, the rotation of the three star clusters is apparent. Moreover, this projection shows that the star clusters are extremely flattened. The reason is that they form from a disk-like rotating gas structure, as shown in Figures~\ref{fig:fig1} and \ref{fig:fig2}. In the first stages of their assembly, the star clusters still retain memory of their parent gas disc.

At this stage ($t=2.5$ Myr) the star clusters are still extremely small and compact. The main structural and kinematic parameters of the simulated star clusters are listed in Table~\ref{tab:tab2}. 
The total bound mass to the star clusters at $t=2.5$ Myr ranges from only $\sim{}15$ M$_\odot$ up to $\sim{}1060$ M$_\odot$ for runs~D and B, respectively. The half mass radius $r_{\rm hm}$ is smaller than $\sim{}0.012$ pc. The virial ratio $Q_{\rm vir}=K/|W|$ (where $K$ and $W$ are the kinetic and potential energy of the stars, respectively\footnote{In the definition of $Q_{\rm vir}$ we do not include the contribution of gas particles, as suggested by \cite{kruijssen2012a} and \cite{farias2015}.})  is $\sim{}0.2-0.3$, indicating that our protoclusters are markedly subvirial. At time $t=2.5$ Myr since the beginning of the simulation, the average age of stars\footnote{The age of a star is defined as the time elapsed since the formation of the sink particle.} in the main star cluster  is $t_{\ast}\sim{}0.7-0.8$ Myr in runs~A, B, and C, and $t_\ast{}\sim{}0.3$ Myr in run~D, with a non-negligible age spread ($\sim{}0.1-0.2$ Myr).

The ellipticity ($\epsilon$) allows us to quantify the flattening of the simulated star clusters. To estimate the ellipticity, we first calculate the rotational inertia matrix of sink particles with respect to their centre of gravity:
 \begin{eqnarray}
I_\text{rot} = \sum_i m_i\begin{bmatrix}
x_i^2 & x_iy_i & x_iz_i  \\
x_iy_i & y_i^2 & y_iz_i  \\
x_iz_i & y_iz_i & z_i^2  \end{bmatrix},
\label{Irot} 
\end{eqnarray}
where $i$ is the index of sink particles. The three eigenvalues of this matrix ($\lambda_1, \lambda_2, \lambda_3$) give the star cluster size in three orthogonal directions:
\begin{eqnarray}
r_i = \beta \sqrt{{5 \lambda_i \over M_\ast}} , \; i = 1, 2, 3,
\label{rsink} 
\end{eqnarray}
where $M_\ast$ is the total mass of sinks. The factor $5$ comes from the assumption that the sink mass is uniformly distributed in space, 
and a correction factor $\beta \geq 1$ accounts for the fact that the mass distribution might not be uniform but rather centrally concentrated (see \citealt{leehennebelle2016a}). The values of $r_1,\,{}r_2,\,{}$ and $r_3$ shown in Table~\ref{tab:tab2} were calculated assuming $\beta{}=1$ for simplicity.

We thus define $\epsilon=1-\sqrt{\lambda_3/\lambda_1}$, where $\lambda_1$ and $\lambda_3$ are the maximum and minimum eigenvalue, respectively. $\epsilon{}=0$ means that the star cluster is spherical, while $\epsilon\sim{}1$ means that the star cluster is nearly flat. For the protostar clusters we find $\epsilon{}\sim{}0.71-0.75$ at $t=2.5$ Myr. Thus, the simulated star clusters are extremely flattened during the early stages of their assembly.

At $t=4$ Myr (i.e. $\sim{}2\,{}t_{\rm ff}$), the signature of rotation is still strong in runs~A, B, and C (Figs.~\ref{fig:fig5} and \ref{fig:fig6}), even if the ellipticity has diminished to $\epsilon{}\sim{}0.4-0.5$ (Table~\ref{tab:tab2}). In these late evolutionary stages, the star clusters have grown by hierarchical assembly to a size of $\approx{}0.5$ pc (Table~\ref{tab:tab2}). The half-mass radius in runs~A, B and C is close to $r_{\rm hm}\sim{}0.05-0.1$ pc, and the virial ratio tends to increase  ($Q_{\rm vir}\sim{}0.3-0.4$), even if the star clusters are still subvirial. The total final star cluster mass ranges from $\sim{}300$ M$_\odot$ to $\sim{}12000$ M$_\odot$ in runs~D and B, respectively. 

At $t=4$ Myr since the beginning of the simulation, the average age of stars in the main star cluster  is $t_{\ast}\sim{}1.3-1.6$ Myr in runs~A, B, and C, and $t_\ast{}\sim{}0.7$ Myr in run~D, with a significant age spread (the standard deviation being $\sim{}0.5-0.7$ Myr). Thus, most simulated stars are still very young objects at the end of the simulation (Table~\ref{tab:tab2}). 

Figure~\ref{fig:fig7} shows the rotation curve of star clusters in runs~A, B, C, and D at the end of the simulation, compared with the one-dimensional velocity dispersion ($\sigma{}_{\rm x}$). The rotation curves of runs~A and B (which are two different random realizations of the same cloud) are very similar to each other, indicating that rotation is a common outcome in star clusters formed from molecular clouds with mass $M\sim{}4\times{}10^4$ M$_\odot$. Rotation is apparent even in the star cluster of run~C, whose mass is $\sim{}1/5$ of the mass of star clusters in runs~A and B.

The star cluster of run~D is the only one which does not show a clear signature of rotation. It assembles later and has a total mass of only $\sim{}300$ M$_\odot$, much smaller than the other star clusters. This indicates that the signature of rotation depends on the total stellar mass and thus on the initial gas mass that is converted into stars. We expect rotation to be negligible for star clusters with mass $M_\ast{}\lesssim{}$ few $\times{}10^2$ M$_\odot$. 

 Stochastic fluctuations might affect this result, because run~D is the less massive cloud we simulated and was sampled with less particles ($N_{\rm gas}=2\times{}10^6$) than the other runs. Thus, we check the importance of stochastic fluctuations by running four additional realisations of run~D, which are discussed in Appendix~\ref{appendix:app1}. Stochastic fluctuations are found to be significant: star clusters with mass ranging from $\sim{}300$ to $\sim{}800$ M$_\odot$ form in the five realisations of run~D. Our main conclusion still holds, because there is no signature of rotation in those realisations of run D where the final star cluster mass is $\lesssim{}500$ M$_\odot$. Only one realisation, with star cluster mass $\sim{}800$ M$_\odot$, shows some hints of rotation. This confirms that rotation is expected to be negligible for star clusters with mass $M_\ast{}\lesssim{}$ few $\times{}10^2$ M$_\odot$. 


\cite{proszkow2009} point out that the collapse of subvirial clusters can produce kinematic signatures similar to rotation, with a red-shifted and a blue-shifted component. The main difference between the signature of genuine rotation and that of subvirial collapse is that the latter has no preferential plane (and ideally zero angular momentum vector), while the former is maximum in the plane perpendicular to the angular momentum vector. 

\begin{figure}
  \center{
    \epsfig{figure=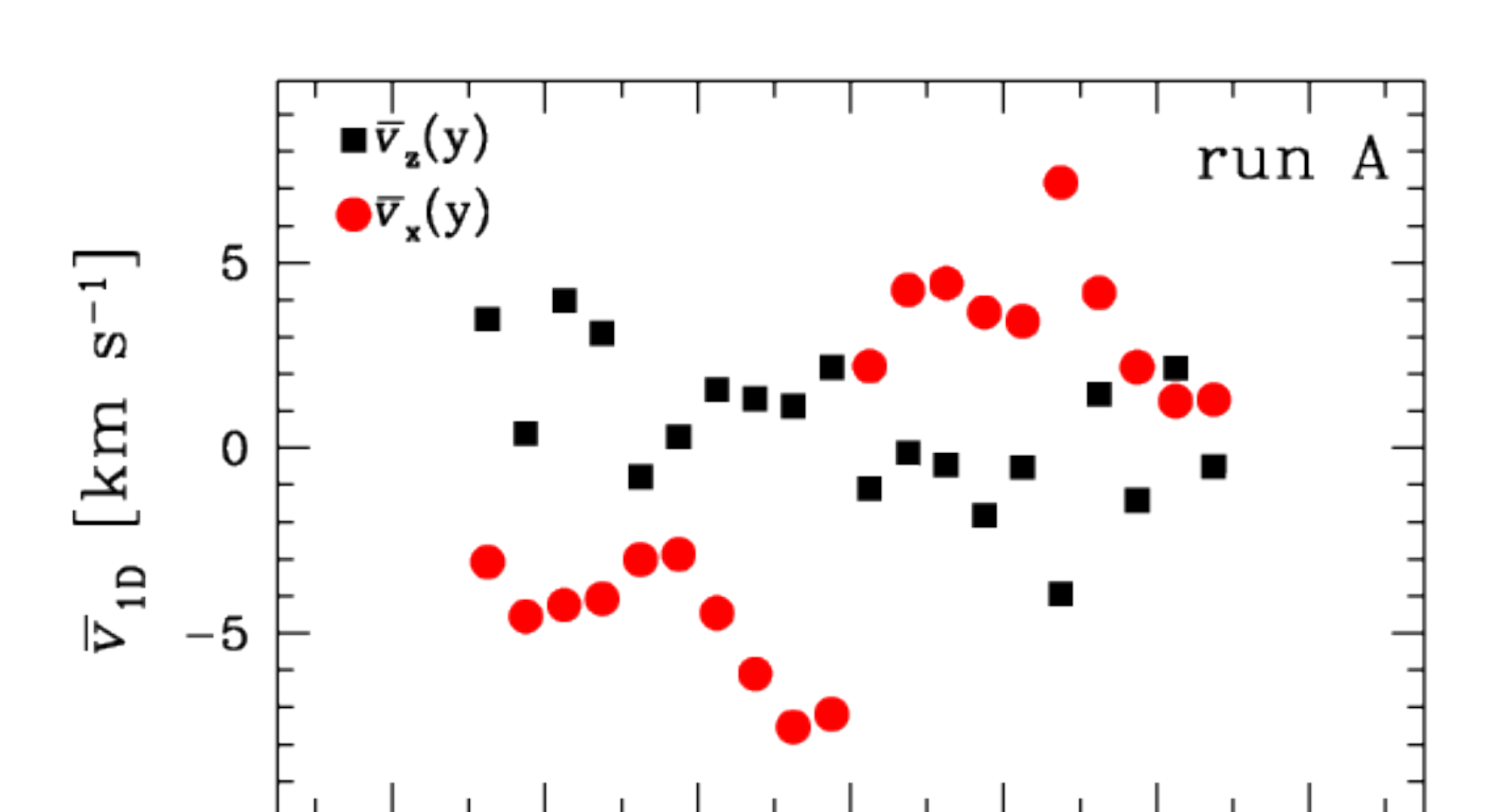,width=6.0cm}
    \epsfig{figure=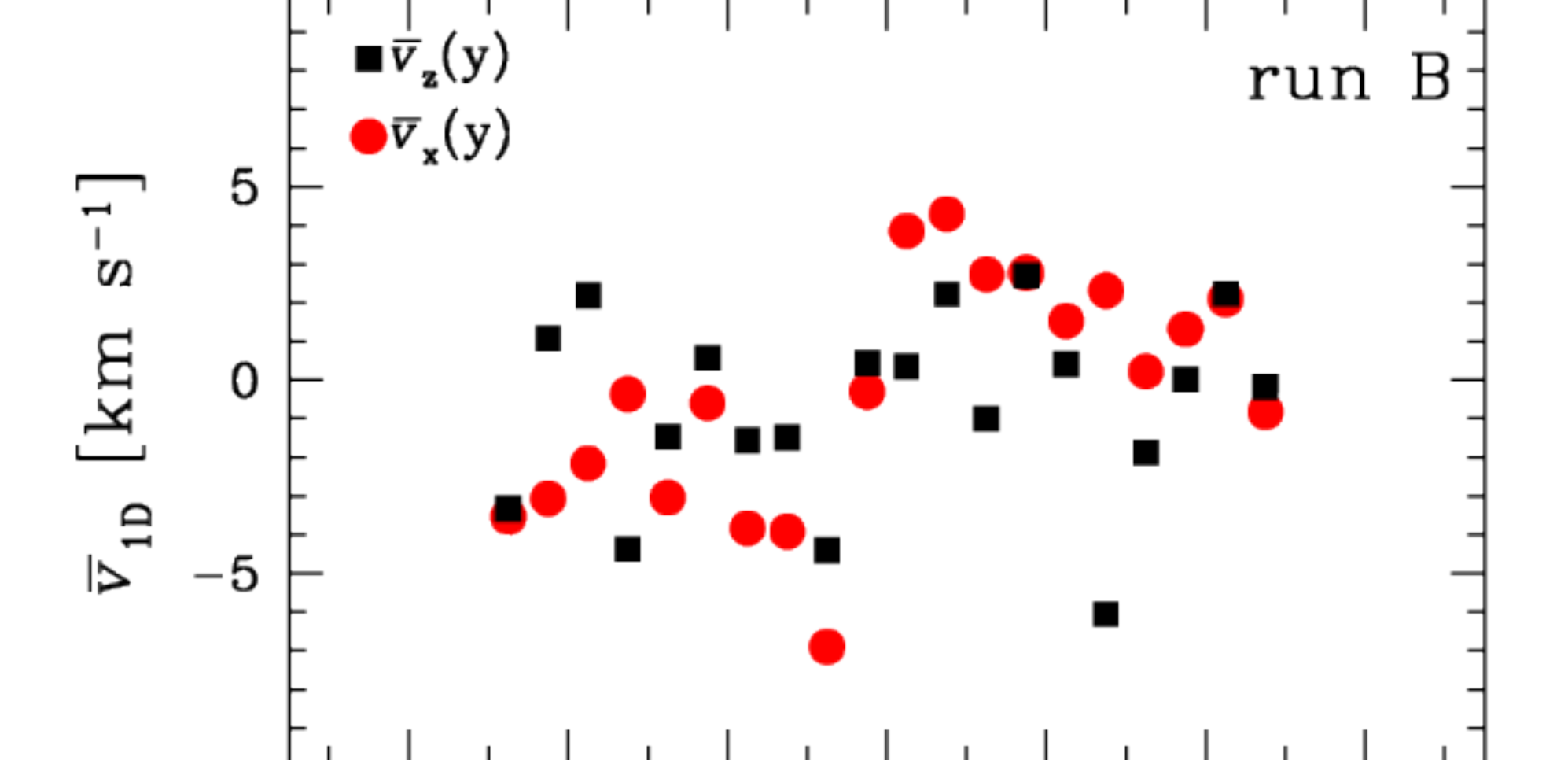,width=6.0cm}
    \epsfig{figure=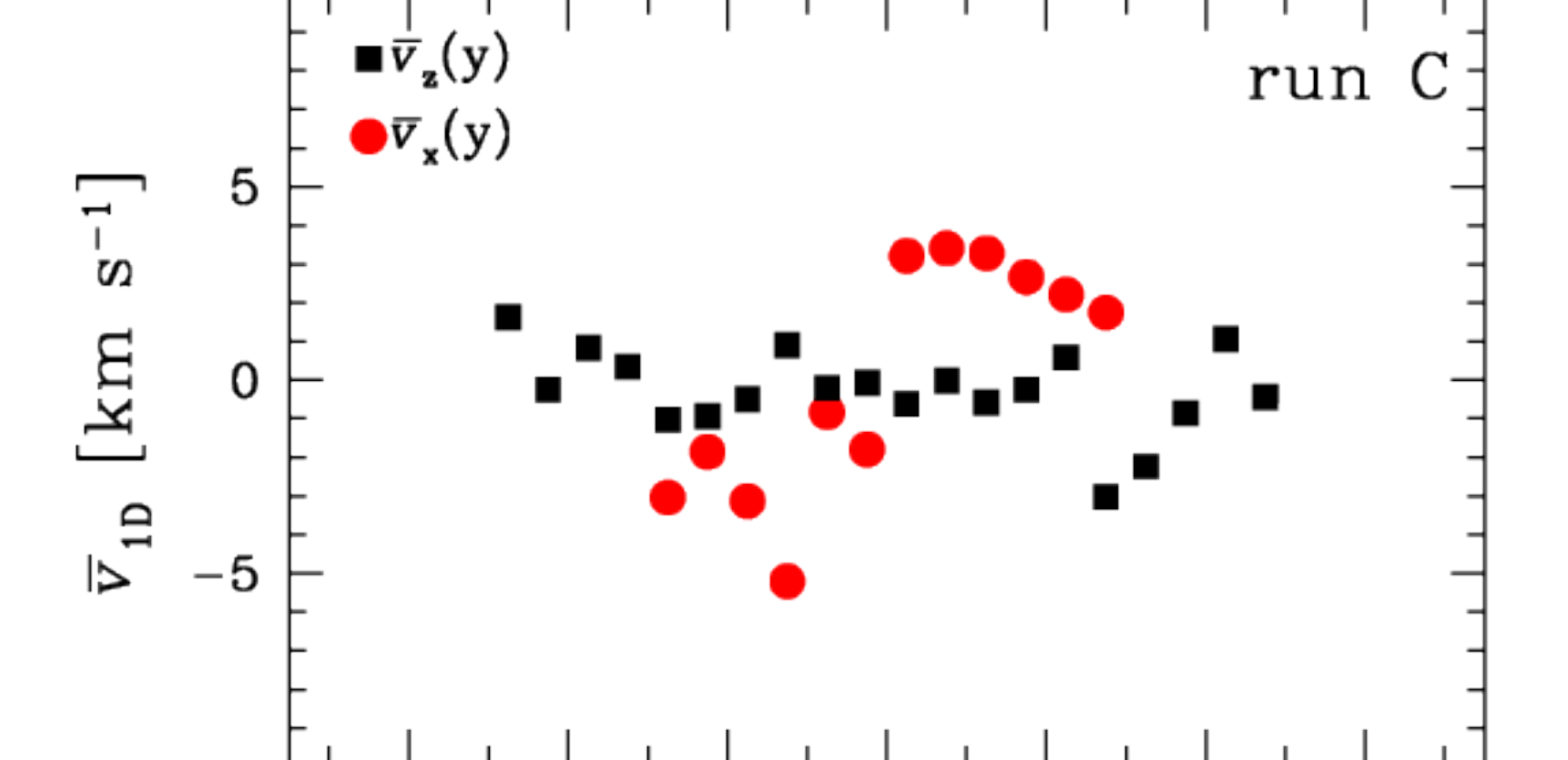,width=6.0cm}
    \epsfig{figure=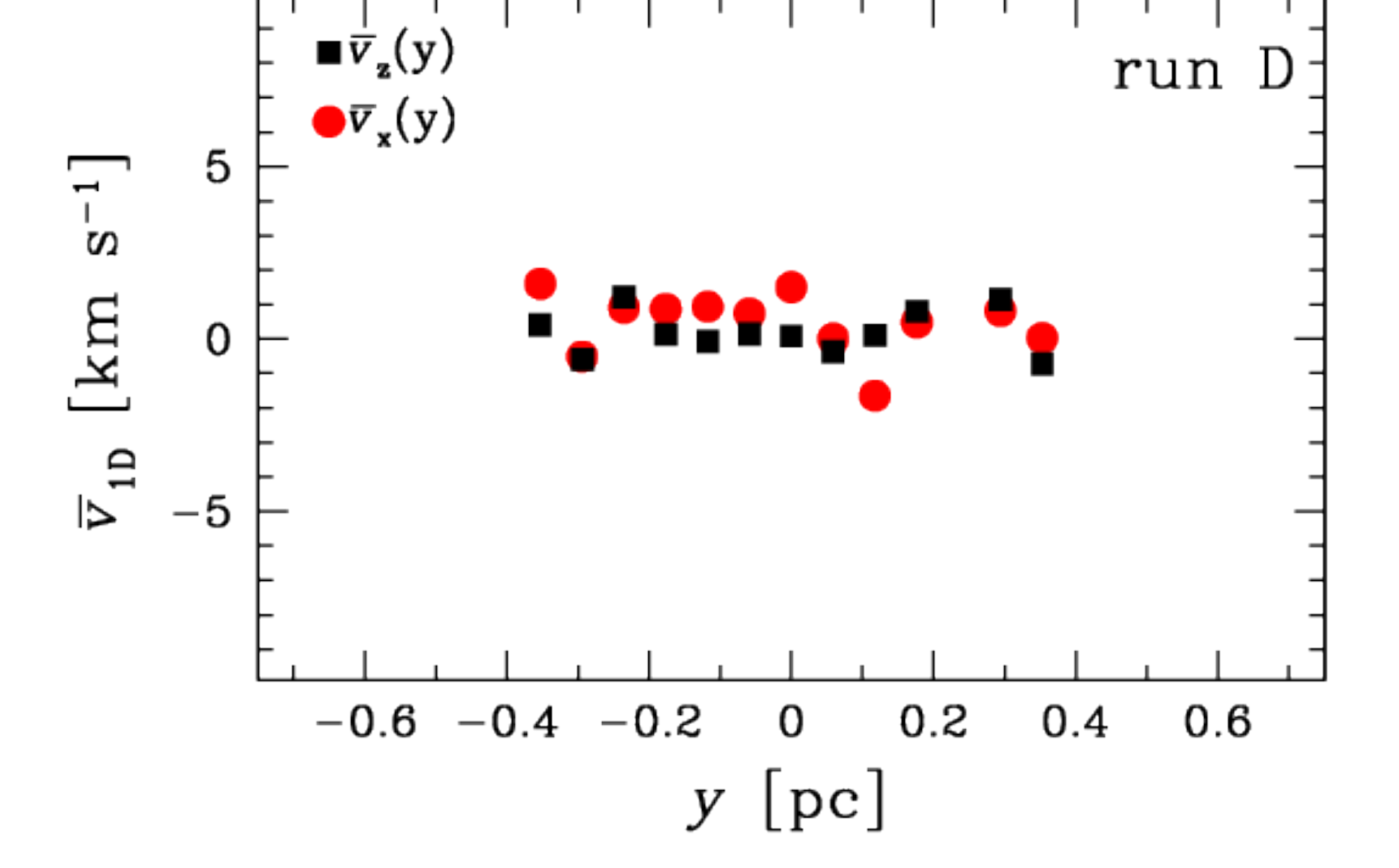,width=6.0cm}
  \caption{  \label{fig:fig8}
Comparison of $\bar{v}_{\rm x}(y)$ (filled red circles) and $\bar{v}_{\rm z}(y)$ (filled black squares) in the simulated star clusters at $t=4$ Myr, where $\bar{v}_{\rm x}(y)$ and $\bar{v}_{\rm z}(y)$ are the velocity along the $x$ axis and the velocity along the $z$ axis  as a function of the position on the $y$ axis, respectively. The values of  $\bar{v}_{\rm x}(y)$  are the same as in Fig.~\ref{fig:fig7}. Both $\bar{v}_{\rm x}(y)$ and $\bar{v}_{\rm z}(y)$ are averaged over  stars with position $y-\Delta{}y/2\le{}y_i<y+\Delta{}y/2$ (where $y$ and $\Delta{}y$ are the middle point and size of the bin, respectively). From top to bottom: run~A, B, C, and D. This Figure shows that the rotation feature is more important than subvirial collapse in all runs but cloud~D. 
}}
\end{figure}

Indeed,  we expect that the effect of subvirial collapse is present even in our simulated star clusters, because they  are subvirial and elongated (Table~\ref{tab:tab2}). The residuals of blue-shifted and red-shifted structures in the colour-coded map of the velocity along the angular momentum vector ($v_{\rm z}$, right-hand panels  of Fig.~\ref{fig:fig5}) are the demonstration of this. However, these residuals are much weaker than the signature of rotation in the plane perpendicular to the angular momentum vector (left-hand and central panels of Fig.~\ref{fig:fig5}). Moreover, Fig.~\ref{fig:fig8} compares  $\bar{v}_{\rm x}(y)$ (i.e. a component of the velocity perpendicular to the angular momentum vector) with $\bar{v}_{\rm z}(y)$ (i.e. the component of the velocity parallel to the angular momentum vector). A velocity gradient (from red-shifted to blue-shifted) is clearly visible only in the $\bar{v}_{\rm x}(y)$ component. 
Thus, genuine rotation is stronger than the effect of subvirial collapse in our runs A, B, and C.


\section{Discussion and {\it caveats}}
Our simulations indicate that natal rotation must be a common feature in massive embedded star clusters. However, there are several possible issues. First, we might wonder whether the simulations are affected by any bias, which makes rotation stronger.  The absence of stellar feedback in our simulations is a limitation for several reasons. Feedback is expected to quench star formation at some point. This affects the final mass of the stellar cluster. The local star formation efficiency (SFE) in our runs is  $\gtrsim{} 50$ per cent; we expect SFE to be significantly lower, if feedback becomes important on a time shorter than (or comparable to) the star formation timescale.  

Also, we expect that the impact of feedback on rotation depends on its timescale. The first supernovae occur when the age of the massive stars is $t_\ast\gtrsim{}3$ Myr, depending on the stellar mass. Similarly, stellar winds by massive stars are particularly strong in the last stages of their life \citep{Chen2015}.  In our simulations, rotation is already well developed when $t_\ast{}\sim{}1-2$ Myr (see Table~\ref{tab:tab2}). Thus, the onset of rotation should not be affected by stellar winds and supernovae. In contrast, photoionization and outflows from protostars affect the evolution of gas even at earlier stages. In particular, outflows from young stars are very common (see e.g. \citealt{bally2016} for a review). They inject energy and momentum into their surroundings. This might counteract the inflow of gas toward the star cluster, reducing the amount of rotation. These possible feedback effects suggest that the minimum star cluster mass to have significant rotation might be higher than what we find in our simulations. Thus, it is essential to include feedback in follow-up simulations. 


Another crucial ingredient that is missing in our simulations is the magnetic field. The effect of magnetic field is known to be crucial on the scale of single cores, where magnetic braking can remove  the angular momentum of the cloud. A very efficient magnetic braking can even prevent the formation of rotationally supported circumstellar discs \citep{galli2006,lizano2015}. Studies on the effects of magnetic fields on star cluster scales (e.g. \citealt{federrath2016}) do not address the issue of rotation. However,  \cite{leehennebelle2016a} include a weak ($3-8\,{}\mu{}{\rm G}$) magnetic field in their simulations and still see a clear feature of rotation in the gas, even if they do not discuss the rotation curve of stars. Forthcoming studies of star cluster rotation including both stellar feedback and magnetic fields are needed to address this topic in detail.

 An additional {\it caveat} we must mention is the importance of initial conditions.  Several properties of the simulated star clusters (e.g. mass function and spatial distribution of stars) are known to depend on the initial conditions, the initial density profile of gas and the initial spectrum of turbulence being among the crucial ingredients \citep{girichidis2011}. Thus, it is important to consider various initial conditions and to check their impact on star cluster rotation.

An important question raised by our results is how long the rotation signature lasts in star clusters. The expulsion of the residual gas, the galactic tidal field, and the dynamics of stars (e.g. two-body relaxation, close encounters between stars and binaries, mass segregation, core collapse, adiabatic cluster expansion) all might affect and possibly quench rotation.  Rotation was found to accelerate the dynamical evolution of a star cluster through the transfer of angular momentum outward: mass segregation tends to be faster, and the core collapse occurs at earlier times \citep{kim2002,kim2004,kim2008}. Under some circumstances, the gravo-gyro catastrophe might occur: the most massive stars, segregated in the cluster core, speed up and rotate faster than the rest of the cluster \citep{hachisu1979,einsel1999,kim2004}. Another important consequence of rotation is that stars with high angular momentum (in the tail of the Maxwellian velocity distribution) are more likely to escape from the star cluster with respect to low angular-momentum stars \citep{agekian1958,shapiro1976}. This leads to a removal of angular momentum from the star cluster,  progressively erasing rotation. 

At the end of our simulations, the gas is nearly exhausted in the central regions of the main star clusters, with a local SFE $>0.5$, in agreement with previous studies \citep{bonnell2011,kruijssen2012a}. Thus, the star clusters are mainly gas-free and their evolution is driven by two-body encounters between stars. The two-body relaxation timescale can be expressed as \citep{portegieszwart2010}
\begin{equation}\label{eq:twobody}
t_{\rm rlx}\sim{}0.5\,{}{\rm Myr}\,{}\left(\frac{M_\ast}{8000\,{}{\rm M}_\odot}\right)^{1/2}\,{}\left(\frac{r_{\rm hm}}{0.1\,{}{\rm pc}}\right)^{3/2}.
\end{equation}
Since $t_{\rm rlx}$ is short, the star cluster will evolve very fast  because of  two-body relaxation and, possibly, binary-single star encounters.  Two-body relaxation causes the angular momentum to diffuse outward, removing it from the system \citep{kim2002}. Moreover,  encounters between single stars and binary systems tend to isotropise the distribution of angular momentum \citep{mapelli2005}. This might be one of the reasons why the star clusters tend to become rounder at the end of the simulations.  Another possible explanation for the loss of flattening is the accretion of new gas and stellar clumps, with a different orientation of angular momentum.

Given these considerations, we expect that stellar dynamics tends to suppress the natal rotation. Thus, it is of foremost importance to investigate the impact of stellar dynamics on the newly born star clusters. However, the adopted tree code (employing a second-order leap-frog scheme) is not particularly suited to investigate close stellar encounters in detail. Only few studies attempt to trace the dynamical evolution of star clusters born from hydrodynamical simulations of molecular clouds \citep{fujii2015a,mcmillan2015,fujii2015b,fujii2016}, because the problem is a numerical challenge (e.g. \citealt{pelupessy2012,hubber2013}). 
In a forthcoming study, we will trace the dynamics of stars after gas expulsion, adopting a suitable direct-summation $N-$body code.

Comparing our results against observations of  young star clusters would be extremely useful to constrain the hierarchical model of star cluster formation.  Unfortunately, a direct comparison between our simulations and the available data is complex for several reasons. First, our simulated star clusters are extremely young embedded objects: in Tab.~\ref{tab:tab2}  we show that $t_\ast{}\sim{}1-2$ Myr at $t=4$ Myr since the beginning of the simulations.  $t_\ast$ is defined as the average time elapsed since the formation of sink particles. The `true' stellar age might be even smaller than $t_\ast$, because the sink particle algorithm cannot distinguish between pre-stellar cores and stars. Moreover, our simulated star clusters are very dense ($>10^4$ M$_\odot$ pc$^{-3}$) and compact ($r_{\rm hm}<0.1$ pc). Only few star clusters with comparable age and density  exist in the Milky Way and in the Magellanic Clouds  (e.g. \citealt{portegieszwart2010}). Moreover, the relatively low line-of-sight velocities ($\sim{}5-10$ km s$^{-1}$) require a very high spectral resolution. Undetected binaries significantly complicate the analysis. For these reasons, observations are rather scanty. 

The youngest massive star cluster for which evidence of rotation was found is R136 in the Large Magellanic Cloud \citep{henault2012}. Our simulated rotation curves (Fig.~\ref{fig:fig7}) are qualitatively similar to that of R136  \citep{henault2012}. However, our simulated star clusters are less massive\footnote{The most massive star cluster in our simulations, run~B, has a maximum mass of $\sim{}10^4$ M$_\odot$, while R136 has a current mass of $\sim{}10^5$ M$_\odot$ \citep{andersen2009}.} and much more compact than R136 (they have not yet started the expansion phase after gravitational collapse).  Moreover, the sample of \cite{henault2012} consists of only 36 apparently single O-type stars, of which only 16 are within 5 pc. Future integral field  spectrographs on 30m-class telescopes (e.g. HARMONI at E-ELT) are needed to make a difference, thanks to their spatial and spectral resolution.

Other young star clusters and star forming regions (e.g. $\rho{}$ Ophiuchi, \citealt{rigliaco2016}; Cha~I, \citealt{tsitali2015}; NGC~1333, \citealt{foster2015}; IC~348 \citealt{cottaar2015}) show velocity gradients  that might be interpreted as radial collapse of subvirial structures, or rotation, or a combination of both. The masses of most such clusters ($\sim{}100-500$ M$_\odot$) are similar  to that of the simulated star cluster in run~D, where we do not see a clear signature of rotation.



 To significantly increase the sample of `real-life' star clusters we can compare our results with, we need to simulate more massive clusters, and/or to evolve them for (at least) several tens of Myr. This can be done only switching to a direct-summation N-body simulation, which treats stellar dynamics in the proper way (this point will be addressed in a follow-up study).

In this respect, our results might be useful to interpret the origin of rotation in the old globular clusters of the Milky Way (e.g. \citealt{bellazzini2012}), but it must be kept in mind that our simulated star clusters are extremely different from globular clusters, as to mass, star formation history, and age. 
To understand whether gas torques can produce the rotation observed in globular clusters, we must focus on larger star cluster masses, and we must account for the extreme conditions of the early Universe (e.g. \citealt{trenti2015,ricotti2016}). Moreover, it is essential to trace the dynamical evolution of simulated star clusters for several Gyr, to constrain what fraction of the pristine rotation can survive to present time.

 An alternative approach to compare our simulations with observational data is to look for rotation in the velocity field of dense gas in massive cluster-forming regions. For example, \cite{liu2015} present Atacama Large Millimeter Array (ALMA) data of the rotating massive molecular clump G33.92+0.11, which contains two central massive ($\sim{}100-300$ M$_\odot$) molecular cores.  \cite{liu2015} interpret the velocity field of G33.92+0.11 as possibly dominated by infall motions on large scale ($\gtrsim{}1$ pc) and by rotational motion on the small scale ($\lesssim{}0.3$ pc). Their figure~6 is quite reminiscent of our Fig.~\ref{fig:fig2}. Other proto-clusters show marginally centrifugally supported accretion flows (e.g. \citealt{galvan2009,liu2010,cesaroni2011}), suggesting that the central $<1$ pc regions of molecular clouds might be dominated by angular momentum. Future ALMA observations will provide invaluable information about rotation in molecular clouds and proto-star clusters.




\section{Summary}
We  ran SPH simulations of turbulence-supported molecular clouds, and we studied the kinematics of the main cluster that forms in each cloud. In agreement with previous studies (e.g. \citealt{bonnell2003,bonnell2011,girichidis2011}), we find that the star clusters assemble hierarchically, by the merger of several gaseous and stellar clumps. The gas component shows significant rotation in the region where the  star cluster assembles (Fig.~\ref{fig:fig2}). This rotation feature arises from large-scale torques in the cloud and from angular momentum conservation.

The stellar component of the embedded star cluster inherits the rotation  from the parent gas (Figs.~\ref{fig:fig3}--\ref{fig:fig6}). Rotation is apparent in the simulated star clusters (Fig.~\ref{fig:fig7}), if their  final stellar mass is $M_\ast{}\gtrsim{}1000$ M$_\odot$. 

Our simulated star clusters form with large ellipticity ($\epsilon{}\sim{}0.70-0.75$ at $t\sim{}1.2$ $t_{\rm ff}$) and are markedly  subvirial ($Q_{\rm vir}=0.2-0.3$  at $t\sim{}1.2$ $t_{\rm ff}$). During their evolution, the simulated star clusters tend to become rounder ($\epsilon{}\sim{}0.4-0.5$ at $t\sim{}2$ $t_{\rm ff}$) and to approach virial equilibrium ($Q_{\rm vir}=0.3-0.4$  at $t\sim{}2$ $t_{\rm ff}$), but their rotation signature is still apparent (Fig.~\ref{fig:fig7}). We find that rotation is more important than radial motions  caused by subvirial collapse (Fig.~\ref{fig:fig8}). This result is a key test to probe the hierarchical formation scenario of star clusters, and might be useful to interpret the observed rotation  signature in young massive star clusters (R136, \citealt{henault2012}) and in old globular clusters (e.g. \citealt{bellazzini2012}). 

\section*{Acknowledgments}
I thank the anonymous referee for their invaluable comments, which helped me improving the manuscript significantly. I warmly thank Francesco Palla for his invaluable suggestions. I also thank Germano Sacco, Elena Sabbi, Hauyu Baobab Liu, Giacomo Beccari, Angela Bragaglia, Eugenio Carretta, Enrico Vesperini, Anna Lisa Varri and Emanuele Ripamonti for useful discussions. The simulations were performed with the  GALILEO cluster at CINECA (through CINECA Award N. HP10CRWH91) and with the SCIGHERA cluster at INAF-Osservatorio Astronomico di Padova.   I acknowledge the CINECA Award N. HP10CRWH91 for the availability of high performance computing resources. I thank the authors of gasoline, especially J. Wadsley, T. Quinn and J. Stadel. I thank Tristan Hayfield for allowing me to use his implementation of cooling for {\sc gasoline}.  To analyze simulation outputs, we made use of the software {\sc tipsy}\footnote{\tt http://www-hpcc.astro.washington.edu/tools/tipsy/\\tipsy.html} and {\sc tipgrid}\footnote{\tt http://www.astrosim.net/code/doku.php?id=home:code:analysistools:misctools}. MM acknowledges financial support from the Italian Ministry of Education, University and Research (MIUR) through grant FIRB 2012 RBFR12PM1F, from INAF through grant PRIN-2014-14 (Star formation and evolution in galactic nuclei), and from the MERAC Foundation.

\appendix
\section{Impact of stochasticity  on low-mass clouds}\label{appendix:app1}
\begin{figure}
  \center{
    \epsfig{figure=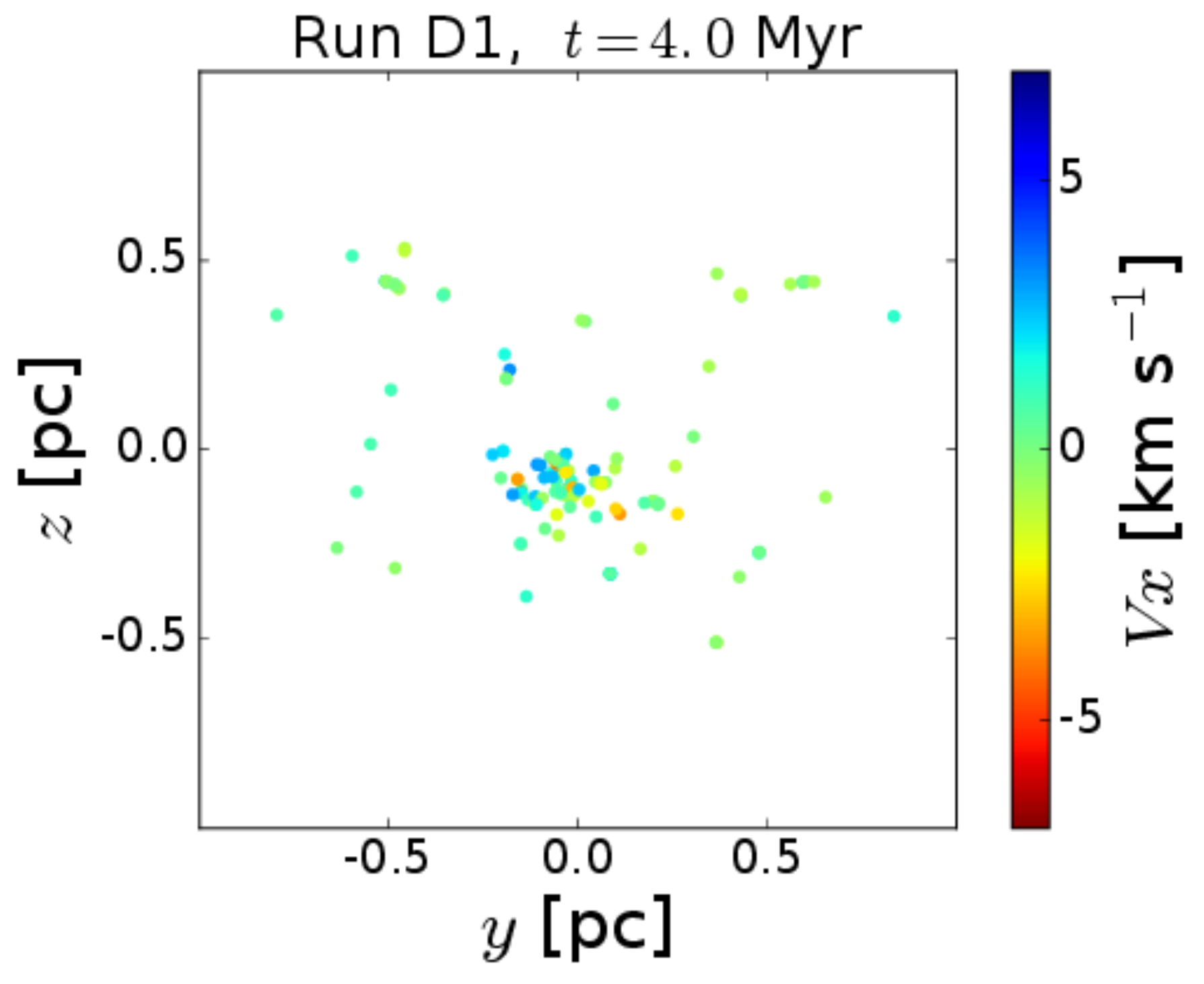,width=6.0cm}
    \epsfig{figure=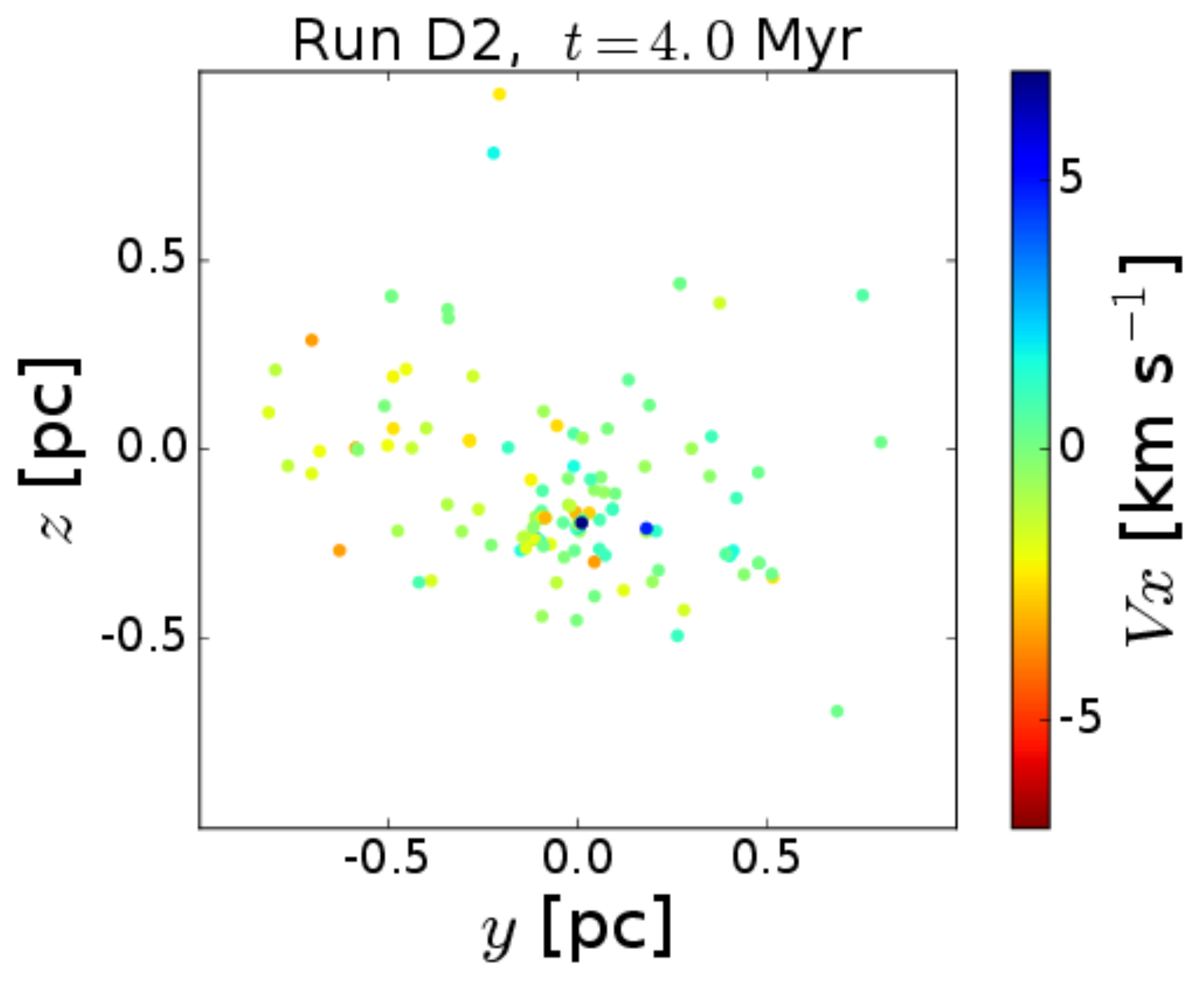,width=6.0cm}
    \epsfig{figure=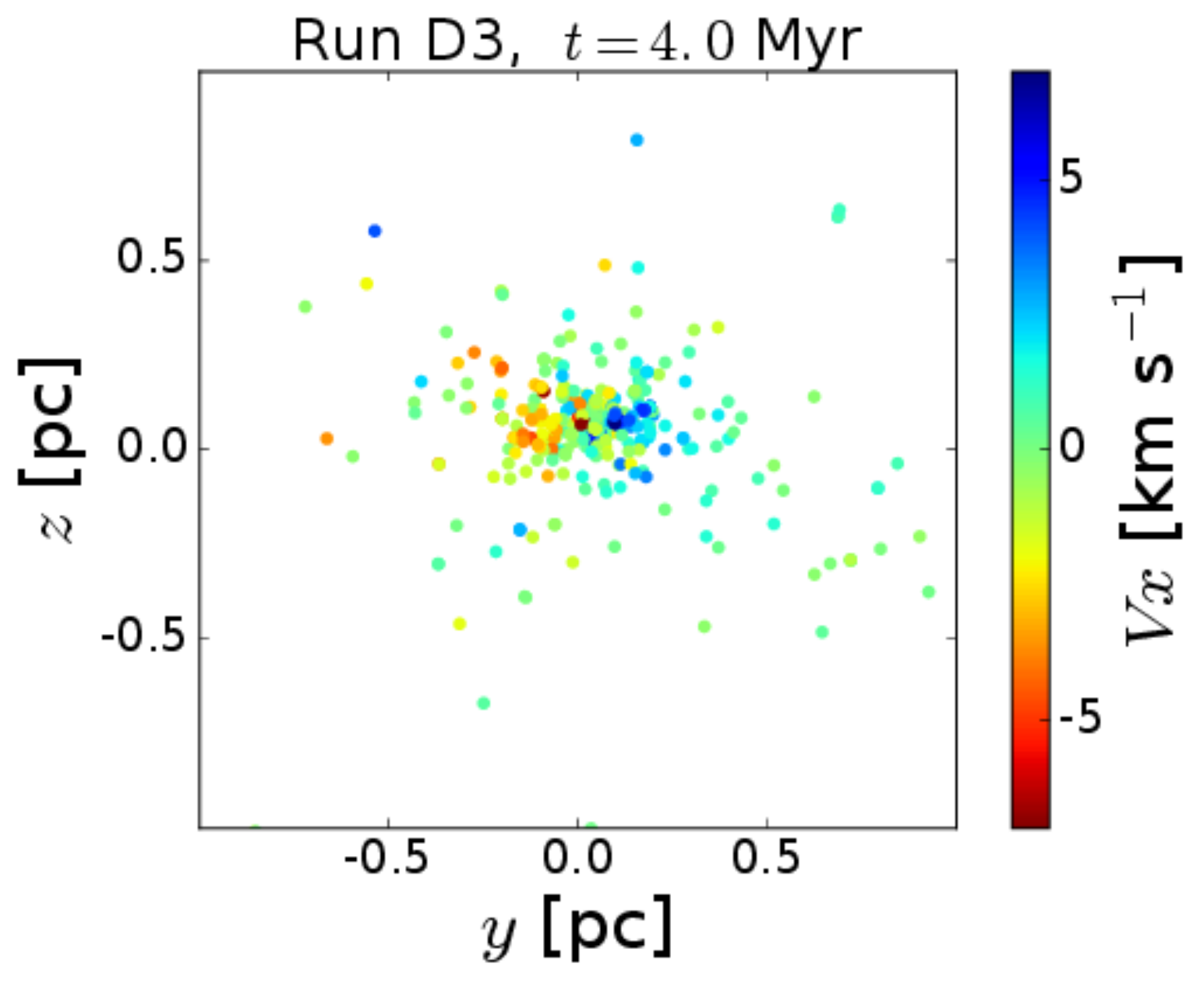,width=6.0cm}
    \epsfig{figure=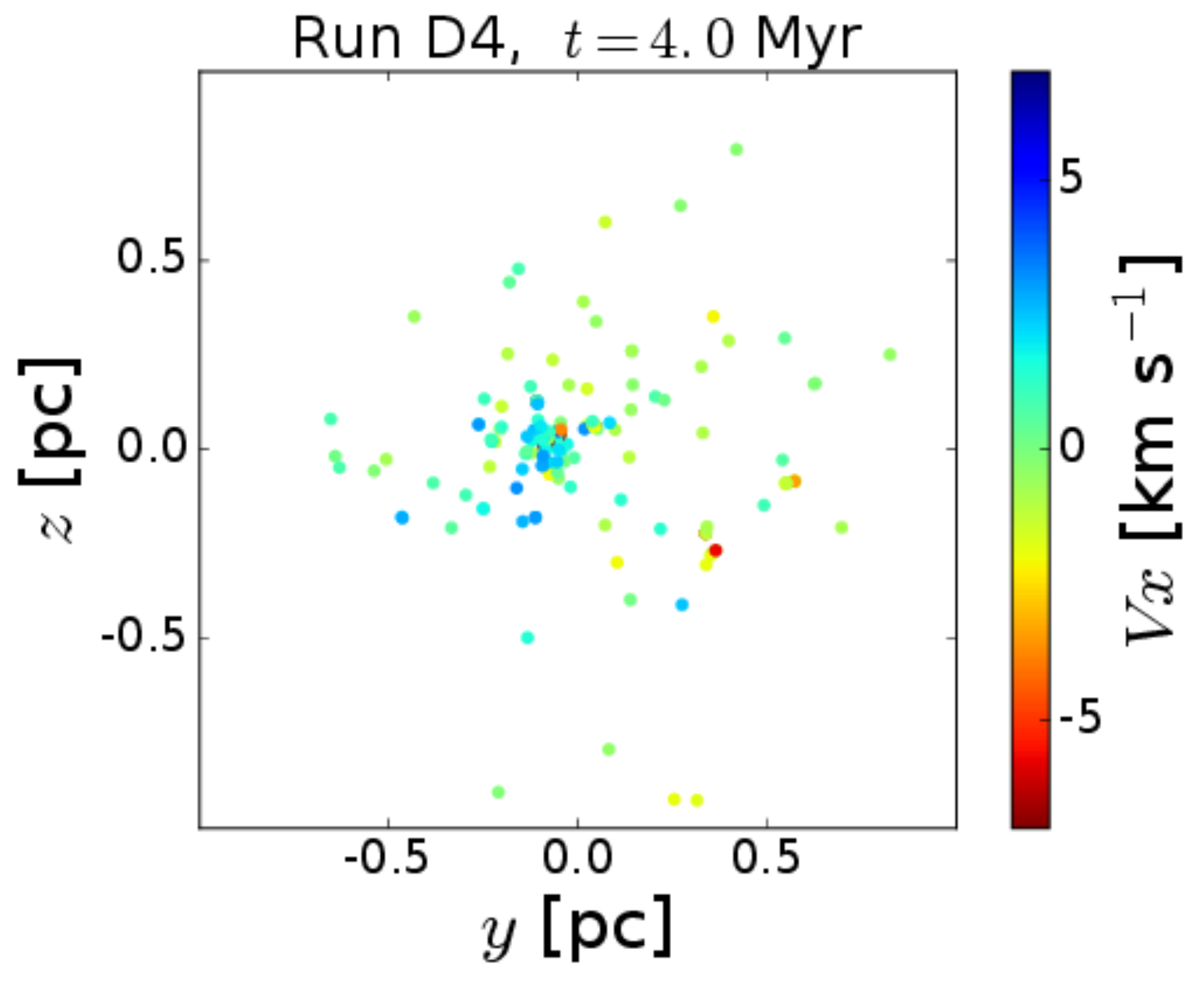,width=6.0cm}
  \caption{  \label{fig:figA1}
Same as Fig.~\ref{fig:fig6} but (from top to bottom) for runs D1, D2, D3, and D4  at $t=4$ Myr. 
}}
\end{figure}

 Stochastic fluctuations might affect our results, especially for the low-mass cloud (run~D). To check the influence of stochasticity we ran four additional realizations of run~D, i.e. four additional molecular clouds with $M=1.7\times{}10^3$ M$_\odot$, $R=3$ pc, and $N_{\rm gas}=2\times{}10^6$, but with different random seeds for the turbulence. We call these four realizations runs~D1, D2, D3, and D4.

Figure~\ref{fig:figA1} is the same as Fig.~\ref{fig:fig6} but for runs D1, D2, D3, and D4. The colour-coded map of Fig.~\ref{fig:figA1} shows the component of the stellar velocity along the $x$ axis at $t=4$ Myr. With respect to the $yz$ plane, the velocity along the $x$ axis ($v_{\rm x}$), shown by the colour-coded map, can be interpreted as the line-of-sight velocity. Only run~D3 shows some signature of rotation in Figure~\ref{fig:figA1}.

\begin{figure*}
  \center{
    \epsfig{figure=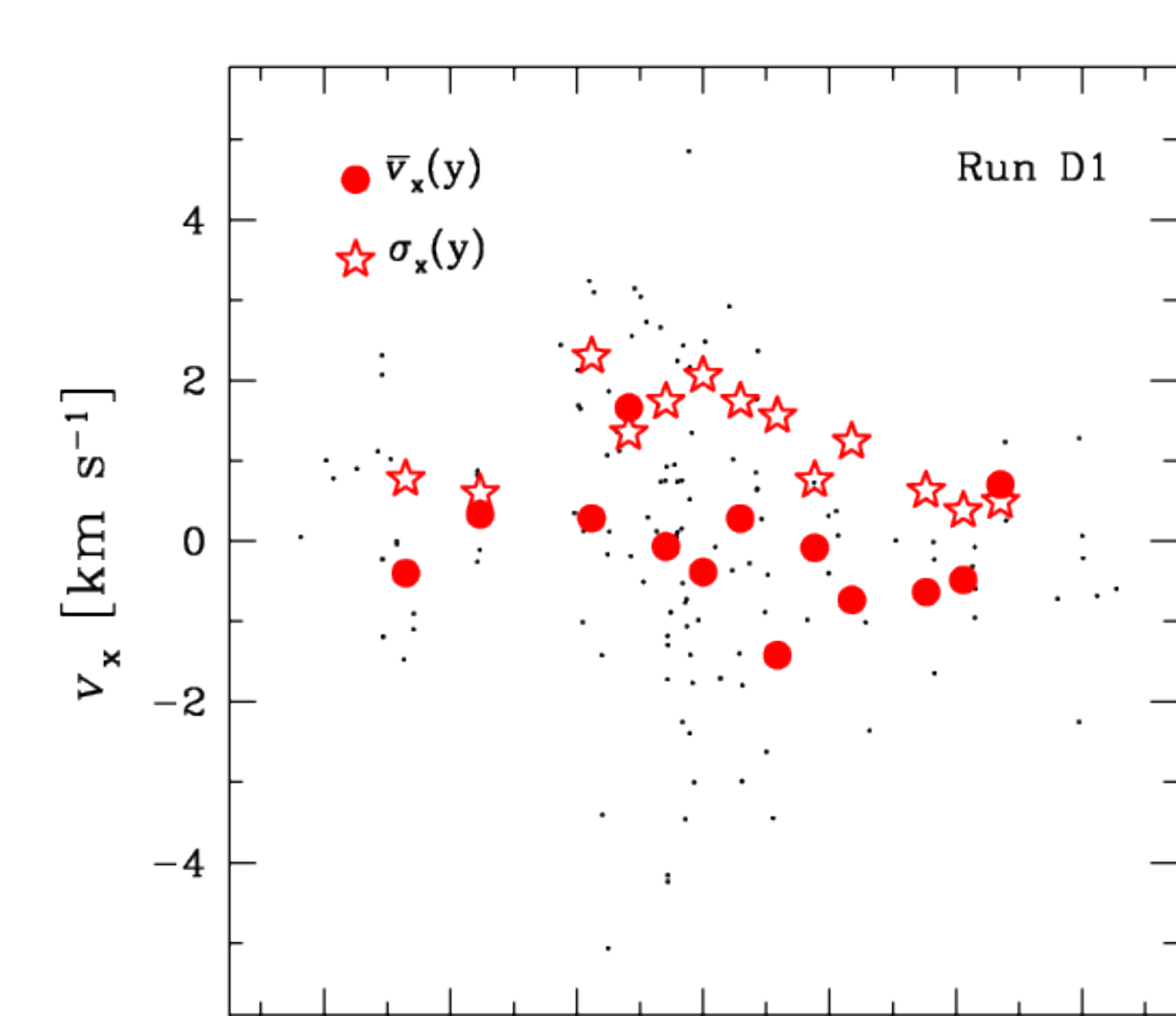,height=6.0cm}
    \epsfig{figure=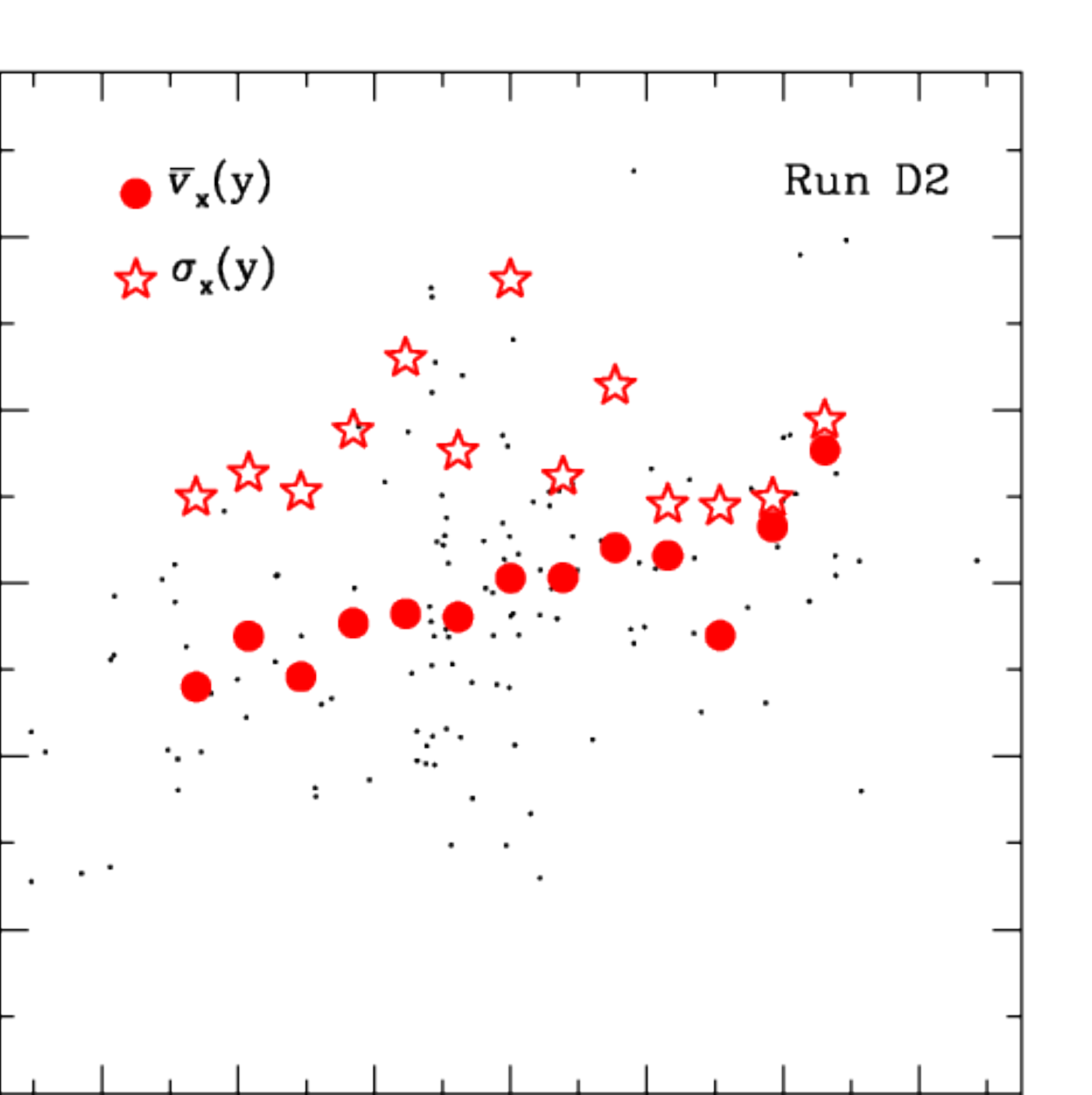,height=6.0cm}
    \epsfig{figure=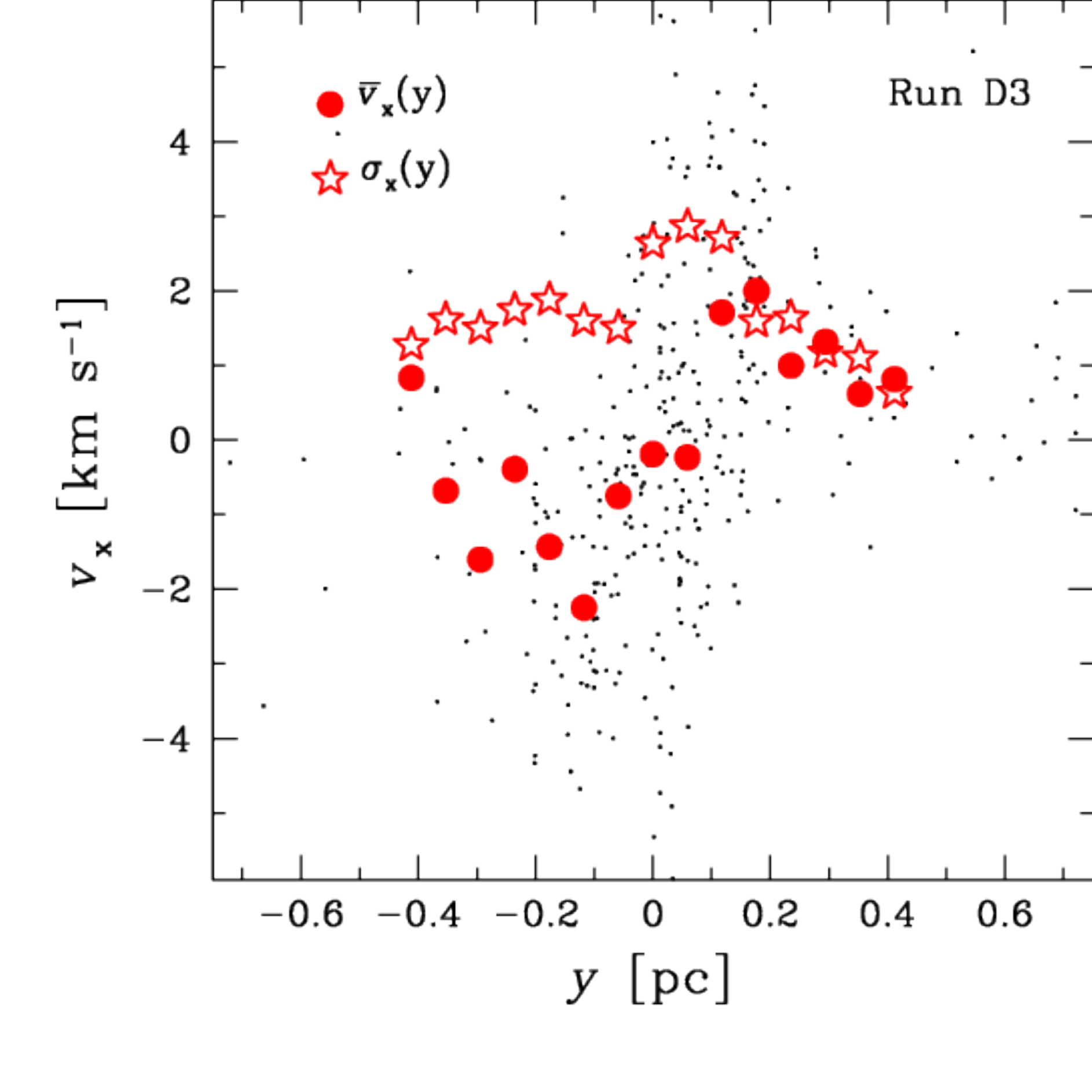,height=6.95cm}
    \epsfig{figure=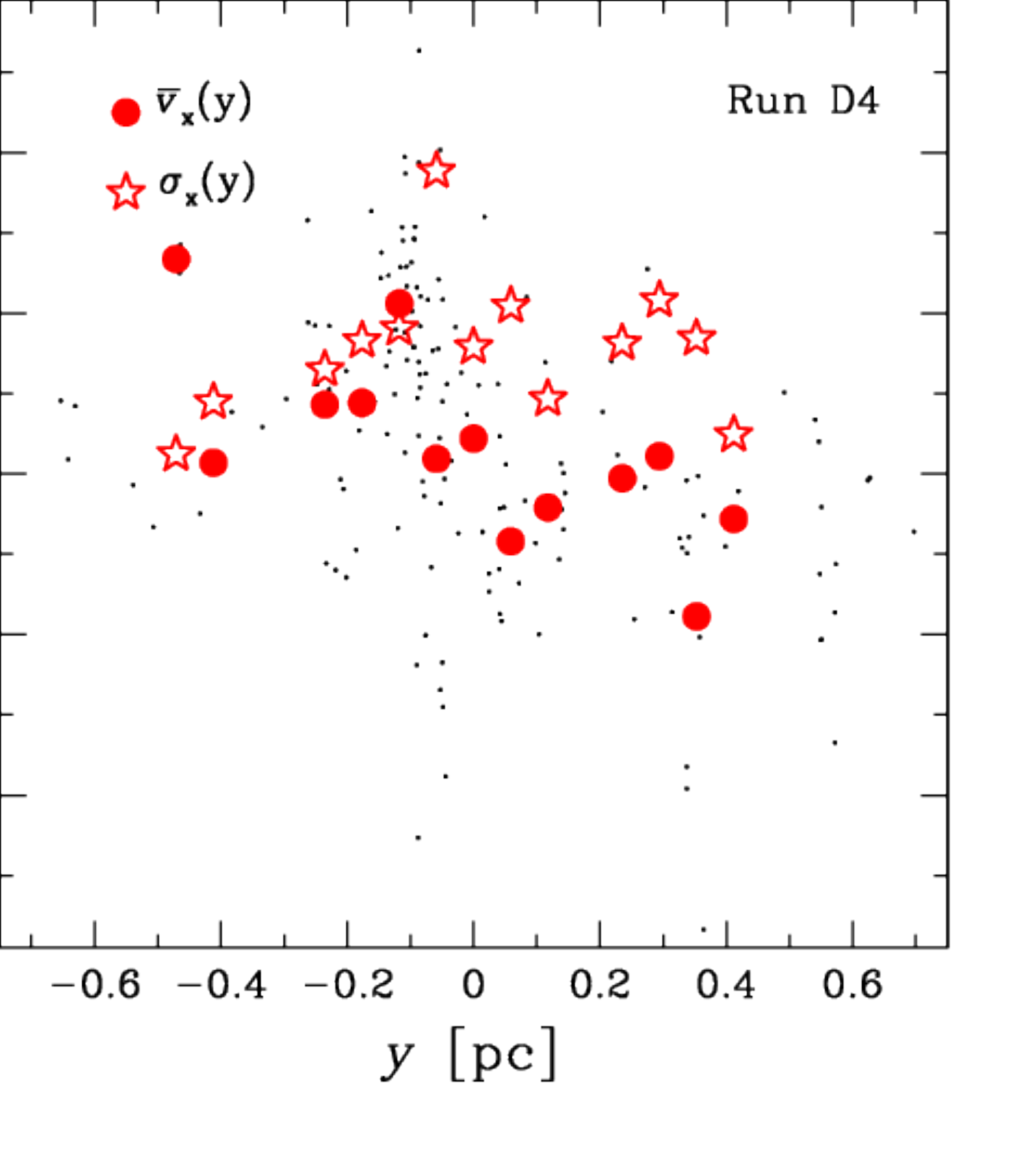,height=6.95cm}
  \caption{  \label{fig:figA2}
Same as Fig.~\ref{fig:fig7} but for runs D1 (top left), D2 (top right), D3 (bottom left), and D4 (bottom right) at $t=4$ Myr. 
}}
\end{figure*}

Fig.~\ref{fig:figA2} is the same as Fig.~\ref{fig:fig7} but for runs D1, D2, D3, and D4, i.e. it shows the rotation curve of the four test runs at $t=4$ Myr. Again, only the rotation curve of run~D3 indicates that the star cluster rotates. The rotation curve of run~D2 is more suggestive of a velocity gradient, whereas the rotation curves of runs~D1 and D4 are very similar to that of run~D (Fig.~\ref{fig:fig7}).

Table~\ref{tab:tabA1} (which is the same as Table~\ref{tab:tab2} but for the test runs) gives some useful hints to interpret Figures~\ref{fig:figA1} and \ref{fig:figA2}. The total mass of a star cluster ($M_\ast$) varies considerably from one realization to the other, indicating that stochastic fluctuations are indeed significant. In particular, the star cluster formed in run~D is also the less massive one among the five different realizations of the same cloud. In contrast, the star cluster in run~D3 is by far the most massive one: its mass ($\sim{}800$ M$_\odot$) is nearly three times larger than that of the star cluster in run~D. This result confirms that the amount of natal rotation strongly correlates with the total mass of the star cluster. Moreover, this comparison also confirms that star clusters with final mass $<$ few $\times{}100$ M$_\odot$ do not show clear indication of rotation. Rotation becomes significant only in star clusters with mass $\gtrsim{}1000$ M$_\odot$.  

\begin{figure}
  \center{
    \epsfig{figure=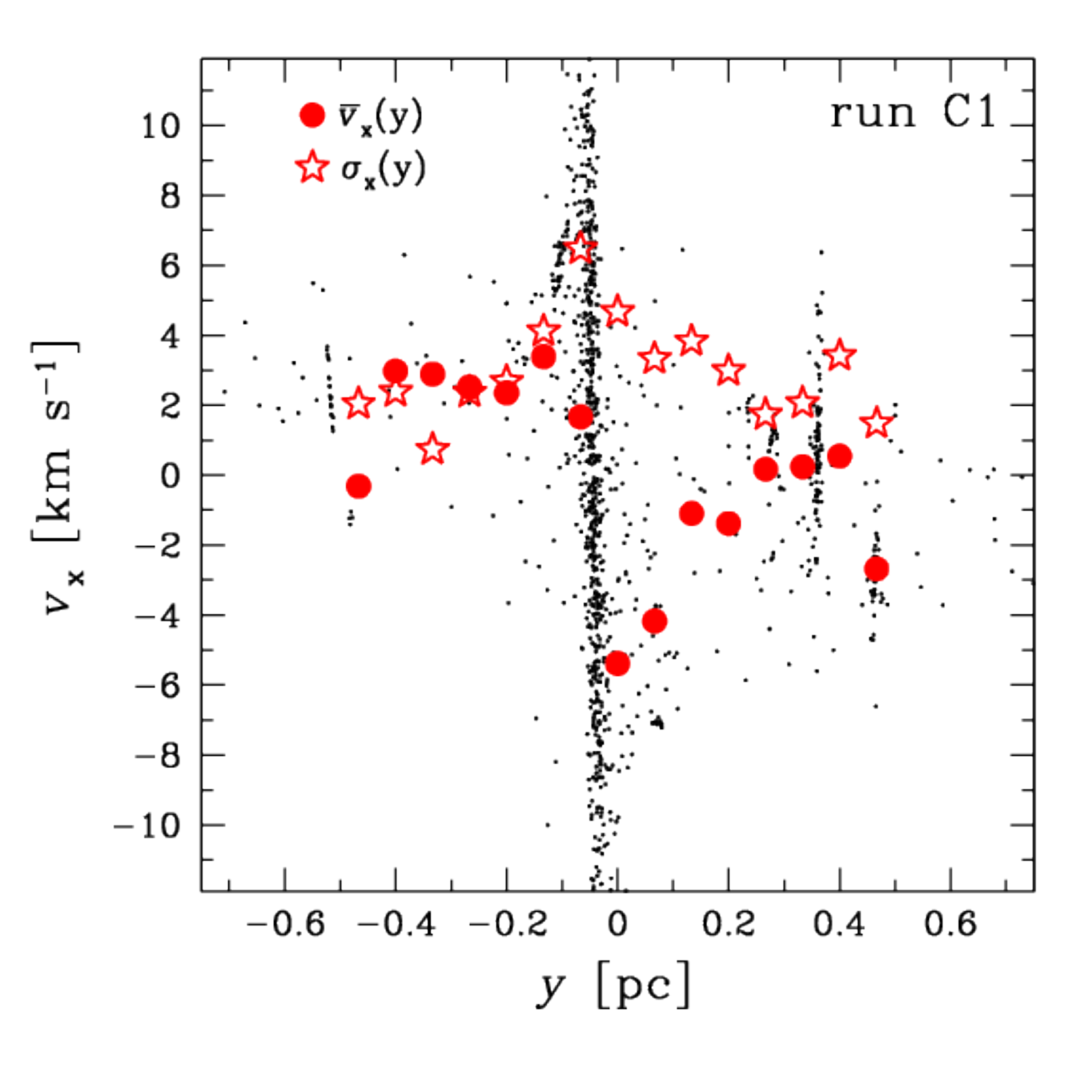,width=6.5cm}
  \caption{  \label{fig:figA3}
Same as Fig.~\ref{fig:fig7} but for run C1 at $t=4$ Myr. 
}}
\end{figure}

Finally, we expect stochastic fluctuations to be less important in run~C, because the number of particles (the mass of the cloud) is  a factor of $\sim{}5$ ($\sim{}6$) larger than it is in run~D. To check this, we ran a second realisation of run~C, called run~C1. Table~\ref{tab:tabA1} shows the comparison between runs~C and C1. The mass of the main star cluster differs by only $\sim{}10$ per cent between run~C and run~C1. The other properties of runs~C and C1 shown in Table~\ref{tab:tabA1} are also similar. The rotation signature of the main star cluster of run~C1 is apparent from Fig.~\ref{fig:figA3}, consistent with what we found for run~C.  
\begin{table*}
\begin{center}
\caption{\label{tab:tabA1} Comparison of runs~C and D with the additional test realizations.} \leavevmode
\begin{tabular}[!h]{lllllllllll}
\hline
Run
&
$t$ 
& 
$t_\ast$ 
&
$M_\ast$ 
&
$r_{\rm hm}$ 
&
$Q_{\rm vir}$
&
$r_{1},$ $r_{2},$ $r_{3}$ 
&
$\epsilon{}$
&
$\sigma_{\rm x}$ 
&
${\langle{}|v_{\rm x}|\rangle{}}/{\sigma{}_{\rm x}}$ 
& $L_{\rm z}$ \\
&
[Myr]
&
[Myr]
&
[M$_\odot{}$]
&
[pc]
&
&
[pc], [pc], [pc]
&
&
[km s$^{-1}$]
&
& [M$_\odot$ km$^2$ s$^{-1}$] \\
\hline
D & 4.0 & $0.7\pm{}0.5$ & 277   & 0.074 & 0.16  & 0.51, 0.41, 0.34    & 0.51 & 3.5   & 0.70 & $6.5\times{}10^{14}$ \\
D1 & 4.0 & $0.9\pm{}0.6$ & 402   & 0.28 & 0.36  & 1.21, 0.58, 0.41    & 0.66 & 1.6   & 0.75 & $3.5\times{}10^{15}$ \\
D2 & 4.0 & $0.8\pm{}0.5$ & 379   & 0.37 & 0.31  & 1.88, 0.54, 0.48    & 0.75 & 2.1   & 0.69 & $2.1\times{}10^{15}$ \\
D3 & 4.0 & $1.1\pm{}0.6$ & 775   & 0.18 & 0.44  & 0.93, 0.68, 0.58    & 0.37 & 2.4   & 0.75 & $6.1\times{}10^{15}$ \\
D4 & 4.0 & $1.1\pm{}0.6$ & 529   & 0.11 & 0.40  & 0.51, 0.44, 0.40    & 0.22 & 2.5   & 0.72 & $2.8\times{}10^{15}$ 
\vspace{0.1cm}\\
C  & 4.0 & $1.3\pm{}0.7$ & 2057  & 0.070 & 0.32  & 0.65, 0.47, 0.40    & 0.39 & 5.0   & 0.76 &  $1.2\times{}10^{16}$ \\
C1 & 4.0 & $1.4\pm{}0.5$ & 1850  & 0.064 & 0.36  & 0.53, 0.41, 0.16    & 0.69 & 5.8   & 0.78 &  $1.6\times{}10^{16}$ \\
\hline
\end{tabular}
\end{center}
\begin{flushleft}
\footnotesize{Same as Table~\ref{tab:tab2} but for runs D, D1, D2, D3, D4, C, and C1.
}
\end{flushleft}
\end{table*}

\bibliography{./bibliography}
\end{document}